\documentclass[11pt]{article}
\usepackage{amsmath, amssymb, amscd, amsthm, amsfonts}
\usepackage{graphicx}
\usepackage{float}
\usepackage{hyperref}
\usepackage[dvipsnames]{xcolor}
\usepackage{tcolorbox}
\usepackage{bm}
\usepackage[mathscr]{euscript}
\usepackage{mathrsfs}
\usepackage{caption}
\usepackage{subcaption}
\usepackage[normalem]{ulem}
\usepackage[style=nature, citestyle=numeric-comp, sorting=none, backend=biber, maxnames = 5]{biblatex}
\newcommand{\rd}{{\rm d}}
\newcommand{\e}{{\cal E}}
\newcommand{\s}{{\cal S}}

\newcommand{\vt}[1]{{\small \bf #1}} 
\newcommand{\Mp}{M_{\rm Pl}} 
\newcommand{\Tr}{{\rm Tr}}

\newcommand{\rred}{\hat{\rho}_{\rm red}}
\newcommand{\Ri}{\tilde{\rho}_{\rm red}}
\newcommand{\dg}{\dagger}

\newcommand{\lh}{{\cal L}}
\newcommand{\zr}{\zeta}
\newcommand{\F}{{\cal F}}
\newcommand{\h}[1]{{\hat #1}}

\newcommand{\tl}[2]{\tilde{#1}_{#2}}
\newcommand{\td}[2]{\tilde{#1}_{#2}^{\dagger}}
\newcommand{\bv}[1]{\overline{#1}}

\begin{document}


{\renewcommand{\thefootnote}{\fnsymbol{footnote}}
        
\begin{center}
{\LARGE The special case of slow-roll attractors in de Sitter:  \\[1.5mm]
Non-Markovian noise and evolution of entanglement entropy} 
\vspace{1.5em}

Suddhasattwa Brahma$^{1}$\footnote{e-mail address: {\tt suddhasattwa.brahma@gmail.com}}, 
Jaime Calder\'on-Figueroa$^{2}$\footnote{e-mail address: {\tt jrc43@sussex.ac.uk}},
Xiancong Luo$^{1}$\footnote{e-mail address: {\tt X.Luo-35@sms.ed.ac.uk}}, and
David Seery$^{2}$\footnote{e-mail address: {\tt D.Seery@sussex.ac.uk}}
\\
\vspace{1.5em}

$^{1}$Higgs Centre for Theoretical Physics, School of Physics and Astronomy,\\ University of Edinburgh, Edinburgh, EH9 3FD, UK\\[2mm]

$^2$Astronomy Centre, University of Sussex, Falmer, Brighton, BN1 9QH, UK\\[2mm]

\vspace{1.5em}
\end{center}
}

\setcounter{footnote}{0}

\begin{abstract}
\noindent We analyse the evolution of the reduced density matrix of inflationary perturbations, coupled to a heavy entropic field via the leading-order term within the Effective Field Theory of Inflation, for two nearly de Sitter backgrounds. We perform a full quantum treatment of the open system and derive a Fokker--Planck equation to describe decoherence and the entanglement structure of the adiabatic perturbations. We find that exotic phenomena, such as \textit{recoherence} and transient \textit{negative} growth of entanglement entropy, appearing for the attractor solution, are absent for the non-attractor background. We comment on the relationship of these to the non-Markovian nature of the system. Finally, we generalise to the case where a few $e$-folds of ultra-slow roll evolution are sandwiched between phases of slow-roll inflation to find its (memory) effects on the curvature perturbation.
\end{abstract}

\section{Introduction}
The
quantum origin of the large-scale structure of the Universe is one of the most striking predictions of inflation.
It links the present-day statistical properties of the Universe with quantum processes
operating during the earliest
stages of its evolution \cite{starobinsky1980new, fang1980entropy, Guth:1980zm, Sato:1981qmu, Linde:1981mu,Albrecht:1982wi,Mukhanov:1982nu}. Moreover, it offers an exciting possibility of finding new physics at high energies beyond the reach of terrestrial colliders. However, unambiguous signatures of this
quantum origin have yet to be detected in the statistical properties
of cosmic structures that are accessible to us. Most recent work in this direction has been based on describing these statistical properties
using cosmological
$n$-point correlation functions \cite{Bernardeau:2001qr,Maldacena:2002vr,,Bartolo:2004if}.
Meanwhile, the possibility that useful information is embedded
within non-unitary evolution of adiabatic perturbation
(which includes the description of how quantum fluctuations
become classical) has remained largely unexplored.

A major reason for this situation is that correlation functions of free,
massless scalars are sufficient to describe the statistical properties
of perturbations produced by slow-roll inflation.
Loop corrections to the tree-level results are under control in the
weakly-coupled
regime~\cite{Maldacena:2002vr,Weinberg:2005vy,Weinberg:2006ac,Seery:2007we,senatore2010loops,Assassi:2012et}.
For a long time, understanding the `quantum-to-classical' transition of these vacuum fluctuations did not involve a discussion of decoherence. Instead, it was understood that curvature fluctuations, after being significantly squeezed outside the horizon, can be interpreted as defining a classical probability distribution. Since the evolution of these adiabatic fluctuations is essentially random after exiting the horizon, their dynamics can be described as a classical, stochastic process.

This description of classicalisation relies on the fact that the commutator between the field creation and annihilation operator decays for super-horizon wavelengths, and is purely a consequence of a gravitational squeezing interaction that appears in the quadratic Hamiltonian. It does not require suppression of the correlations that are characteristic of quantum states. Genuine suppression of such correlations is typically mediated by interaction with an environment, increasing entanglement and effectively transforming an initial pure state into a highly mixed one. Recent efforts to study this process during inflation have adopted tools from out-of-equilibrium physics \cite{calzetta2009nonequilibrium} that capture the non-unitary evolution of a quantum state as it becomes classical. On the one hand, these tools are the same ones needed to study dissipative and diffusive effects such as those appearing in (for example) warm inflation \cite{Berera:1995ie} and stochastic inflation \cite{Starobinsky:1979ty, Morikawa:1989xz,nambu1988stochastic,nambu1989stochastic,kandrup1989stochastic,nakao1988stochastic,mollerach1991stochastic,linde1994big,starobinsky1994equilibrium,Andersen:2021lii}. On the other hand, rapid advancements in quantum computing have driven the application of similar techniques in quantum information theory, with a focus on minimizing or more accurately accounting for the effects of dissipation in laboratory setups. As a result, cosmology—and high-energy theory more broadly—now leverages concepts like complexity \cite{Susskind:2014rva, Nielsen:2005mkt}, entanglement entropy \cite{Ryu:2006bv}, and chaos to probe fundamental aspects of physics. These efforts are further supported by advances in our understanding of how these concepts integrate within the framework of quantum field theory \cite{Balasubramanian:2011wt,Agullo:2024har,Jefferson:2017sdb}.

These tools are especially helpful when treating inflationary perturbations as part of an
open quantum system. The physical reason lies in our ignorance about UV physics
and how adiabatic modes interact with heavy degrees of freedom (``dofs''), in addition to
self-interactions between different modes due---at least---to the non-linear nature of gravity.
The traditional understanding of the effective field theory of inflationary
perturbations~\cite{Cheung:2007st} has been to account for the influence of the heavy
environment by systematically adding operators, order by order, in the local,
unitary effective action for the light modes. By construction this excludes non-unitary
effects such as energy and information exchange which can, however, become important in gravitational
setups~\cite{calzetta2009nonequilibrium,breuer2002theory,Burgess:2022rdo,Colas:2022hlq,Colas:2022kfu,Colas:2023wxa,
Colas:2024lse,Colas:2024xjy,Colas:2024ysu,Salcedo:2024smn,Boyanovsky:2015xoa,Boyanovsky:2015jen,
Boyanovsky:2015tba,hollowood2017decoherence, shandera2018open, Choudhury:2018ppd,Brahma:2020zpk,Brahma:2021mng,
Brahma:2022yxu,Burgess:2022nwu,banerjee2023thermalization,Kaplanek:2022xrr,Cao:2022kjn,Prudhoe:2022pte,Kading:2022hhc,
Kading:2022jjl,Kading:2023mdk,Alicki:2023rfv,Alicki:2023tfz,Creminelli:2023aly,Keefe:2024cia,Bowen:2024emo,
crossley2017effective, Colas:2022hlq, Burgess:2015ajz,Colas:2024xjy,brandenberger1990classical,Calzetta:1995ys,
barvinsky1999decoherence,lombardo2005decoherence,lombardo2005influence,martineau2007decoherence,
prokopec2007decoherence,burgess2008decoherence,sharman2007decoherence,campo2008decoherence,
anastopoulos2013master,Nelson:2016kjm,martin2018non,Oppenheim:2022xjr,DaddiHammou:2022itk,
Sharifian:2023jem,Ning:2023ybc,joos2014decoherence,Burgess:2024eng,Bhattacharyya:2024duw,Colas:2024lse,Burgess:2022nwu,
Micheli:2023qnc,Kaplanek:2019dqu,DaddiHammou:2022itk,hollowood2017decoherence,burgess2008decoherence,
Chandran:2018wwc,Bhattacharya:2022wpe,Bhattacharya:2023twz,Bhattacharya:2023xvd,Sou:2022nsd,Sou:2024tjv}.
To this end, since our main goal in this paper is to better understand entanglement
and decoherence due to diffusive effects during inflation,
an open EFT approach is appropriate. At a technical level, a master equation approach to this problem
is well-suited to treat late-time secular divergences that are present even
for tree-level computations involving massless scalars in de Sitter (dS).
In this sense it is complementary to the dynamical renormalisation group approach~\cite{Boyanovsky:1996rw,Dias:2012qy},
even when the interaction arises from gravitational non-linearities  \cite{Brahma:2021mng}.  

In this paper, we restrict ourselves to the Gaussian sector of the open EFT of inflationary perturbations, keeping the leading order interaction term between the adiabatic ``system'' mode and the entropic ``environment''. Such a setup has been shown to exhibit novel phenomena, namely \textit{recoherence} and \textit{purity-freezing}, when considering  slow-roll inflation \cite{Colas:2022kfu,Colas:2024xjy}. We study the  behaviour of diffusion and dissipation of this system for a non-attractor ultra slow-roll background and compare it to the corresponding attractor solution. The noise and drift terms are identified through the formulation of the Fokker--Planck equation, allowing us to interpret the coefficients of the master equation. It is well-established that the dynamics of this open system is non-Markovian and we study the implications of this for quantum information markers such as entanglement entropy and purity. Most importantly, we show that a na\"ive expectation that the non-attractor background, being closer to pure dS since the evolution takes place on the flatter part of the potential, does \textit{not} reflect similar features in the evolution of these quantities.  Interestingly, slow-roll attractors are found to exhibit recoherence and a transient negative growth in entanglement entropy for a specific range of parameters. We explore the physical reason behind this and start a new line of inquiry regarding the possible (transient) violations of the second law in sub-systems, corresponding to a given Hubble patch, in the early universe. As we shall show, the non-Markovianity of the noise kernel plays a pivotal role in this discussion which necessitates the effect of such corrections for stochastic inflation. Although effects of the background on decoherence have been considered in the past, it has been in the context of contrasting with bouncing cosmological backgrounds \cite{Colas:2024xjy}, and not for different phases of dS expansion.  A few $e$-foldings of ultra-slow roll (USR) evolution, followed by a standard slow-roll (SR) expansion, is nowadays considered a plausible background for abundant production of primordial black holes and here we examine the effect of hidden environments on the adiabatic perturbation in a fully quantum treatment of such models.

\section{A two-field model for inflation}
An effective description of any physical system relies on appropriately separating relevant and irrelevant degrees of freedom. Various criteria can guide this separation, such as significant differences in particle masses or the typical timescales of their evolution. In the context of inflationary cosmology, current cosmic surveys, particularly observations of the cosmic microwave background (CMB), provide strong evidence that the primordial spectrum is predominantly sourced by adiabatic fluctuations rather than isocurvature modes. Consequently, it is logical to associate the observable degrees of freedom---the system---with the adiabatic sector, while the environment consists of fields in the entropic (hidden) sector.

We restrict attention to the leading-order interactions, corresponding to quadratic mixing between the adiabatic
mode $\zr$ and some isocurvature mode $\F$.
An economical implementation of this scenario is given by the two-field Lagrangian 
\begin{equation}
    {\cal L} = a^2 \epsilon \Mp^2 \zr'^2 - a^2 \epsilon \Mp^2 (\partial_i \zr)^2 + \frac{1}{2}a^2 \F'^2 - \frac{1}{2}a^2 (\partial_i \F)^2 - \frac{1}{2} m^2 a^4 \F^2 
    + \lambda a^3 \sqrt{2\epsilon} \Mp \zr' \F\;,
\end{equation}
Here, $a$ denotes the scale factor, $\tau$ the conformal time, and $\epsilon \equiv -\dot{H}/H^2$ is the slow-roll parameter.
We denote the Hubble parameter as $H$.
Although Lagrangians with this structure can emerge from specific two-field models (see e.g., Refs.~\cite{Tolley:2007nq,Chen:2009zp}), or more generally in effective field theories with a shift symmetry \cite{Assassi:2013gxa}, we shall assume the Lagrangian as given, with the background quantities evolving as per the chosen (attractor \textit{vs.} non-attractor) solution. 

Defining $z^2 \equiv 2\epsilon a^2 \Mp^2$, and introducing the Mukhanov-Sasaki variables for both sectors $v = z \zeta$ and $u = a \F$, we rewrite the Lagrangian as
\begin{equation}
    \lh = \frac{1}{2}\left[ \left( v' - \frac{z'}{z} v  \right)^2  - (\partial_i v)^2 + \left( u' - \frac{a'}{a}u \right)^2 - (\partial_i u)^2 - m^2 a^2 u^2 \right] + \lambda a \left( v' - \frac{z'}{z} v \right) u\;,
\end{equation}
where the terms inside the square brackets correspond to the free-theory evolution of each field (which already includes the usual squeezing term due to gravity), whereas the last one is the interaction between them. It is straightforward to see that the conjugate momenta are given by
\begin{equation}
    \pi_{(\s)} \equiv \pi  = v' - \frac{z'}{z}v + \lambda a u\;, \qquad \pi_{(\e)} = u' - \frac{a'}{a} u\;.
\end{equation}
In this way, the Hamiltonian density ${\mathscr H} = \pi v' + \pi_{(\e)} u' - \lh$ can easily be obtained,
\begin{align}\label{eq:Hml}
    {\mathscr H} & = \frac{1}{2} \left[ \pi^2 + (\partial_i v)^2 + \frac{z'}{z} \left\{ \pi, v \right\} \right] + \frac{1}{2} \left[ \pi_{(\e)}^2 + (\partial_i u)^2 + (m^2 + \lambda^2) a^2 u^2 + \frac{a'}{a} \left\{\pi_{(\e)},u \right\} \right] - \lambda a \pi u\;.
\end{align}

\subsection{Heisenberg Equations of Motion}
Before taking the quantum corrections from the interaction term into account, note that the terms inside the first brackets in eq.~\eqref{eq:Hml} govern the (free) unitary evolution of the system's ($\s$) degrees of freedom. The second set of terms governs the unitary evolution of the environment's ($\e$) degrees of freedom, while the last set is the coupling between $\s$ and $\e$. Thus, from the quadratic Hamiltonians, we get the equations of motion
\begin{equation}
    v_k'' + \left(k^2 - \frac{z''}{z}\right)v_k = 0\;, \qquad u_k'' + \left(k^2 + m_{\rm eff}^2 a^2 - \frac{a''}{a}\right)u_k = 0\;,
\end{equation}
where we have defined $m_{\rm eff}^2 = m^2 + \lambda^2$. From here onwards, we shall remove the subindex, and $m$ will refer to the effective mass shown in the Hamiltonian. 

We focus on dS backgrounds, for slow-roll (SR) and ultra--slow-roll (USR) solutions. For either case, we have that $z''/z = a''/a =  2/\tau^2$, so the solutions to the Heisenberg equations of motion, assuming Bunch-Davies initial states, are given by\footnote{\label{eps_def}This can be seen by defining the slow–roll parameters as $\epsilon_{i+1} = \rd \ln \epsilon_i/\rd N$, and identifying $\epsilon = \epsilon_1$. We can then write $$\frac{z''}{z} = \frac{1}{\tau^2} \left( 2 - \epsilon_1 + \frac{3}{2}\epsilon_2 + \frac{1}{4}\epsilon_2^2 - \frac{1}{2}\epsilon_1 \epsilon_2 + \frac{1}{2}\epsilon_2 \epsilon_3 \right)\;.$$ SR solutions correspond to the combination $\epsilon_i \sim 0$, while the USR solutions correspond to $\epsilon_2 = -6$ with the rest of $\epsilon_i$ being small. In both cases, $z''/z = 2/\tau^2$, which explains why the equations of motion look the same for the attractor and non-attractor models we are studying.}
\begin{align}
    v_k (\tau) & = \frac{1}{2} e^{i \frac{\pi}{2} (\nu_\s + \frac{1}{2})} \sqrt{-\pi \tau} H_{\nu_\s}^{(1)}(-k \tau)\;, \qquad \nu_\s^2 = \frac{9}{4}\;.          \\
    u_k (\tau) & = \frac{1}{2} e^{i \frac{\pi}{2} (\nu_\e + \frac{1}{2})} \sqrt{-\pi \tau} H_{\nu_\e}^{(1)}(-k \tau)\;, \qquad \nu_\e^2 = \frac{9}{4} - \frac{m^2}{H^2}\;. \label{eq:envmf}
\end{align}
It is a well-known peculiarity that the mode functions appear to be insensitive to the different backgrounds, particularly for $\s$, where we would expect to notice a difference. However, a distinction arises in the conjugate momentum, which is given by:
\begin{equation}
    \pi_k = v_k' - \frac{z'}{z} v_k = v_k' - \frac{a'}{a}\left(1 + \frac{\epsilon_2}{2}\right) v_k\;,
\end{equation}
with $\epsilon_2 = 0$ for SR and $\epsilon_2 = -6$ for USR (see footnote \ref{eps_def}). The explicit expressions are: 
\begin{equation}
    \pi^{\rm SR}_k = -i \sqrt{\frac{k}{2}} e^{-i k\tau}\;, \qquad \pi^{\rm USR}_k = -i \sqrt{\frac{k}{2}} e^{-i k\tau} \left(1 - \frac{3i}{k\tau} - \frac{3}{(k\tau)^2} \right)\;.
\end{equation}

\section{An open EFT description for adiabatic perturbations}
Having chosen a set of degrees of freedom to be tracked, we now focus on providing an effective description of their dynamics. The dynamics of the system, given some coarse-graining of the environment, can be described by a master equation (ME) for the system's density matrix (in this context referred to as `reduced' density matrix, $\Ri$). These master equations come in various forms. Which of these is most
suitable
depends on the properties of $\s$ and $\e$, and the interactions between them. 


The time evolution of the density matrix for the full system
is given by the von Neumann equation 
\begin{equation}
    \frac{\rd \tilde{\rho}}{\rd \tau} = -i [\tilde{V}(\tau), \tilde{\rho}(\tau)]\;,
\end{equation}
where $\tilde{\rho}$ represents the density matrix of the entire system $\s + \e$, which undergoes unitary evolution. Here, tilded operators ($\tilde{O}$) are in the interaction picture, whereas hatted operators ($\hat{O}$) are in the Schr\"odinger picture. 
To describe the evolution of the $\s$ degrees of freedom alone, we trace over $\e$ on both sides of the equation. It is typically assumed that the state of the environment is stationary,\footnote{The validity of this assumption depends on the typical timescales of the processes involving the environment and will decide if the system's influence on the environment can be ignored or not.} allowing the master equation to be written as:
\begin{equation}\label{eq:Born}
    \frac{\rd}{\rd \tau}\Ri = - \int_{\tau_0}^{\tau} \rd \tau'\ \Tr_\e \Big[ \tilde{V}(\tau), \big[ \tilde{V}(\tau'), \tilde{\rho}_\s(\tau') \otimes \tilde{\rho}_\e \big] \Big]\;,
\end{equation}
which is the so-called Born master equation which captures the leading order behaviour assuming weak-coupling.%
    \footnote{In principle, there is also a term arising from $\Tr_\e[\tilde{V}(\tau), \tilde{\rho}(\tau_0)]$.
    However, one can always remove it by an appropriate field redefinition or by normal ordering.}
Moreover, $\Ri = \Tr_\e (\tilde{\rho}_\s \otimes \tilde{\rho}_\e)$. 
Next, we introduce the Redfield master equation,
\begin{equation}\label{eq:Red}
    \frac{\rd}{\rd \tau}\Ri = - \int_{\tau_0}^{\tau} \rd \tau'\ \Tr_\e \Big[ \tilde{V}(\tau), \big[ \tilde{V}(\tau'), \tilde{\rho}_\s(\tau) \otimes \tilde{\rho}_\e \big] \Big]\;,
\end{equation}
which differs from the Born equation by the absence of a time convolution of the density matrix.
The Redfield equation typically approaches the correct dynamics
if the environment is sufficiently weakly coupled.

More general master equations can be derived using projection operator techniques. One example is the Nakajima--Zwanzig (NZ) equation, which is exact and capable of capturing non-Markovian behaviour. The Born ME, eq.~\eqref{eq:Born}, represents the leading order approximation of the NZ equation. Similarly, the Redfield equation, eq.~\eqref{eq:Red}, serves as the leading order approximation of the time convolutionless (TCL) master equation. The latter removes the time convolution present in the NZ equation by employing back-propagation operators. The leading order TCL equation, or TCL$_2$ ME, has been benchmarked for their application in inflationary cosmology \cite{Colas:2022hlq}, showing remarkable agreement with exact results from the cosmological Caldeira--Leggett model. This, combined with their use in other models, has demonstrated that the TCL$_2$ equation provides an excellent open effective description of inflationary perturbations.

Before moving on to write the TCL$_2$ equation for the model introduced in the previous section, let us rewrite eq.~\eqref{eq:Red} in a way more suitable for calculations. For this, we write the interaction Hamiltonian as 
\begin{equation}
    \tilde{V}(\tau) = \int \rd^3 x\ \tilde{J}_\s (t, x) \otimes \tilde{J}_\e (t, x)\;.
\end{equation}
such that, upon tracing over $\e$, we get the ME: 
\begin{align}
    \frac{\rd}{\rd \tau}\Ri = - \int \rd \tau' & \int \rd^3 x \int \rd^3 y\ \Big\{ \big[ \tilde{J}_\s (\tau,x) \tilde{J}_\s (\tau',y) \Ri (\tau) - \tilde{J}_\s (\tau',y) \Ri (\tau) \tilde{J}_\s (\tau,x) \big] {\cal K}^{>} (x,y) \nonumber \\
    & - \big[ \tilde{J}_\s (\tau,x) \Ri (\tau) \tilde{J}_\s (\tau',y) - \Ri (\tau) \tilde{J}_\s (\tau',y) \tilde{J}_\s (\tau,x) \big] [{\cal K}^{>} (x,y)]^*\Big\}\;. \label{me0}
\end{align}
Here, we have defined a {\it memory kernel}:
\begin{equation}
    {\cal K}^> (x,y) \equiv \Tr_\e \left[ J_\e (\tau,x) J_\e (\tau',y) \rho_\e (\tau_0) \right]\;,
\end{equation}
which captures the correlations within the environment. For our simple Gaussian setup, both $\tilde{J}_\s$ and $\tilde{J}_\e$ are linear in the field operators.

\subsection{The TCL$_2$ Master Equation}
For the model at hand, the TCL$_2$ equation has the following structure:
\begin{align}\label{eq:medd}
    \frac{\rd \Ri}{\rd \tau}  = \sum_{\vt k} & \Big\{ -\frac{i}{2} \Delta_{ij}(\tau) \big[\tilde{q}_i (\tau) \tilde{q}^\dg_j (\tau), \Ri (\tau) \big] - \frac{1}{2} D_{ij} (\tau) \left[ \tilde{q}_i (\tau), \big[ \tilde{q}^\dg_j (\tau), \Ri(\tau) \big]\right] \nonumber \\
    & + \frac{i}{2} \Delta_{12} (\tau) \omega_{ij} \left[ \tilde{q}_i (\tau), \big\{ \tilde{q}^\dg_j (\tau), \Ri(\tau) \big\} \right] \Big\}\;,
\end{align}
where $\tilde{q}_i$ represents a phase-space vector with components $\{ \tilde{v}_{\vt k}, \tilde{\pi}_{\vt k} \}$, and similarly in the Schr\"odinger picture representation. Further, $D_{ij}$ and $\Delta_{ij}$ represent $2\times 2$ symmetric matrices with real entries, whereas $\omega_{ij}$ is the two-dimensional Levi-Civita symbol. 

A physical interpretation of the equation is more accessible in the Schr\"odinger picture, where the ME is given by
\begin{align}\label{sme}
    \frac{\rd \rred}{\rd \tau} =\sum_{\vt k} & \Big\{-i \left[H_{ij}^{(2)} \h{q}_i \h{q}_j^\dg\,, \rred (\tau) \right] - \frac{1}{2} D_{ij} (\tau) \left[ \h{q}_i\,, \big[ \h{q}_j^\dg\,, \rred (\tau) \big]\right] \nonumber \\
    & + \frac{i}{2}\Delta_{12} \ \omega_{ij} \left[ \h{q}_i\,, \big\{ \h{q}_j^\dg\,, \rred (\tau) \big\} \right] \Big\}\;,
\end{align}
where (the Fourier mode of) the resulting quadratic Hamiltonian is explicitly given by
\begin{align}
   H^{(2)}_{ij} \h{q}_i \h{q}_j^\dg = \frac{1}{2} \left[ (1+\Delta_{22})\h{q}_2 \h{q}^\dg_2 + (k^2 + \Delta_{11}) \h{q}_1 \h{q}^\dg_1 + \left(\frac{z'}{z} + \Delta_{12} \right) \left( \h{q}_1 \h{q}_2^\dg + \h{q}_2 \h{q}_1^\dg \right)  \right]\;.
\end{align}
This shows how the elements of $\Delta_{ij}$ correspond to a Lamb shift of the Hamiltonian
governing the unitary part of the system evolution. On the other hand, the elements of $D_{ij}$ correspond to diffusion terms \cite{Caldeira:1982iu,Dekker77}, while $\Delta_{12}$ plays a role in the dissipative dynamics of the system. It can also be shown that the TCL$_2$ equation can be cast in a Lindblad-like form,
\begin{equation}\label{eq:MELf}
    \frac{\rd \rred}{\rd \tau} = \sum_{\vt k}  \Big\{ -i H_{ij}^{(2)} \big[ \h{q}_i \h{q}_j^\dg \,, \rred (\tau) \big] + \gamma_{ij} (\tau) \left( \h{q}_i \rred (\tau) \h{q}_j^\dg - \frac{1}{2} \{ \h{q}_j^\dg \h{q}_i\,, \rred (\tau) \} \right) \Big\}\;,
\end{equation}
where $\gamma_{ij} (\tau) \equiv D_{ij} (\tau) +i \Delta_{12} (\tau) \omega_{ij}$ is known as the dissipator matrix. These quantities will play a central role in our discussion of (non-)Markovianity later on. 

Finally, let us reiterate that it is the action of the inflationary perturbations, particularly the interaction terms, that ultimately determine the structure of the ME. For instance, in models such as those studied in \cite{Colas:2022hlq} and \cite{Brahma:2024yor}, there is a minus sign in front of $\Delta_{12}$, either in the definition of the dissipator matrix $\gamma_{ij}$ or in the ME written as in eq.~\eqref{sme}. This is because, in these cases, the interaction involves only the fields and not their derivatives, and thus not their conjugate momenta.

\subsubsection{Diffusion and Dissipation: Exchange of information and energy}
In order to obtain the elements of the dissipator matrix, we start from the ME in the form of eq.~\eqref{me0}. At this stage, the equation is not yet suitable for writing it as the TCL$_2$ equation, since some of the system operators are evaluated at $\tau'$ instead of $\tau$. We can express these operators in terms of operators at local time as follows:
\begin{align}\label{eq::ttr}
    \tilde{v}_{\vt k} (\tau') & = 2\ {\rm Im}\left[v_k (\tau') \pi_k^* (\tau)\right] \tilde{v}_{\vt k} (\tau) - 2\ {\rm Im}\left[v_k (\tau') v_k^*(\tau)\right] \tilde{\pi}_{\vt k} (\tau)\,, \\
    \tilde{\pi}_{\vt k} (\tau') & = 2\ {\rm Im}\left[ \pi_k^*(\tau) \pi_k (\tau') \right] \tilde{v}_{\vt k} (\tau) - 2\ {\rm Im}\left[ v_k^* (\tau) \pi_k (\tau') \right] \tilde{\pi}_{\vt k} (\tau)\;.
\end{align}
These identities can be derived from Bogolyubov transformations by comparing the creation and annihilation operators at the initial time with those at times $\tau'$ and $\tau$. This establishes a relationship between these operators and the phase-space fields. 
In the end, we will only need the last equation, which we write in terms of the phase-space coordinates as
\begin{align}
    \tl{q}{2} (\tau') & = 2\ {\rm Im}\left[ \pi_k^*(\tau) \pi_k (\tau') \right] \tl{q}{1} (\tau) - 2\ {\rm Im}\left[ v_k^* (\tau) \pi_k (\tau') \right] \tl{q}{2} (\tau) \nonumber \\
    & \equiv A_k(\tau, \tau') \tl{q}{1} (\tau) + B_k(\tau,\tau') \tl{q}{2} (\tau)\;.
\end{align}
Substituting these expressions into eq.~\eqref{me0} we obtain
\begin{align}
    \frac{\rd}{\rd \tau}\Ri  = - & \sum_{\vt k} \int_{\tau_0}^\tau\rd \tau' \Big[ A_k(\tau,\tau') {\cal K}^{>}_k (\tau, \tau') \left(\tl{q}{2} (\tau) \td{q}{1} (\tau) \Ri (\tau) - \td{q}{1} (\tau) \Ri (\tau) \tl{q}{2} (\tau) \right) \nonumber \\
    & + B_k(\tau,\tau') {\cal K}^{>}_k (\tau, \tau') \left( \tl{q}{2}(\tau) \td{q}{2} (\tau) \Ri(\tau) - \td{q}{2}(\tau) \Ri (\tau) \tl{q}{2} (\tau) \right) \nonumber \\
    & - A_k(\tau, \tau') [{\cal K}^{>}_k (\tau,\tau')]^* \left( \tl{q}{2}(\tau) \Ri (\tau) \td{q}{1} (\tau)  - \Ri(\tau) \td{q}{1}(\tau) \tl{q}{2}(\tau) \right) \nonumber \\
    & - B_k(\tau, \tau') [{\cal K}^{>}_k (\tau,\tau')]^* \left( \tl{q}{2}(\tau) \Ri(\tau) \td{q}{2}(\tau) - \Ri(\tau) \td{q}{2}(\tau) \tl{q}{2}(\tau) \right) \Big]\;.
\end{align}
We compare this to eq.~\eqref{eq:medd}, which we have expanded, and reorganized, as follows:
 \begin{align}
    \frac{\rd \Ri}{\rd \tau} & = - \frac{1}{2} \sum_{\vt k} \Big\{ [D_{11} + i \Delta_{11}] (\tilde{q}_1 \tilde{q}_1^\dg \Ri - \tilde{q}_1^\dg \Ri \tilde{q}_1) - [D_{11} - i \Delta_{11}](\tilde{q}_1 \Ri \tilde{q}_1^\dg - \Ri \tilde{q}_1^\dg \tilde{q}_1) \nonumber \\
    & + [D_{22} + i \Delta_{22}] (\tl{q}{2}\td{q}{2}\Ri - \td{q}{2}\Ri \tl{q}{2}) - [D_{22} - i \Delta_{22}] (\tl{q}{2}\Ri \td{q}{2} - \Ri \td{q}{2}\tl{q}{2}) \nonumber\\
    & + 2 [D_{12} + i \Delta_{12}](\tl{q}{2}\td{q}{1}\Ri - \td{q}{1}\Ri \tl{q}{2}) - 2 [D_{12} - i \Delta_{12}](\tl{q}{2}\Ri \td{q}{1} - \Ri \td{q}{1}\tl{q}{2})    \Big\}
\end{align}
Finally, the functional coefficients appearing in the ME can be written as
    \begin{align}
    D_{11} = 0\;, & \quad \Delta_{11} = 0\;,\label{eq:DDs} \\
    D_{12}  =  \lambda a(\tau) \int_{\tau_0}^{\tau} \rd\tau' \lambda a(\tau') A_k(\tau, \tau') {\rm Re}[{\cal K}^{>}_k (\tau,\tau')]\;, & \quad 
    \Delta_{12} = \lambda a(\tau)\int_{\tau_0}^{\tau} \rd\tau' \lambda a(\tau') A_k(\tau, \tau') {\rm Im}[{\cal K}^{>}_k (\tau,\tau')]\;, \nonumber \\
    D_{22} = 2 \lambda a(\tau)\int_{\tau_0}^{\tau} \rd\tau' \lambda a(\tau') B_k (\tau,\tau') {\rm Re}[{\cal K}^{>}_k (\tau,\tau')]\;, & \quad 
    \Delta_{22} = 2 \lambda a(\tau)\int_{\tau_0}^{\tau} \rd\tau' \lambda a(\tau') B_k (\tau,\tau') {\rm Im}[{\cal K}^{>}_k (\tau,\tau')]\;. \nonumber
\end{align}
The memory kernel is given by
\begin{equation}
    {\cal K}^>_k (\tau,\tau') = u_k (\tau) u_k^* (\tau') = \frac{\pi \alpha}{4k} \sqrt{-k\tau} H_{\nu}^{(1)} (-k \tau)\left[ \sqrt{-k\tau'} H_{\nu}^{(1)} (-k \tau')\right]^*\;,
\end{equation}
where we have defined $\alpha = e^{-\pi \mu}$ for $\nu \equiv i \mu$, and $\alpha = 1$ for $\nu \in \mathbb{R}$. In general, the memory kernel involves a convolution over different momenta, capturing effects similar to loop corrections. While these technical complexities are absent in our current context, the integrals in eqs.~\eqref{eq:DDs} can still be challenging to evaluate, especially for the USR background. Nonetheless, analytical expressions can be derived and are provided in Appendix \ref{ApA}.

\begin{figure}[h!]
     \centering
     \begin{subfigure}[b]{0.48\textwidth}
         \centering
         \hspace*{-1.5cm}
         \includegraphics[width=1.3\textwidth]{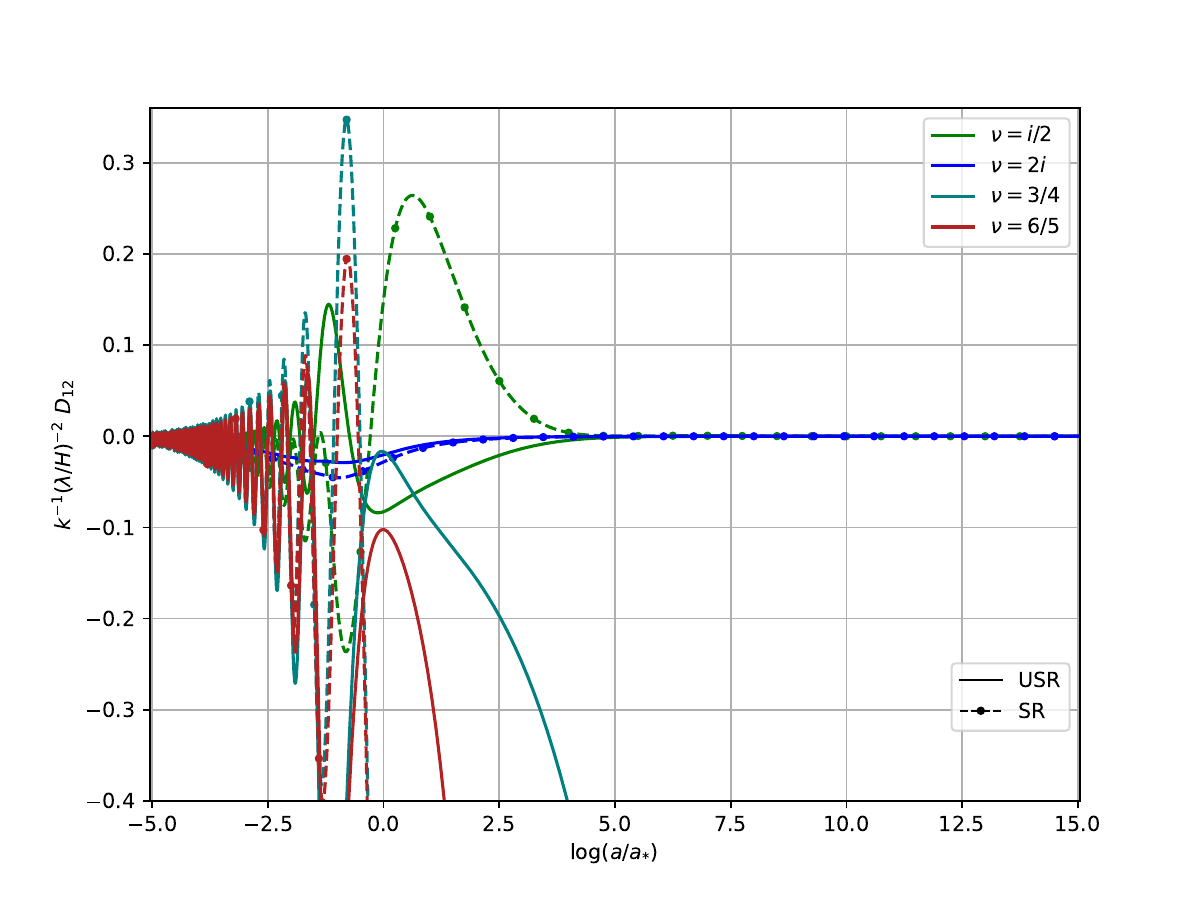}
         \caption{}
     \end{subfigure}
     \hfill
     \begin{subfigure}[b]{0.48\textwidth}
         \centering
         \hspace*{-0.8cm}
         \includegraphics[width=1.3\textwidth]{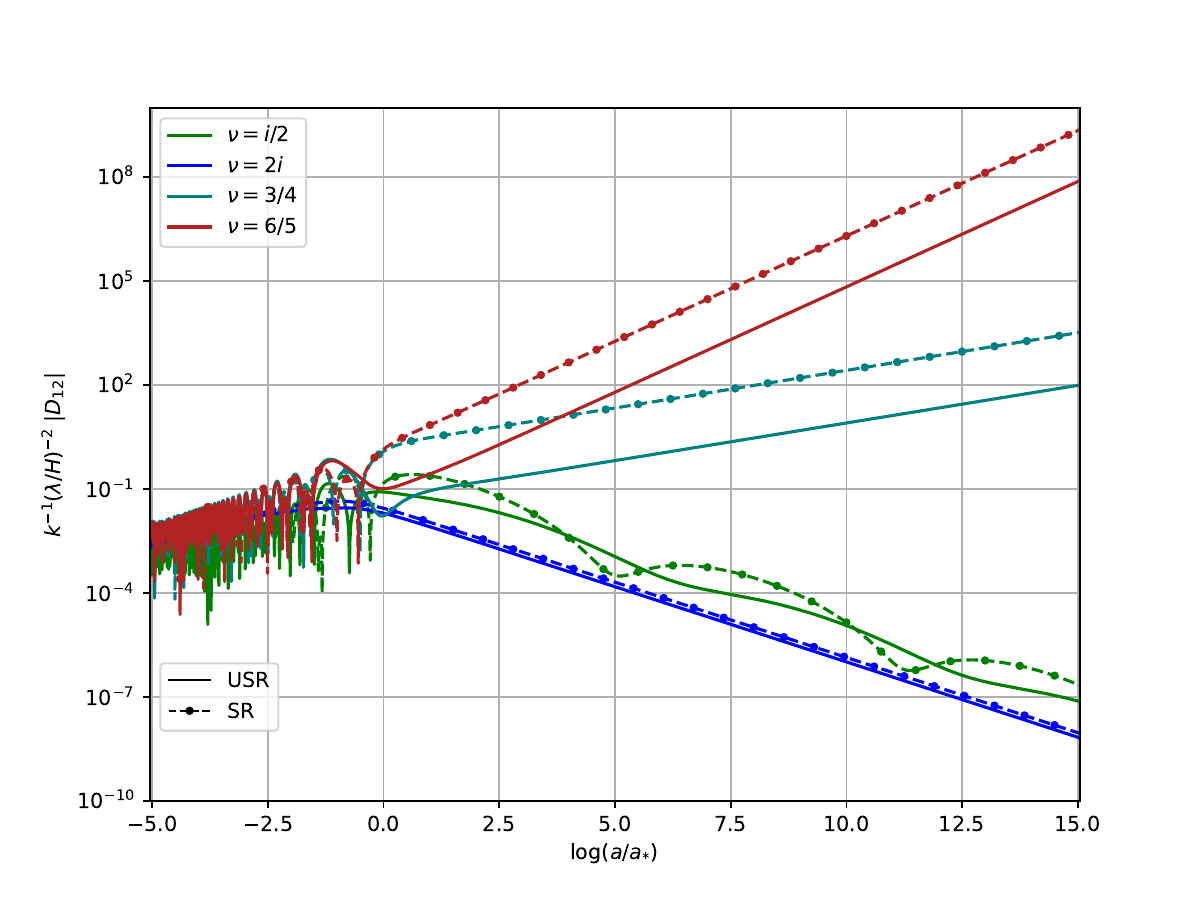}
         \caption{}
     \end{subfigure}
        \caption{Left: Rescaled $D_{12}$ for various $\nu$ values. The evolution under a SR background is represented by dashed lines with bullets, while the evolution under a USR background is shown by solid lines. Right: Magnitude of $|D_{12}|$ in a logarithmic scale for the same $\nu$ values and backgrounds.}
        \label{D12}
\end{figure}

\begin{figure}[h!]
     \centering
     \begin{subfigure}[b]{0.48\textwidth}
         \centering
         \hspace*{-1.5cm}
         \includegraphics[width=1.3\textwidth]{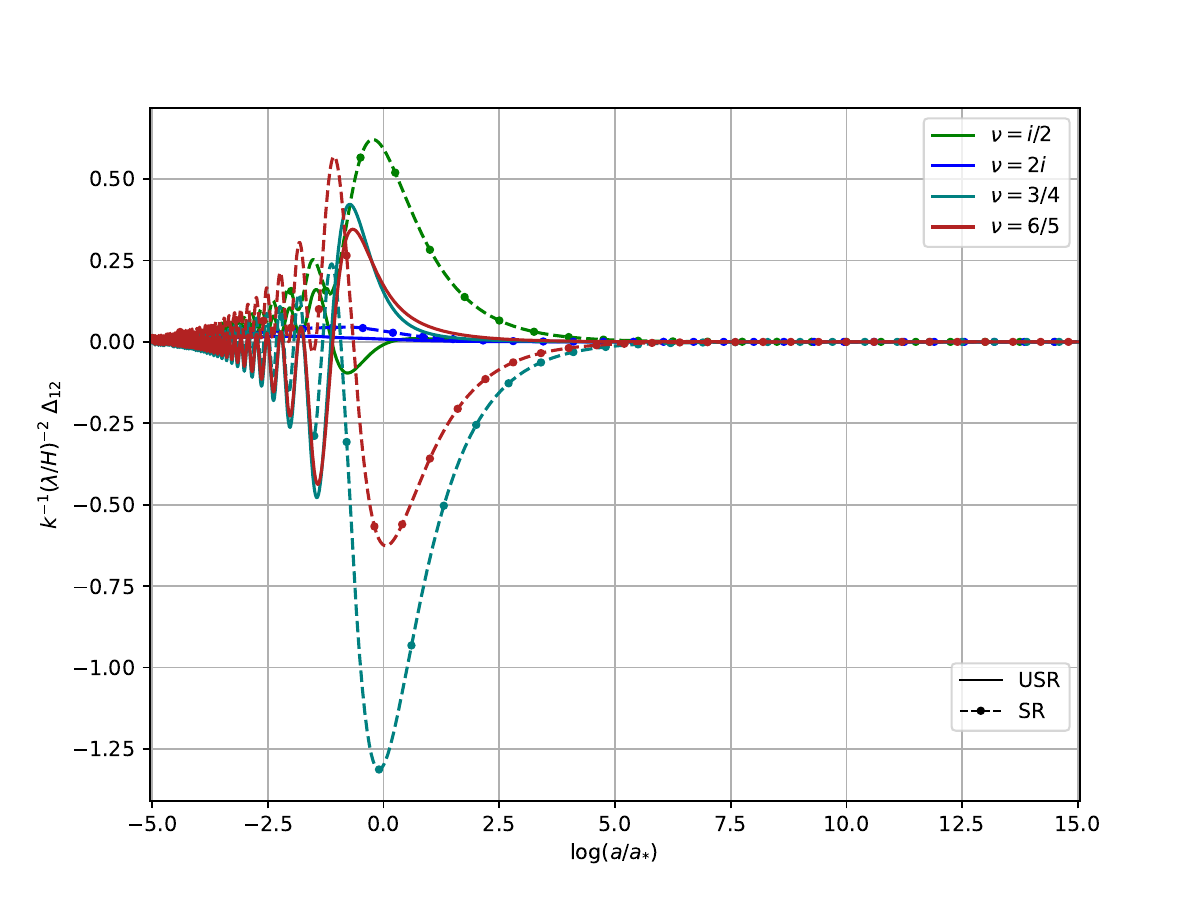}
         \caption{}
     \end{subfigure}
     \hfill
     \begin{subfigure}[b]{0.48\textwidth}
         \centering
         \hspace*{-0.8cm}
         \includegraphics[width=1.3\textwidth]{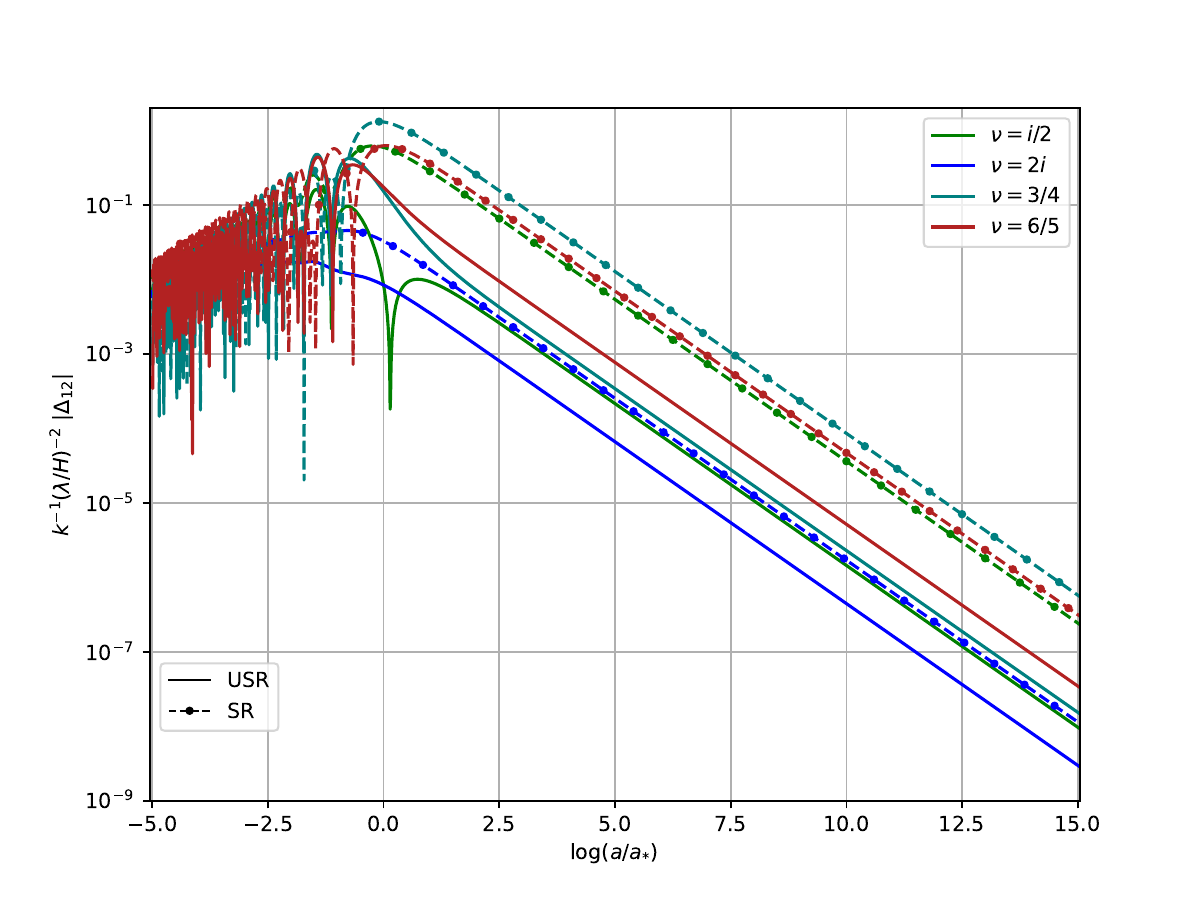}
         \caption{}
     \end{subfigure}
        \caption{Left: Rescaled $\Delta_{12}$ for the same $\nu$ values. The evolution under a SR background is represented by dashed lines with bullets, while the evolution under a USR background is shown by solid lines. Right: Magnitude of the same functions in a logarithmic scale.}
        \label{L12}
\end{figure}

\begin{figure}[h!]
     \centering
     \begin{subfigure}[b]{0.48\textwidth}
         \centering
         \hspace*{-1.5cm}
         \includegraphics[width=1.3\textwidth]{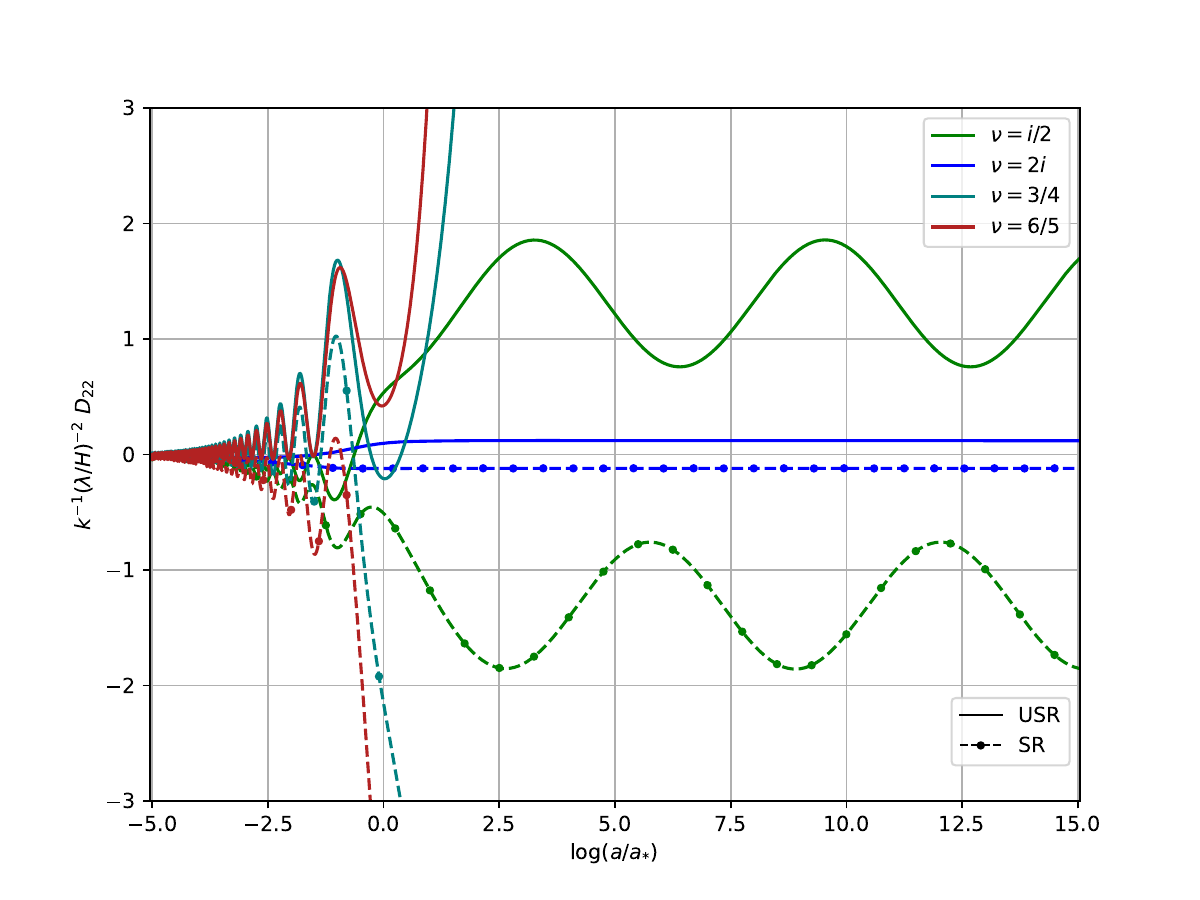}
         \caption{}
     \end{subfigure}
     \hfill
     \begin{subfigure}[b]{0.48\textwidth}
         \centering
         \hspace*{-0.8cm}
         \includegraphics[width=1.3\textwidth]{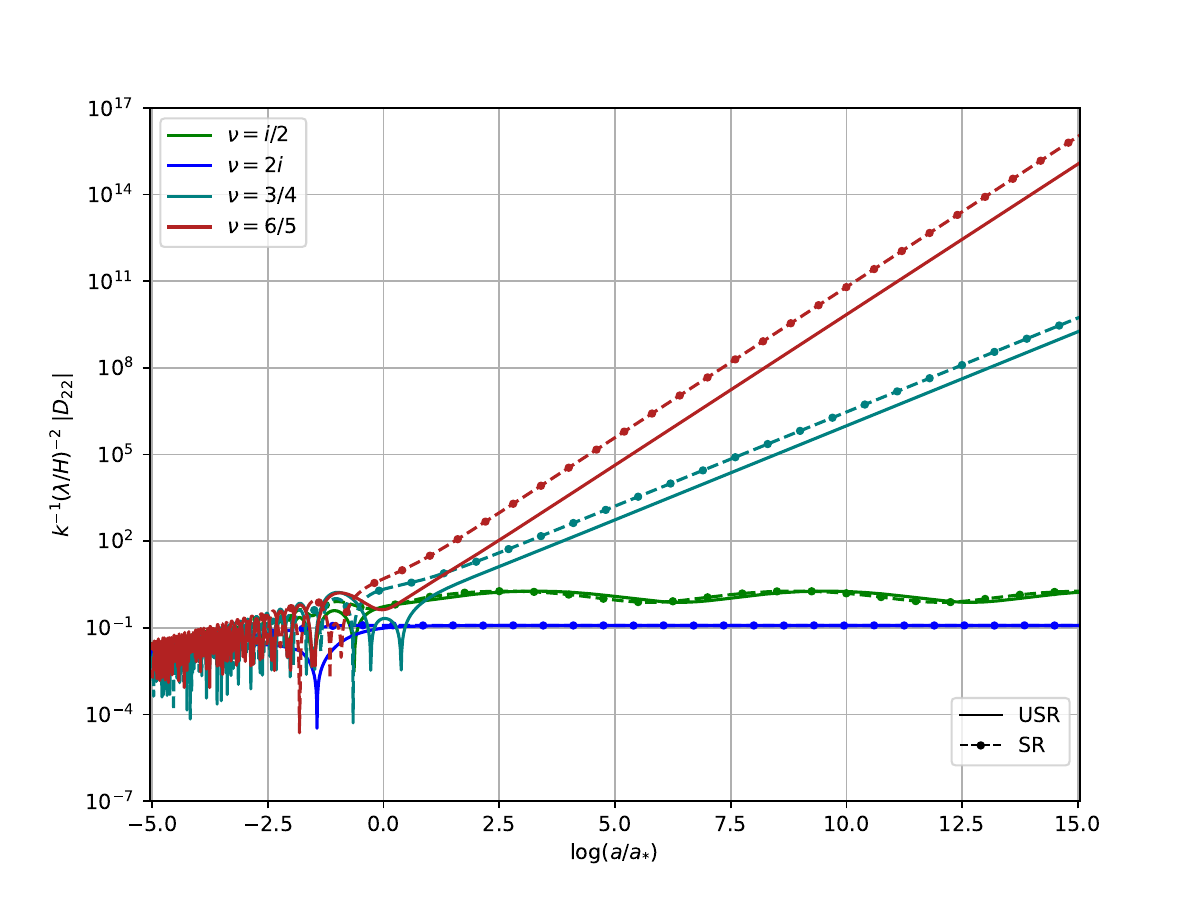}
         \caption{}
     \end{subfigure}
        \caption{Left: Rescaled $D_{22}$ for various $\nu$. The evolution under a SR background is represented by dashed lines with bullets, while the evolution under a USR background is shown by solid lines. Right: Magnitude of $|D_{22}|$ in a logarithmic scale for the same $\nu$ values and backgrounds.}
        \label{D22}
\end{figure}

\begin{figure}[h!]
    \centering
    \includegraphics[width=0.8\textwidth]{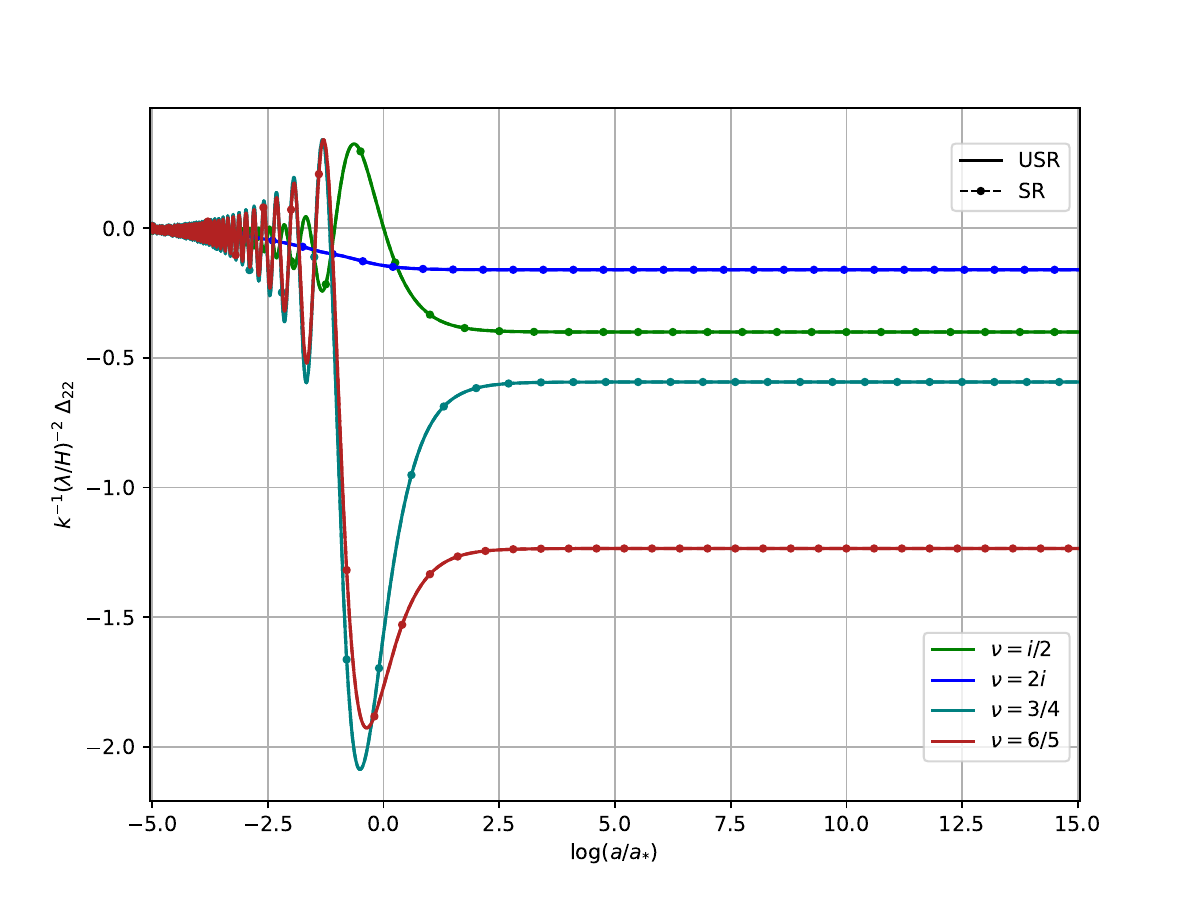}
    \caption{Evolution of the rescaled $\Delta_{22}$ for different $\nu$ in a SR (dashed lines with bullets) and an USR (solid lines) background. The overlap between solid and dashed lines indicates no noticeable difference in this function between the backgrounds.}
    \label{L22}
\end{figure}

We illustrate the behaviour of these dissipation and diffusion functions for various parameter choices in figs.~\ref{D12}--\ref{L22}.
We also compare with the SR background for the corresponding parameters.
Of particular interest are the curves corresponding to $\nu = i/2$ and $\nu = 2i$, for which the phenomenon of recoherence \cite{Colas:2022kfu} is observed (as we shall show later), the effect being stronger for the latter value. These choices correspond to effective masses%
    \footnote{Recall that the effective masses also include a contributions from the $\lambda^2$ term.}
$m^2/H^2$ of $5/2$ and $25/4$, respectively, as can be seen from eq.~\eqref{eq:envmf}. As a reminder, note that real values of $\nu$ are associated with lighter effective masses, with $m^2/H^2 = 27/16$ for $\nu = 3/4$ and $m^2/H^2 = 81/100$ for $\nu = 6/5$.

Several key observations can be made. All of these diffusion and dissipation functions oscillate before the mode crosses the horizon, as expected. For most choices of parameters, $D_{12}$ and $\Delta_{12}$ tend to approach zero after horizon crossing. The rate of this approach depends on the specific parameters and the background, with the SR solution exhibiting a slower approach after horizon exit compared to the non-attractor case. An exception to this general behaviour is seen for $\nu = 3/4$, which is representative of the parameter regime when the effective mass becomes comparable to (or smaller than) the Hubble parameter. Therefore, for very light entropic fields, the diffusion term corresponding to the off-diagonal term in the noise kernel takes large negative values after horizon-crossing. On the other hand, the dissipative term $\Delta_{12}$, which renormalises the comoving Hubble scale, tends to go to zero for all choices of $\nu$. In contrast to this, $\Delta_{22}$ shows no significant difference between the SR and USR backgrounds, with the corresponding curves almost perfectly overlapping. Notably, for all the parameters sampled, $\Delta_{22}$ stabilises at a roughly constant value after horizon crossing.

The most interesting behaviour is observed in the momentum-momentum diffusion term, $D_{22}$, which constitutes the only non-zero diagonal term in the noise kernel. As we will discuss later, $D_{22}$ plays a crucial role in determining the (non-unitary) evolution of the system. Depending on the couplings, the magnitude of $D_{22}$ can either increase over time or oscillate around, and stabilise at, a constant value. Additionally, its sign is influenced by the background, being positive in the USR case and negative in the SR case. Although not clear at the moment, the physical interpretation of the behaviour of $D_{22}$ will be shown to influence the decoherence (and the entanglement entropy) of the system later on.


\subsection{TCL$_2$ master equation in Field Basis}
In this section, we express the ME in a different representation, which we refer to as the field basis. This approach is reminiscent of the position basis in quantum mechanics and helps build intuition regarding the role of each term in the master equation. Additionally, the solution to the ME in this basis can be written in a relatively simple manner, facilitating further manipulation of the reduced density matrix in subsequent calculations.

The projection to the field basis follows the steps outlined in \cite{Hollowood:2017bil} and references therein. We start with the ME in its form given by eq.~\eqref{sme}, focusing on the contributions from the $\pm \mathbf{k}$ modes, which are the coupled modes in our model. For more complicated systems, at the TCL$_2$ level, these correspond to interactions linear in the sub-system coordinates.

In the field basis, the reduced density matrix is represented as $\varrho(\chi_{\mathbf{k}}, \varphi_{\mathbf{k}}) \equiv \langle \chi_{\mathbf{k}} | \hat{\rho}_\mathrm{red} (\mathbf{k}, -\mathbf{k}) | \varphi_{\mathbf{k}} \rangle$. It is important to note that $\chi$ and $\varphi$ are not different fields but rather represent different values that $v_k (\tau)$ can take. Thus, these are complex numbers, accounting for contributions from both momenta, ${\bf k}$ and $-{\bf k}$. To avoid clutter, we omit the $k$ subscript in what follows, keeping in mind that it is the only Fourier mode we are considering. 

Using the operator identities
\begin{align}\label{eq:opid}
    \langle \chi | \hat{M} \hat{\pi}^\dagger | \varphi \rangle = i \frac{\partial}{\partial \varphi} \langle \chi | \hat{M} | \varphi \rangle\;, & \qquad \langle \chi | \hat{M} \hat{\pi} | \varphi \rangle = i \frac{\partial}{\partial \varphi^*} \langle \chi | \hat{M} | \varphi \rangle \;, \nonumber \\
    \langle \chi | \hat{\pi}^\dagger \hat{M} | \varphi \rangle = -i \frac{\partial}{\partial \chi} \langle \chi | \hat{M} | \varphi \rangle\;, & \qquad \langle \chi | \hat{\pi} \hat{M} | \varphi \rangle = - i \frac{\partial}{\partial \chi^*} \langle \chi | \hat{M} | \varphi \rangle \;,
\end{align}
we can write the ME in the field basis as:
\begin{align}\label{eq:MEF}
    \frac{\mathrm{d} \varrho}{\mathrm{d} \tau} & = -i k^2 (|\chi|^2 - |\varphi|^2) \varrho + i (1 + \Delta_{22}) \left( \frac{\partial^2 \varrho}{\partial \chi \partial \chi^*} - \frac{\partial^2 \varrho}{\partial \varphi \partial \varphi^*} \right)  \nonumber \\
    & - \left( \frac{z'}{z} + \Delta_{12} \right) \bigg[ \chi \frac{\partial \varrho}{\partial \chi} + \chi^* \frac{\partial \varrho}{\partial \chi^*} + \varphi \frac{\partial \varrho}{\partial \varphi} + \varphi^* \frac{\partial \varrho}{\partial \varphi^*} + 2\varrho \bigg] \nonumber \\
    & + i D_{12} \bigg[ (\chi - \varphi) \left(\frac{\partial \varrho}{\partial \chi} +  \frac{\partial \varrho}{\partial \varphi} \right) + (\chi^* - \varphi^*) \left(\frac{\partial \varrho}{\partial \chi^*} +  \frac{\partial \varrho}{\partial \varphi^*} \right)  \bigg] \nonumber \\
    & - \Delta_{12} \bigg[ \chi \frac{\partial \varrho}{\partial \varphi} + \varphi \frac{\partial \varrho}{\partial \chi} + \chi^* \frac{\partial \varrho}{\partial \varphi^*} + \varphi^* \frac{\partial \varrho}{\partial \chi^*} + 2\varrho  \bigg] \nonumber \\
    & + D_{22} \bigg[ \frac{\partial^2 \varrho}{\partial \chi \partial \chi^*} + \frac{\partial^2 \varrho}{\partial \varphi \partial \varphi^*} + \frac{\partial^2 \varrho}{\partial \chi \partial \varphi^*} + \frac{\partial^2 \varrho}{\partial \chi^* \partial \varphi}   \bigg]\;.
\end{align}

\subsubsection{A Gaussian ansatz for the density matrix}
As previously mentioned, one of the advantages of projecting the master equation onto a particular basis is the ability to write down an explicit formula for the density matrix in that representation. The feasibility of this task depends on the specific model and the level of approximation used. In our case, the following Gaussian ansatz is effective:
\begin{equation}\label{eq:rpdf}
    \varrho (\chi, \varphi) = C \exp\left( -\Omega |\chi|^2 - \Omega^* |\varphi|^2 - \frac{\xi}{2} |\chi - \varphi|^2 \right)\;,
\end{equation}
where we have adopted the parametrization shown in \cite{Hollowood:2017bil}, with $\Omega \in \mathbb{C}$ and $\xi \in \mathbb{R}$.

Next, we will trade the ME for differential equations for the parameters in our Gaussian function. For this, take the derivative with respect to time, 
\begin{equation}
    \frac{\rd \varrho}{\rd \tau} = \left( \frac{1}{C} \frac{\rd C}{\rd \tau}  -  |\chi|^2 \frac{\rd \Omega}{\rd \tau} - |\varphi|^2 \frac{\rd \Omega^*}{\rd \tau} - \frac{1}{2}|\chi-\varphi|^2 \frac{\rd \xi}{\rd \tau} \right) \varrho\;,
\end{equation}
and match the resulting factors of $|\chi|^2$, $|\varphi|^2$, $|\chi-\varphi|^2$, and those independent of them. Writing $\Omega = \Omega_R + i \Omega_I$, a straightforward calculation leads to:
\begin{align}
    \frac{1}{C} \frac{\rd C}{\rd \tau} & = 2 (1+ \Delta_{22})\Omega_I - 2 \left( \frac{z'}{z} + 2 \Delta_{12} \right) - 2 D_{22} \Omega_R\;, \label{eq:fops0} \\
     \frac{\rd \Omega_R}{\rd \tau} & = 2 (1+\Delta_{22})\Omega_R \Omega_I -2 \left(\frac{z'}{z} + 2 \Delta_{12}\right) \Omega_R - 2 D_{22} \Omega_R^2\;, \label{eq:fops1} \\
      \frac{\rd \Omega_I}{\rd \tau} & = k^2 - (1+\Delta_{22})(\Omega_R^2 - \Omega_I^2 + \xi \Omega_R) - 2 \left(\frac{z'}{z} + \Delta_{12}\right) \Omega_I + 2 D_{12} \Omega_R -2 D_{22} \Omega_R \Omega_I\;, \label{eq:fops2} \\
    \frac{\rd \xi}{\rd \tau} & = 2(1+\Delta_{22})\Omega_I \xi -2 \frac{z'}{z}\xi -4 D_{12}\Omega_I + 4 \Delta_{12} \Omega_R + 2 D_{22} |\Omega|^2\;. \label{eq:fops3}
\end{align}

In principle, the system is determined only by three equations since the normalization of $\varrho$ imposes: 
\begin{equation}
   1 = \Tr_\s(\Ri) = \int_{-\infty}^{\infty} [\rd\chi] \varrho (\chi,\chi) = C \frac{\pi}{2\Omega_R} \quad \implies \quad C = \frac{2\Omega_R}{\pi}\;,
\end{equation}
where $[\mathrm{d} \chi]$ represents the differential element over the real and imaginary parts of $\chi$. Notice that the differential equations of $C$ (eq.~\eqref{eq:fops0}) and $\Omega_R$ (eq.~\eqref{eq:fops1}) are consistent with this constraint.

\subsection{Cosmological Observables and Transport Equations}
Another way to characterize the dynamics of the system is through its covariance matrix, with entries defined as:
\begin{equation}
    \Xi_{ab} = \frac{1}{2} \left( \hat{q}_a \hat{q}_b^\dagger + \hat{q}_b \hat{q}_a^\dagger \right)\;.
\end{equation}
Observables are associated with the expectation values of these elements and are computed as follows:
\begin{equation}
    \Sigma_{ab} \equiv \langle \Xi_{ab} \rangle = \Tr_\mathcal{S} (\hat{\Xi}_{ab} \hat{\rho}_\mathrm{red} ) \;.
\end{equation}
The differential equations governing the time evolution of the expectation values of the covariance matrix are called \textit{transport equations}. A straightforward approach to find them is to substitute the right-hand side of the ME in any of the forms we have presented and manipulate the expressions to write them exclusively in terms of the different elements of $\Sigma_{ab}$. (This forms an exact, closed system of equations only in the Gaussian case.) Given the equations we have already derived, let us follow a slightly different approach, which has the added benefit of drawing a comparison between the different parameters of the density matrix, eq.~\eqref{eq:rpdf}, and the covariance matrix elements.

First, let us start with $\Sigma_{11}$, which is readily found through
\begin{align}
    \Sigma_{11} = \Tr ( \hat{\chi} \hat{\chi}^\dag \hat{\rho}_\mathrm{red}) & = \int_{-\infty}^{\infty} [\mathrm{d} \chi] |\chi|^2 \varrho (\chi,\chi) \nonumber \\
    & = C \int_{-\infty}^{\infty} [\mathrm{d} \chi] |\chi|^2 \exp(-2\Omega_R |\chi|^2) \nonumber \\
    & = \frac{1}{2 \Omega_R}\;. \label{S11_0} 
\end{align}
 Here, in going from the second to third line we have used standard Gaussian integration formulas. Notice how the structure of this equation indicates that the diagonal elements of $\varrho$ define a probability distribution function (PDF), a fact we will use in the next section.

The other expectation values are similarly found with the help of the identities in eq.~\eqref{eq:opid}. For instance, for the off-diagonal term, we have
\begin{align}
    \Sigma_{12} = \Sigma_{21} = \frac{1}{2}\Tr((\hat{\chi}\hat{\pi}^\dag + \hat{\pi}\hat{\chi}^\dag)\hat{\rho}_\mathrm{red}) & = \frac{i}{2} \int_{-\infty}^{\infty} [\mathrm{d} \chi] \left[\left(\varphi \frac{\partial}{\partial\varphi} - \chi^* \frac{\partial}{\partial\chi^*} \right) \varrho (\chi, \varphi)\right]_{\varphi = \chi} \nonumber \\
    & = - \Omega_I \int_{-\infty}^{\infty} [\mathrm{d} \chi] |\chi|^2 \varrho(\chi,\chi) \nonumber \\
    & = -\frac{1}{2} \frac{\Omega_I}{\Omega_R}\;.
\end{align}
Here, the integrand in the first line corresponds to the field representation of the overall operator inside the trace. Finally, we compute $\Sigma_{22}$ in a similar fashion,
\begin{align}
    \Sigma_{22} = \Tr(\hat{\pi}\hat{\pi}^\dag \hat{\rho}_\mathrm{red}) & = - \int_{-\infty}^{\infty} [\mathrm{d} \chi] \left[\frac{\partial^2}{\partial \varphi \partial\varphi^*} \varrho(\chi,\varphi) \right]_{\varphi=\chi} \nonumber \\
    & = \int_{-\infty}^{\infty} [\mathrm{d} \chi] \left[ \Omega^* (1- \Omega^* |\chi|^2 ) + \frac{\xi}{2} \right] \varrho(\chi,\chi) \nonumber \\
    & = \frac{\Omega_R}{2}\left[ 1 + \frac{\xi}{\Omega_R} + \left( \frac{\Omega_I}{\Omega_R} \right)^2 \right]\;.
\end{align}

Having identified how the covariance matrix elements depend on the parameters of the density matrix in the field representation, we now simply need to substitute these expressions in eqs.~\eqref{eq:fops0}-\eqref{eq:fops3}. After a slightly long but straightforward calculation, one finally arrives at 
\begin{align}
    \frac{\mathrm{d}}{\mathrm{d} \tau} \Sigma_{ab} = 
     \begin{pmatrix}
2(z'/z + 2\Delta_{12})\Sigma_{11} + 2(1+\Delta_{22}) \Sigma_{12} + D_{22} & (1+\Delta_{22})\Sigma_{22} - p^2 \Sigma_{11} - D_{12} + 2 \Delta_{12} \Sigma_{12} \\
(1+\Delta_{22})\Sigma_{22} - p^2 \Sigma_{11} - D_{12} + 2 \Delta_{12} \Sigma_{12} & -2 p^2 \Sigma_{12} - 2 (z'/z) \Sigma_{22}
\end{pmatrix}\;,\label{eq:Teqs}
\end{align}
as expected \cite{Brahma:2024yor, Colas:2022hlq}. One can easily check that the USR correlation functions, with Bunch-Davies initial conditions,
\begin{align}\label{eq:BDUSR}
    \Sigma_{11}^{\rm free} (\tau) = \frac{1}{2p} \left( 1 + \frac{1}{(p \tau)^2}\right)\;, \quad 
    \Sigma_{12}^{\rm free} (\tau) = - \frac{1}{p\tau} - \frac{3}{2(p\tau)^3}\;, \quad 
    \Sigma_{22}^{\rm free} (\tau) = \frac{p}{2} \left[ 1 + \frac{3}{(p\tau)^2} + \frac{9}{(p\tau)^4} \right]\;,
\end{align}
satisfy the free-theory transport equations. 

Let us reiterate that both the set of transport equations we just derived and eqs.~\eqref{eq:fops0}-\eqref{eq:fops3} contain the same dynamical information about the system. Essentially, one is a repackaging of the other due to the Gaussian nature of the interaction, and each can be more useful depending on the circumstances. For instance, the transport equations make it more transparent how the ME coefficients $(D_{ij}, \Delta_{ij})$ influence the evolution of the correlation functions. In contrast, the parameters of the field representation provide little insight in that regard. Conversely, the field representation can be more intuitive when analysing quantities such as the purity. Purity is directly related to the density matrix and quantifies how much the quantum state deviates from its initial (pure) state, making it a very useful marker for decoherence. The purity $\gamma$ is computed as\footnote{Sometimes, the purity is defined in the literature as $\gamma = 1/\sqrt{4 {\rm det}\Sigma}$. We will adhere to the definition provided here.}
\begin{align}
    \gamma = \Tr (\hat{\rho}_\mathrm{red}^2) & = \int_{-\infty}^{\infty} [\mathrm{d}\chi] \left[\int_{-\infty}^{\infty} [\mathrm{d} {\cal X}] \varrho(\chi,{\cal X})\varrho({\cal X},\varphi)\right]_{\varphi=\chi} \nonumber \\
    & = \left(1 + \frac{\xi}{\Omega_R}\right)^{-1} = \frac{1}{4 \det \Sigma_{ab}}\;. \label{eq:pur}
\end{align}
This emphasizes how $\xi$, and in turn purity, is related to the decoherence of the quantum state. In fact, for standard quantum mechanics, the $\xi$ term is known as the localization factor, as it suppresses the contribution to the density matrix from off-diagonal elements in position space. Such an interpretation is not directly apparent from just looking at the determinant of the covariance matrix.

\section{Fokker--Planck Equation}
It is now clear that possessing complete knowledge of the density matrix equates to having comprehensive information about the system; and in particular, about its probabilistic aspects. In quantum mechanics, however, there is no unambiguous notion of probability.
Instead, observables must typically be described by quasiprobability distributions that can become negative in some regions.
The equation governing these distributions can be interpreted as a Fokker--Planck equation, a concept well-known in stochastic classical systems for describing the time evolution of probability distributions.
In quantum mechanics there are three primary quasiprobability distributions of interest.
These are the Wigner quasiprobability distribution, the Glauber--Sudarshan $P$-representation, and the Husimi $Q$-representation. Our approach aligns more closely with the $P$-representation, as Gaussian states can be expanded using an (overcomplete) basis of coherent states. 

To extract the PDF, we examine the diagonal elements of the reduced density matrix. In fact, equation \eqref{S11_0} already demonstrates how the (configuration space) power spectrum is derived through a weighted average, with the diagonal
elements $\varrho(\chi,\chi)$ serving as the PDF. Following a method similar to that outlined in \cite{Burgess:2014eoa, Burgess:2015ajz}, we seek an equation of the form
\begin{equation}
\frac{\rd P}{\rd \tau} = \mathcal{N} \nabla^2 P - \mathcal{D} \nabla \cdot (\mathbf{F} P)\;,
\end{equation}
where $P$ denotes the PDF, $\mathcal{N}$ is the noise term, $\mathcal{D}$ is the drift term, and $\mathbf{F}$ represents a force term. Although we know that $P[\chi] \equiv \varrho(\chi,\chi) = C \exp(-2\Omega_R |\chi|^2)$, our objective is to derive the noise and drift terms from the master equation. This approach allows us to elucidate how the underlying dynamics (SR vs USR) comes into play in an effective description of the reduced sub-system. 

First, to obtain the noise term, we focus on the diagonal elements in eq.~\eqref{eq:MEF} containing second derivatives:
\begin{align}
    \left.\frac{\rd \varrho}{\rd \tau}\right|_{\chi=\varphi} & \sim 2 \Omega_I (1+ \Delta_{22}) (1-2\Omega_R |\chi|^2) \varrho (\chi,\varphi = \chi) - 2 \Omega_R D_{22} (1-2\Omega_R |\chi|^2) \varrho (\chi,\varphi = \chi) \\
    & = \left[ D_{22} - \frac{\Omega_I}{\Omega_R} (1+\Delta_{22}) \right] (-2 \Omega_R) (1-2 \Omega_R |\chi|^2) \varrho (\chi,\varphi = \chi) \\
    & = \left[ D_{22} - \frac{\Omega_I}{\Omega_R} (1+\Delta_{22}) \right] \frac{\partial^2 P[\chi]}{\partial \chi \partial \chi^*}\;.
\end{align}
Thus, we identify the noise term as
\begin{equation}\label{Noise_term}
    {\cal N} = D_{22} - \frac{\Omega_I}{\Omega_R} (1 + \Delta_{22}) = D_{22} + 2 (1+ \Delta_{22}) \Sigma_{12}\;.
\end{equation}

Next, the drift term can read off from the diagonal terms containing a single derivative: 
\begin{align}
    \left.\frac{\rd \varrho}{\rd \tau}\right|_{\chi=\varphi} & \sim - \left( \frac{z'}{z} + 2 \Delta_{12} \right) 2 (1 - 2 \Omega_R |\chi|^2) \varrho(\chi,\varphi = \chi) \\
    & = - \left( \frac{z'}{z} + 2 \Delta_{12} \right) (2 + \chi \partial_{\chi} + \chi^* \partial_{\chi^*}) P[\chi] \\
    & = - \left( \frac{z'}{z} + 2 \Delta_{12} \right) \nabla \cdot ((\chi,\chi^*) P[\chi])\;,
\end{align}
such that
\begin{equation}
    {\cal D} = \frac{z'}{z} + 2 \Delta_{12}\;.
\end{equation}
Finally, after collecting terms, the Fokker--Planck equation can be written
\begin{equation}
    \frac{\rd P[\chi]}{\rd \tau} = \left[ D_{22} - \frac{\Omega_I}{\Omega_R} (1 + \Delta_{22}) \right] \nabla^2 P[\chi] - \left( \frac{z'}{z} + 2 \Delta_{12} \right) \nabla \cdot ((\chi,\chi^*) P[\chi])\;.
\end{equation}
First, we note that both the drift and noise terms manifest in the same way in the transport equation for $\Sigma_{11}$ (eq.~\eqref{eq:Teqs}), such that $\Sigma_{11}^\prime \sim \mathcal{N} + \mathcal{D} \,\Sigma_{11}$, where one must remember that $2 \Sigma_{12} = - \Omega_I/\Omega_R$. It is reassuring that the Fokker--Planck equation for the PDF, derived using the field-space representation, resembles that for the configuration-space power spectrum. Second, $2 \Sigma_{12}$ term is the only one that appears in the free theory noise, which in the context of stochastic inflation would be interpreted as being local in time and emerging from the sub-horizon fluctuations. Here, we see explicitly the corrections (in the form of $D_{22}$ and $\Delta_{22}$ in eq.~\eqref{Noise_term}) due to the system-environment interaction through the ME coefficients which are manifestly time non-local. 

\begin{figure}
     \centering
     \begin{subfigure}[b]{0.48\textwidth}
         \centering
         \hspace*{-1.5cm}
         \includegraphics[width=1.3\textwidth]{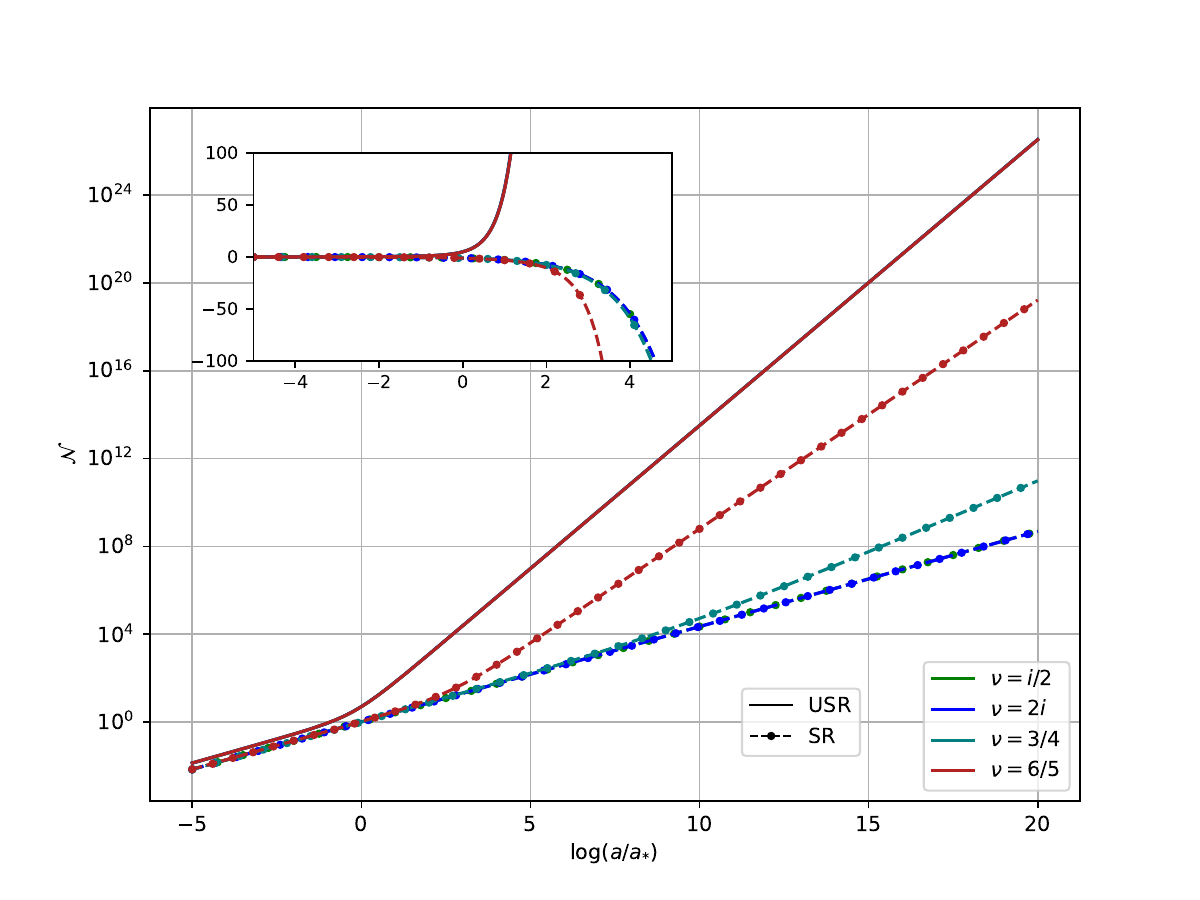}
         \caption{}
         \label{fig:noise_a}
     \end{subfigure}
     \hfill
     \begin{subfigure}[b]{0.48\textwidth}
         \centering
         \hspace*{-0.8cm}
         \includegraphics[width=1.3\textwidth]{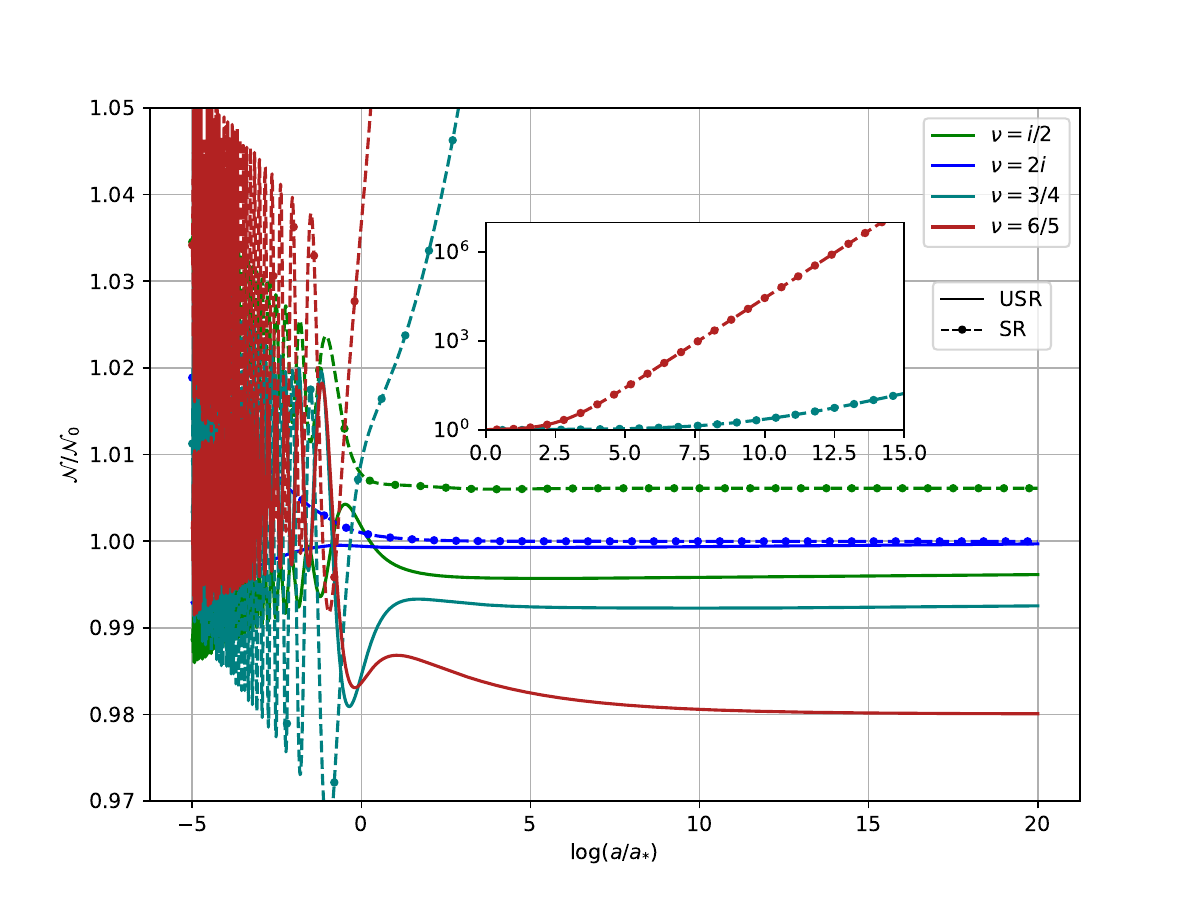}
         \caption{}
         \label{fig:noise_b}
     \end{subfigure}
        \caption{Left: Noise for various $\nu$ values at a representative coupling strength $\lambda/H =0.1$, on log scale. The inset illustrates that USR (SR) curves evolve towards positive (negative) values. Right: Ratio of the computed noise to the corresponding free-theory prediction. The box zooms in on the SR curves departing significantly from a ratio of 1.}
\end{figure}

Fig. \ref{fig:noise_a} shows the evolution of the noise term for different couplings. There is significant degeneracy in the noise term for an USR background. Fig.~\ref{fig:noise_b} presents the ratio between the noise term and its value in the free theory. It is observed that the noise term is slightly larger in the SR case and smaller in the USR scenario. These ratios stabilise at approximately constant values after horizon crossing. 

Finally, it is worth noting that the observed behaviour of the noise can be somewhat misleading, particularly for the ultra slow-roll scenario, where the physics is better understood by examining the physical fields, namely, $\zeta$. Therefore, we now derive the Fokker--Planck equation for these fields. We define the new probability distribution function as ${\cal P}[\zeta] = z^2 P[\zeta] = 2 \epsilon a^2 \Mp^2 P[\zeta]$, where the $z^2$ term originates from the Jacobian of the PDF integral. Considering this, and how the derivatives with respect to the fields change, we obtain
\begin{equation}
    \frac{\rd {\cal P}[\zeta]}{\rd \tau} = 2 \frac{z'}{z} {\cal P}[\zeta] + \frac{1}{z^2}\left[ D_{22} - \frac{\Omega_I}{\Omega_R} (1 + \Delta_{22}) \right] \nabla^2_{(\zeta)} {\cal P}[\zeta] - \left( \frac{z'}{z} + 2 \Delta_{12} \right) \nabla_{(\zeta)} \cdot ((\zeta,\zeta^*) {\cal P}[\zeta])\;,
\end{equation}
where the ansatz of the PDF now reads
\begin{equation}
    {\cal P}[\zeta] = z^2 C \exp \left( -2 z^2 \Omega_R |\zeta|^2 \right)\;,
\end{equation}
a function appropriately normalised for $\zeta$. 

Notice how the noise term has been rescaled by $z^{-2}$, which shows significantly different behaviour depending on the dynamics of the background. In the free-theory, the noise term in SR would be suppressed as $e^{-N}$, whereas for USR the amplitude of the noise increases as $e^{7N}$ (or $e^{-2N}$ vs. $e^{6N}$ if $t$ is the time variable).

\section{Markers of the Quantum State Evolution}\label{sec:QSE}
We are now able to leverage the formalism assembled in the previous sections.
One of the defining characteristics of a quantum state is its ``mixedness'' or how entangled it is. Recently there has been particular emphasis on examining the purity \cite{Colas:2022kfu,Colas:2024xjy,Burgess:2024eng} and entanglement entropy \cite{Brahma:2023hki, Brahma:2020zpk,Colas:2024ysu} of the quantum state of inflationary perturbations. This has the dual goal of discovering new ``smoking gun'' signals for the quantum origin of these fluctuations as well as explaining how they classicalise after horizon-crossing. However, a less appreciated aspect of these computations involve understanding the effect such non-unitary effects (namely, decoherence) has on cosmological observables \cite{martin2018non,Boyanovsky:2015tba}. 

For our model involving the leading order term in the open EFT of inflation, it has been established the dynamics are generically non-Markovian. We will show that this conclusion remains unaltered for USR backgrounds as well (see Sec.~\ref{sonm} for details). For non-Markovian systems, decoherence of the initial state does not follow on expected lines, and often reveal new intriguing phenomena such as recoherence, where a state regains purity post-horizon crossing after an initial loss in coherence. This section will show that recoherence does not take place in USR backgrounds, contrasting with SR scenarios, even though the system remains non-Markovian, leading us explore the physical reason behind this in the next section.

Another important goal of this section is to explore the evolution of entanglement entropy for a non-Markovian open EFT. Classical notions of non-Markovianity, as time-locality alone, loses its interpretation in quantum mechanics and properties of quantum non-Markovian systems often differ from their classical counterparts. By examining the eigenvalues of the dissipator matrix, as well as the time evolution of the entanglement entropy, we shall elucidate the role of diffusion in this story. 

\subsection{Purity}
\begin{figure}[H]
     \centering
     \begin{subfigure}[b]{0.48\textwidth}
         \centering
         \hspace*{-1.5cm}
         \includegraphics[width=1.3\textwidth]{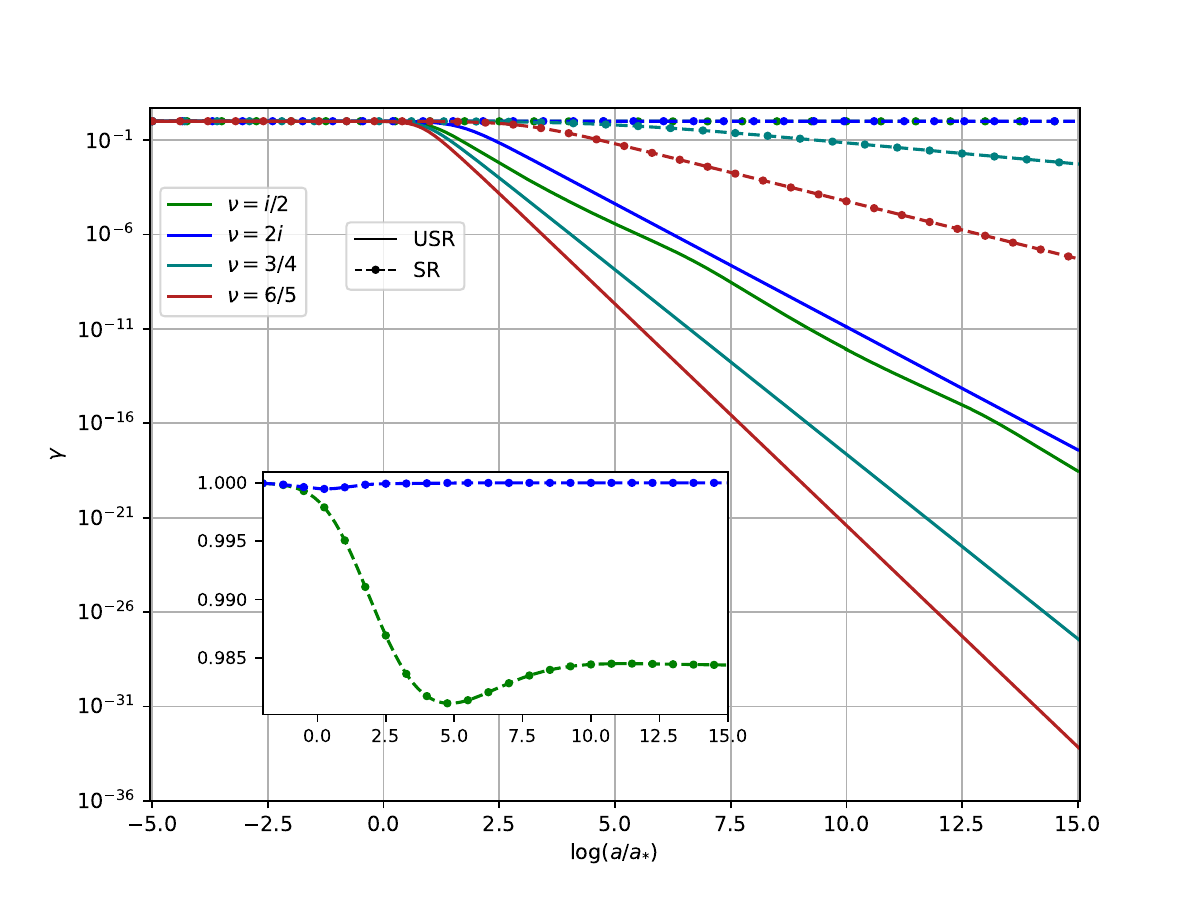}
         \caption{}
     \end{subfigure}
     \hfill
     \begin{subfigure}[b]{0.48\textwidth}
         \centering
         \hspace*{-0.8cm}
         \includegraphics[width=1.3\textwidth]{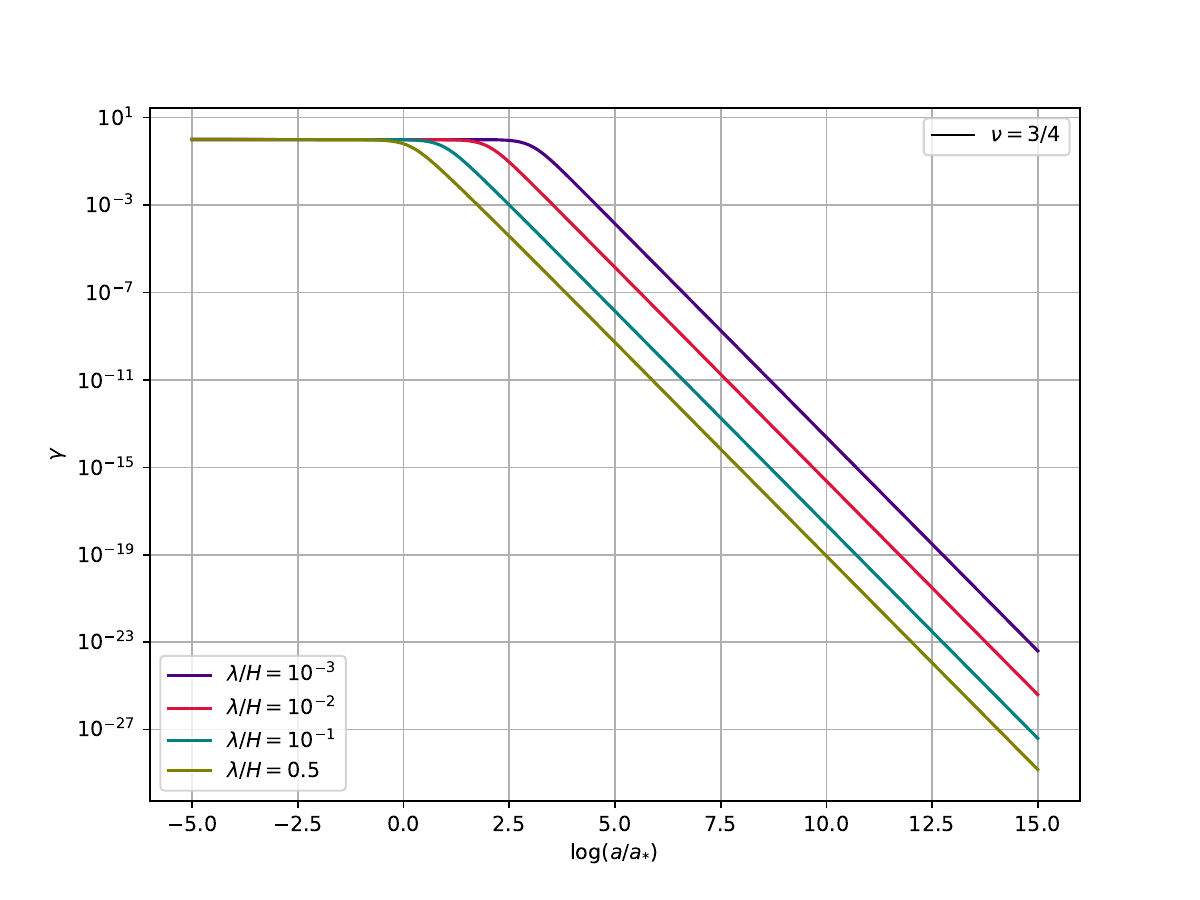}
         \caption{}
     \end{subfigure}
        \caption{Left: Purity evolution for different parameters for a system under a SR (dashed lines) and USR (solid lines) background for different values of $\nu$. The inset zooms into the SR curves showing recoherence. Right: Purity evolution under an USR background, for different $\lambda/H$ values, with fixed $\nu = 3/4$.}
        \label{fig:pur}
\end{figure}

Let us start by evaluating the purity for the different scenarios we have considered so far. We have already computed this quantity in eq.~\eqref{eq:pur}, where we showed that
\begin{equation}\label{eq:purxi}
    \gamma_k = \left(1 + \frac{\xi}{\Omega_R}\right)^{-1} = \frac{1}{4\ {\rm det}\ \Sigma_{ab}}\;,
\end{equation}
where $\xi(\tau_0) = 0$ for initially pure states, and the ratio $\xi/\Omega_R$ increasing over time if decoherence takes place, leading to $1 \geq \gamma_k > 0$. 
We show in fig. \ref{fig:pur} the values of purity for the same regions of parameter space studied so far.

The SR case has been extensively studied in previous works \cite{Colas:2022kfu,Colas:2024xjy}. However, it is worth mentioning that we have sampled parameters that highlight the distinctive features found in this model, namely (near-)complete recoherence for $\nu = 2i$, purity freezing for $\nu = i/2$, and decoherence for $\nu = 3/4$ and $6/5$. In contrast, we do not see the same variety in the USR regime. Instead, we clearly observe that decoherence occurs for the same parameters, and does so more efficiently than their SR counterparts. In order to gain some insight into the behaviour of these curves, it can be illustrative to write down the equation of motion of the (inverse of the) purity, which is given by
\begin{align}
    \frac{\rd}{\rd \tau} \left( 1 + \frac{\xi}{\Omega_R}\right) & = 4 \Delta_{12} \left( 1 + \frac{\xi}{\Omega_R}\right) - 4 D_{12} \frac{\Omega_I}{\Omega_R} + 2 D_{22} \Omega_R \left( 1 + \frac{\xi}{\Omega_R} + \frac{\Omega_I^2}{\Omega_R^2} \right)\;,
\end{align}
or, in terms of the covariance matrix elements,
\begin{align}
     \frac{\rd}{\rd \tau} ({\rm det}\ \Sigma_{ab}) = 4 \Delta_{12} ({\rm det}\ \Sigma_{ab}) + 2 D_{12} \Sigma_{12} + D_{22} \Sigma_{22}\;.
\end{align}
An increase in the determinant corresponds to a decrease in purity, indicating decoherence. These equations effectively illustrate the types of processes at play. The first term on the rhs is mediated by the dissipation term, $\Delta_{12}$. Fig.~\ref{L12} clearly demonstrates that this function takes relatively small values and rapidly decreases after horizon exit. This suggests that, for the problem at hand, dissipative dynamics is not the primary driver of decoherence. Consequently, both decoherence and recoherence (when it occurs in SR backgrounds) are primarily driven by diffusive processes. One can readily check this despite the absence of analytical results, provided that the correlation functions do not significantly deviate from their free-theory values. Under these circumstances, it can be seen that the momentum-momentum diffusion term dominates, both in the SR and USR regimes. 

For example, for $\nu = i\mu$, $D_{22}$ is an oscillating function, and thus for USR, $({\rm det}\ \Sigma_{ab})' \sim \Sigma_{22} \sim (p\tau)^{-4}$. This translates into a purity evolution of the form $\gamma \sim \exp(-3N)$, which is the overall behaviour of the blue and green curves post-horizon exit in fig.~\ref{fig:pur}. On the other hand, for $\nu \in \mathbb{R}$, $D_{22} \sim (-p\tau)^{-2\nu}$, so that $({\rm det}\ \Sigma_{ab})' \sim (-p\tau)^{-2\nu - 4}$, or $\gamma \sim \exp((-2\nu - 3)N)$. This simple analysis predicts a purity varying as $e^{-9N/2}$ for $\nu = 3/4$ and as $e^{-27N/5}$ for $\nu = 6/5$, which matches the behaviour of the corresponding curves in fig.~\ref{fig:pur}. This discussion reinforces previous indications \cite{Brahma:2024yor} that recoherence is \textit{not} a necessary implication of a non-Markovian open quantum system. 

Finally, it is worth introducing another parametrisation of purity that the reader can find useful when looking at decoherence/recoherence. Notice that the definition of the determinant of the covariance matrix can be written as
\begin{equation}\label{eq:nparam}
    {\rm det}\ \Sigma_{ab} \sim \langle \hat{v}^2 \rangle \langle \hat{\pi}^2 \rangle - \left(\frac{1}{2} \left\{\hat{v}, \hat{\pi}\right\} \right)^2 = \left(\bar{n}_k + \frac{1}{2}\right)^2\;,
\end{equation}
where one can simply identify 
\begin{equation}\label{eq:defn}
    \bar{n}_k = \frac{1 - \sqrt{\gamma_k}}{2\sqrt{\gamma_k}}\;.
\end{equation}
The advantage of this parametrization is that eq.~\eqref{eq:nparam} resembles the uncertainty relation for a simple harmonic oscillator, with $\bar{n}_k$ denoting the occupation number.\footnote{This interpretation warrants careful consideration. One might be inclined to associate $\bar{n}_k$ with the expected occupation number of the $k$-mode in a squeezed state, which evolves according to $(2k\tau)^{-2}$ in pure de Sitter. However, it is important to note that this expression is valid only in the context of free-theory evolution of the inflationary perturbation state. In such a scenario, $\bar{n}_k = 0$, since $\gamma_k = 1$. In other words, $\bar{n}_k > 0$ only in the presence of interactions, and could be interpreted as excitations superposed on top of those associated to squeezed states.} In this framework, decoherence is associated with an increase in the population of the mode, while recoherence corresponds to its depopulation. (See also the discussion of
Ref.~\cite{Anderson:2005hi}.)
This interpretation will be useful to understand the evolution of entanglement entropy in the next section.

\subsection{Entanglement Entropy}
\begin{figure}[h!]
    \centering
    \includegraphics[width=\textwidth]{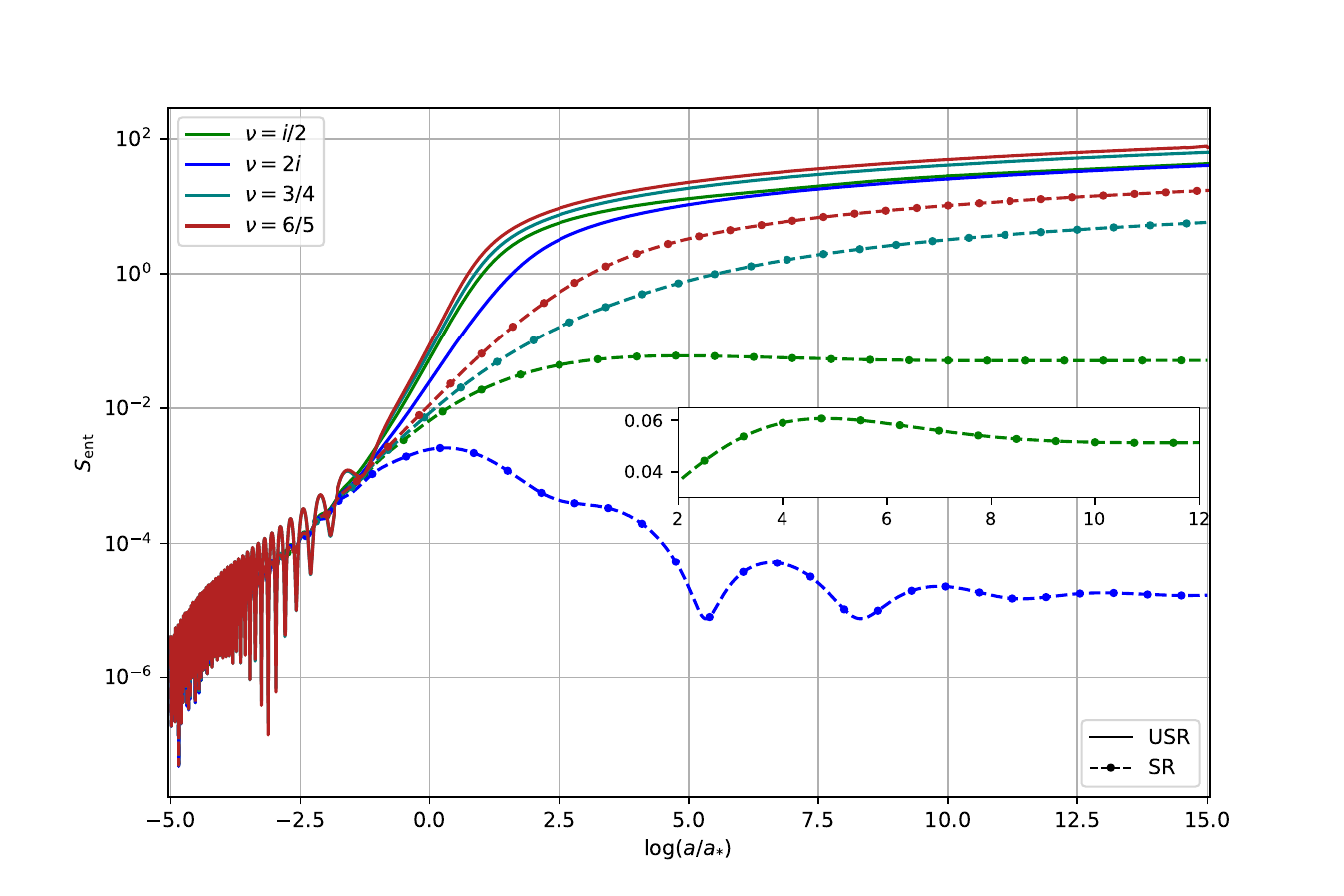}
    \caption{Von Neumann entropy for the studied system under a SR (dashed lines) and USR (solid lines) background. The box zooms into the behaviour of the SR curve corresponding to $\nu = i/2$, where a small decrease in entropy takes place before freezing.}
    \label{fig:vN}
\end{figure}

\begin{figure}[h!]
     \centering
     \begin{subfigure}[b]{0.48\textwidth}
         \centering
         \hspace*{-2cm}
         \includegraphics[width=1.25\textwidth]{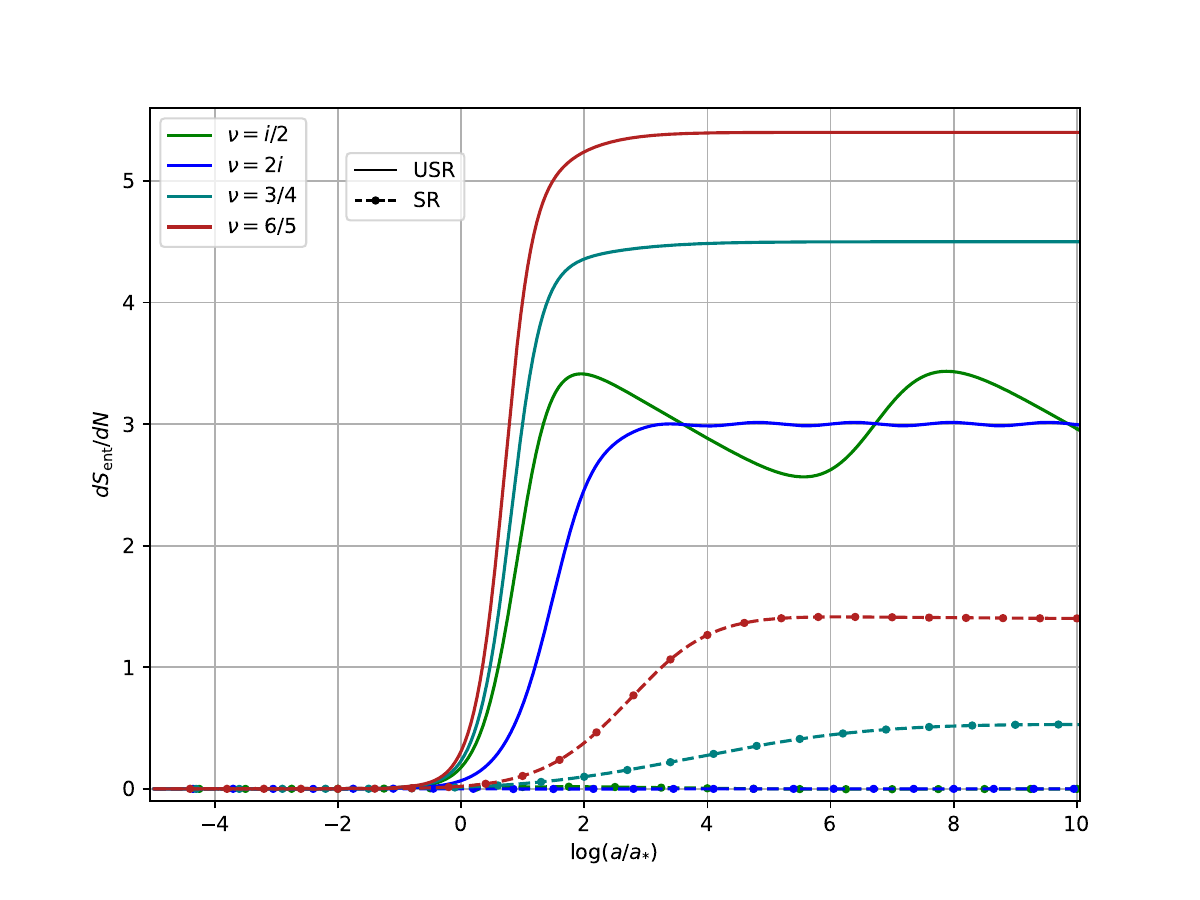}
         \caption{}
     \end{subfigure}
     \hfill
     \begin{subfigure}[b]{0.48\textwidth}
         \centering
         \hspace*{-1.2cm}
         \includegraphics[width=1.3\textwidth]{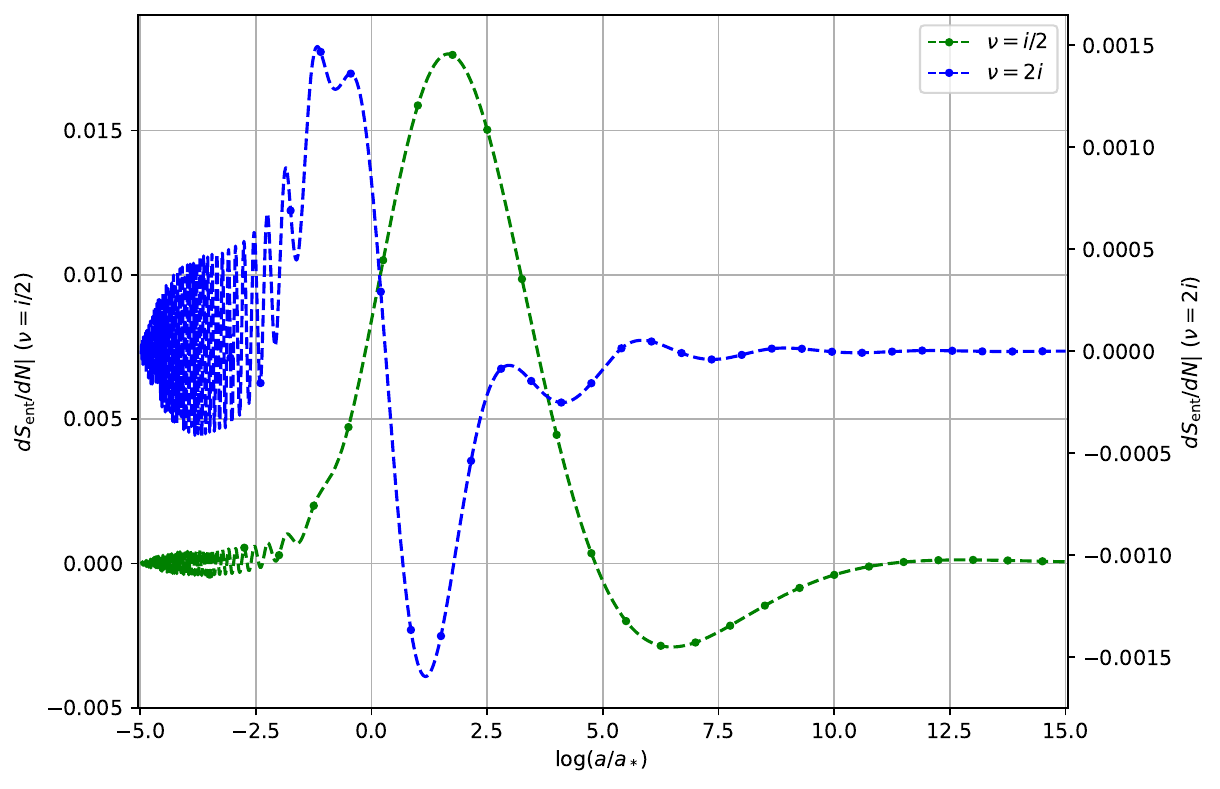}
         \caption{}
     \end{subfigure}
        \caption{Derivative of the von Neumann entropy illustrated in fig.~\ref{fig:vN}. The left plot displays this function for all considered parameters, while the right plot zooms in on the curves exhibiting recoherence. Note the different scales on the axes for both curves for the plots on the right panel.}
        \label{fig:ds}
\end{figure}

Another interesting quantum informatic measure to examine is the von Neumann entanglement entropy of the system density matrix. Following \cite{Hollowood:2017bil, Joos:1984uk}, we first factorize the density matrix as $\varrho( \chi, \varphi) = \varrho_R (\chi_R, \varphi_R) \ \varrho_I (\chi_I, \varphi_I)$ to subsequently diagonalise each bit. The eigenvalue equation is:
\begin{equation}
    p_n \psi_n^R (\chi) = \int \rd \chi' \rho_R(\chi,\chi') \psi_n (\chi')\;, \qquad \psi_n (\chi) = N_{\psi} H_n (\alpha \chi) \exp(-\beta \chi^2)\;,
\end{equation}
where $H_n$ are Hermite polynomials and $\alpha, \beta$ are functions of $\Omega$ and $\xi$ given by
$$\alpha^2 = 2 \Omega_R \sqrt{1 + \frac{\xi}{\Omega_R}}\;, \qquad \beta = i \Omega_I + \frac{\alpha^2}{2}\;.$$
The eigenstates $\psi_n (\chi)$ are defined by Hermite polynomials because of the Gaussian form of the density matrix. In this way, performing the integrals one finds that
\begin{equation}\label{eq:p0}
    p_0 = \frac{2}{1 + \sqrt{1+\xi/\Omega_R}}\;, \qquad p_n = p_0 (1-p_0)^n\;.
\end{equation}
Consequently, the von Neumann entropy is given by
\begin{equation}
    S_{\rm ent} = - 2 \sum_{n=0}^{\infty} p_n \ln p_n = - 2 \left[ \ln p_0 + \frac{1-p_0}{p_0} \ln (1-p_0) \right]\;,
\end{equation}
where the factor of $2$ comes from adding up the contribution from the real and imaginary contributions of the density matrix. This is a standard result for the von Neumann entropy for Gaussian systems \cite{PhysRevE.66.036102,Serafini_2005,horhammer_2008,Lally:2019dit} (see \cite{Ribes-Metidieri:2024vjn} for a recent discussion on how this translates for local observables in the case of a free massive scalar field).
We illustrate in fig.~\ref{fig:vN} the evolution of the von Neumann entropy. As expected, in the USR regime, the entropy increases rapidly across all parameters. In the SR background, a similar albeit less pronounced increase is observed, except in cases exhibiting recoherence. In these cases, the entropy decreases while oscillating near horizon exit, before stabilizing at a constant value. Just as for recoherence, this peculiar evolution of the entanglement entropy is also a result of the underlying non-Markovianity of the system.  To illustrate this more clearly, we show the rate of growth of entropy in fig.~\ref{fig:ds} which clearly turns negative (transiently) for the parameters in the SR case that correspond to recoherence or purity-freezing (right panel).


Finally, from eqs.~\eqref{eq:purxi}, \eqref{eq:defn}, and \eqref{eq:p0}, it follows that the entanglement entropy can be expressed in terms of $\bar{n}_k$ as
\begin{equation}
    S_{\rm ent} = 2 \left[ (1 + \bar{n}_k) \ln(1 + \bar{n}_k) - \bar{n}_k \ln (\bar{n}_k) \right]\;,
\end{equation}
which is recognized as (twice) the entropy of a system of $\bar{n}_k$ non-interacting bosons. As noted in \cite{glavan_notes}, $\bar{n}_k$ then represents the number of uncorrelated regions in a Gaussian state, reinforcing the idea that decoherence corresponds to the regime $\bar{n}_k \gg 1$.

\subsection{On (non-)Markovianity}\label{sonm}

A concept that naturally arises when discussing open quantum systems is Markovianity (a comprehensive reference on the various types and definitions of non-Markovianity is \cite{Li_2018}). In classical physics, this concept is associated with random forces acting on a system, where the amplitude is independent of the system's history. In other words, the random noise acting on a system at any given time is uncorrelated with the noise at earlier times. Hence, time-locality is often considered synonymous with Markovianity. This notion is sometimes extended to, say, the Redfield master equation \eqref{eq:Red}, where time-locality and history independence imply a master equation of the form
$$
\Ri' = - \int_{0}^{\infty} \rd t'\ \Tr_\e \Big[ \tilde{V}(t), \big[ \tilde{V}(t-t'), \tilde{\rho}_\s(t) \otimes \tilde{\rho}_\e \big] \Big].
$$
This is commonly referred to as the Markovian approximation and is justified for scenarios involving weak system-environment couplings and where typical environmental timescales cannot be resolved \cite{banerjee2018open}. Notice that the argument of the interaction Hamiltonian, as well as the integration limits, are chosen to satisfy a `memoryless' condition. 

However, Markovianity is more rigorously formalised based on the properties of quantum dynamical semigroups. If there is a dynamical map $W(t)$ such that, given $\rho_\s (t) = W(t) \rho_\s (0)$, the system exhibits Markovian behaviour if the semigroup property $W(t_1) W(t_2) = W(t_1 + t_2)$ holds for $t_1, t_2 \geq 0$. In this case, the ME can be written in the Gorini--Kossakowski--Sudarshan--Lindblad (``GKS--Lindblad'') form: 
\begin{equation}
    \h{\rho}_\s' = {\cal L}\h{\rho}_\s = -i [\h{H}, \h{\rho}_\s] + \sum_j \left(\h{A}_j \h{\rho}_\s \h{A}_j^\dg - \frac{1}{2} \{\h{A}_j^\dg \h{A}_j, \h{\rho}_\s \}\right)\;,
\end{equation}
where $\h{A}_j$ are the Lindblad operators. The approximations mentioned before should thus lead to a ME of this type, which is also referred to as a Quantum Markovian master equation. This concept of Markovianity is generalised by allowing the non-unitary part of the ME to be of the form $\sum_j \gamma_j (t) \left(\h{A}_j \h{\rho}_\s \h{A}_j^\dg - \frac{1}{2} \{\h{A}_j^\dg \h{A}_j, \h{\rho}_\s \}\right)$, with $\gamma_j (t) \geq 0$. Consequently, quantum non-Markovianity is associated with the existence of at least one of the eigenvalues $\gamma_j(t) <0$ \cite{Li_2018,Hall_2014}. This more general form is the one closely associated to our case under study, described by the ME eq.~\eqref{eq:MELf}.

\begin{figure}[h!]
     \centering
     \begin{subfigure}[b]{0.48\textwidth}
         \centering
         \hspace*{-1.5cm}
         \includegraphics[width=1.3\textwidth]{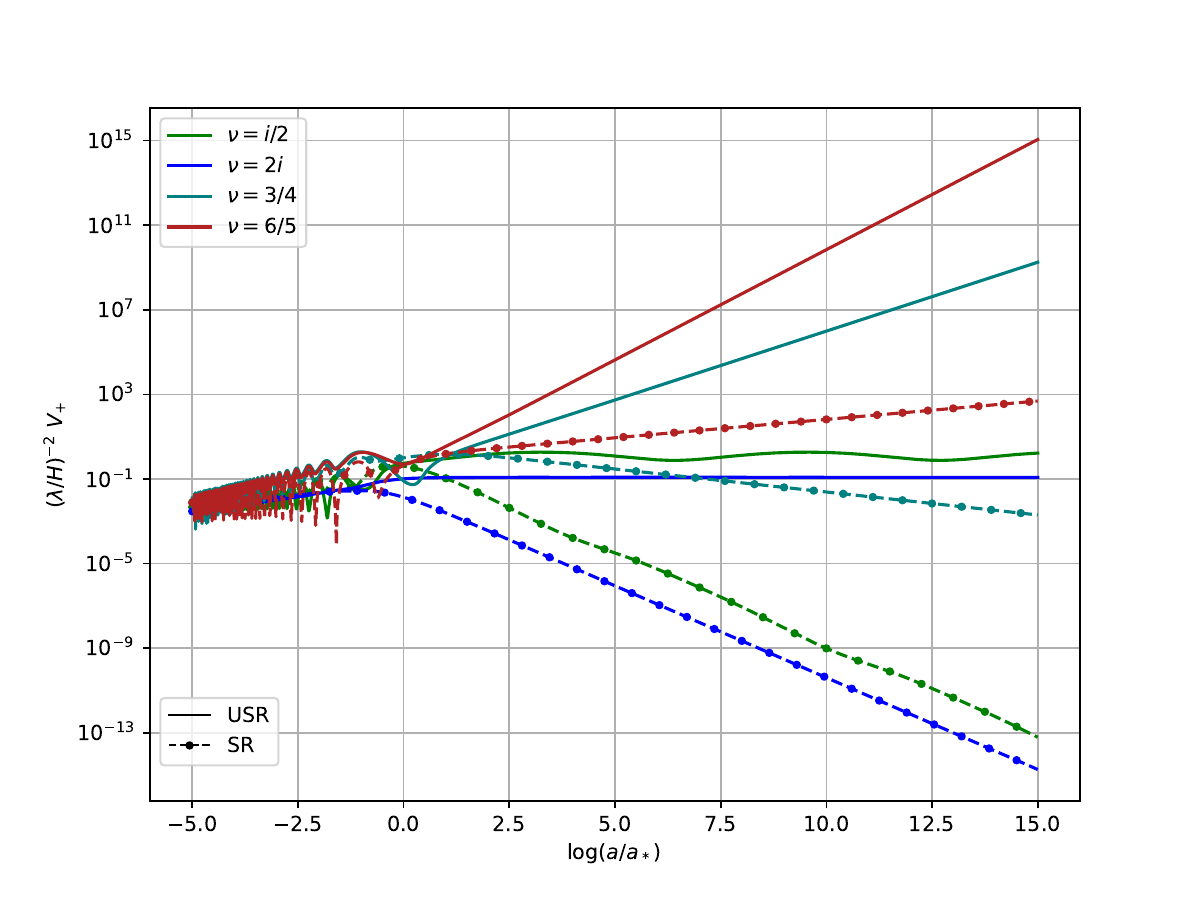}
         \caption{}
     \end{subfigure}
     \hfill
     \begin{subfigure}[b]{0.48\textwidth}
         \centering
         \hspace*{-0.8cm}
         \includegraphics[width=1.3\textwidth]{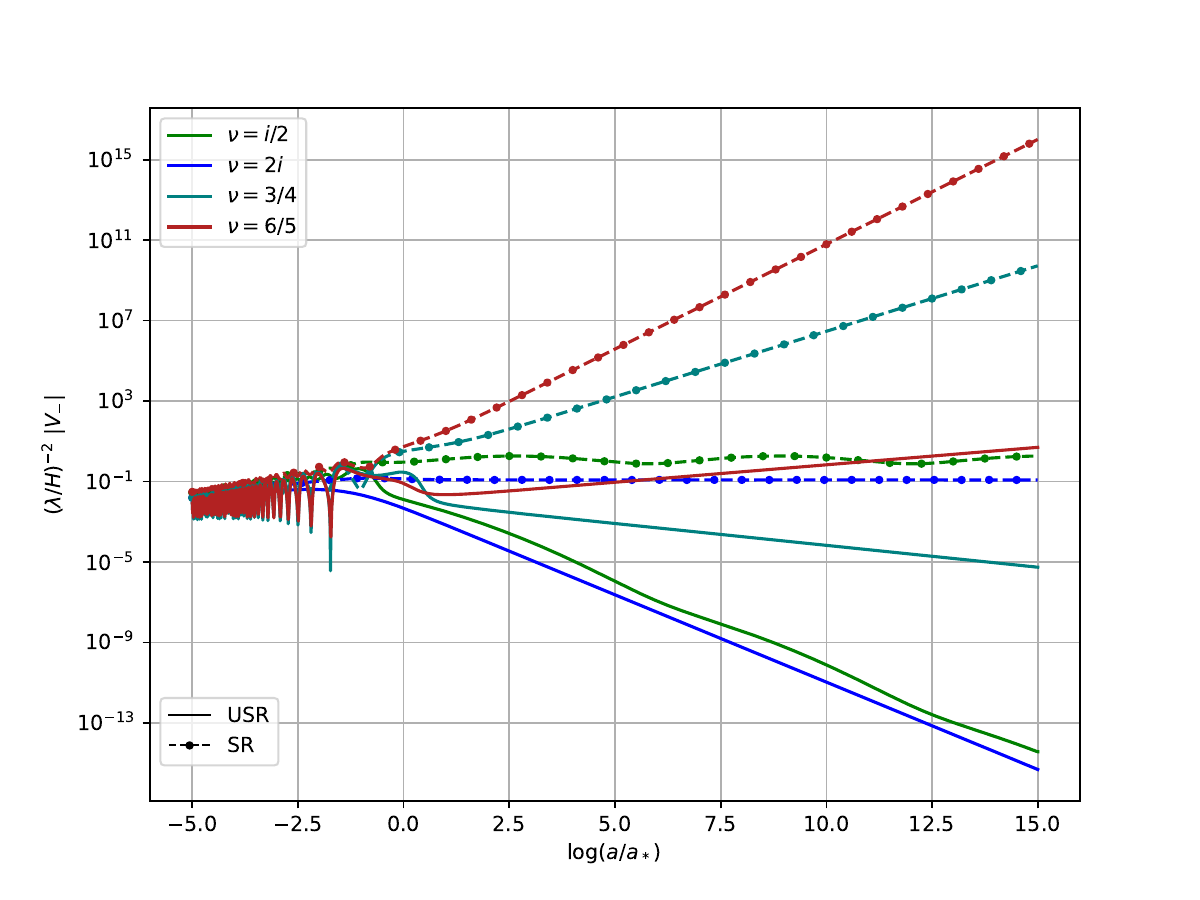}
         \caption{}
     \end{subfigure}
        \caption{Rescaled eigenvalues of the dissipator matrix. The left plot shows the positive eigenvalue, while the right plot displays the absolute value of the negative eigenvalue.}
        \label{fig:EVs}
\end{figure}

In fig.~\ref{fig:EVs} we show the eigenvalues of the dissipator matrix associated to eq.~\eqref{eq:MELf}. One of them is always positive (left panel) and the other always negative (right panel). Somewhat surprisingly, the magnitude of the negative eigenvalues is larger than the positive ones for SR, whereas for USR they decrease with time for most cases except for $\nu = 6/5$. Nevertheless, even for this parameter choice, the positive eigenvalue is dominant over the negative one for USR. A noteworthy case is that of complex $\nu$ for SR inflation. Here, the magnitude of the negative eigenvalue oscillates around a constant value, whereas the positive one decreases rapidly after horizon crossing. This behaviour is particularly significant as it correlates with recoherence and purity-freezing. In these instances, the negative eigenvalue, indicative of non-Markovianity, maintains its oscillation around a constant value.

It is crucial to observe that this behaviour closely resembles that of $D_{22}$ (see fig.~\ref{D22}). For SR inflation, in which case the dominance of the negative eigenvalue is ubiquitous across all considered parameters, $D_{22}$ serves as an excellent distinguishing feature of such exotic phenomena. Conversely, for USR inflation, the magnitude of the positive eigenvalue consistently dominates over the negative one. In these cases, $D_{22}$ also effectively describes the evolution of the dominant eigenvalue.
This difference in eigenvalue behaviour between SR and USR inflation appears to explain why recoherence occurs only in SR scenarios and not in USR ones. The persistent non-Markovianity in SR, characterized by the oscillating negative eigenvalue, seems to facilitate a late regain in coherence. In contrast, the dominance of the positive eigenvalue in USR inflation likely inhibits such effects.

Having described possible correlations between diffusion terms and recoherence, recall that the same choice of parameters also lead to a transient decrease in the entanglement entropy of the subsystem mode. A physical way to understand this is that the coupling term between the adiabatic and the entropic mode is proportional to the decaying mode in the SR case, and this can get turned-off  (quenched) dynamically for certain parameter ranges. This leads to incomplete decoherence or a loss of entanglement between $\s$ and $\e$. On the other hand, for the non-attractor solution, the momentum mode (corresponding to the adiabatic perturbation) has a growing mode after horizon exit, such that the interaction is never turned off. This leads to the system quickly evolving to a highly entangled state and results in efficient decoherence.

\section{Multiple-phase scenario}
In this section, we explore the role of the memory of the adiabatic mode by examining a scenario transitioning from SR to USR, followed by another SR phase. This investigation aims to enrich our understanding of how system-memory interactions contribute to the broader discussion on quantum state evolution.

It is natural to ask how the results obtained thus far would fit within a model that considers transitions between attractor and non-attractor backgrounds. Even though it would be interesting to tackle this problem using the TCL technology, obtaining the ME coefficients beyond the perfect dS approximation is highly complex. Thus, dealing with the peculiarities of the interface(s) between backgrounds becomes very challenging. However, a viable approach to obtaining some results remains available, because of the Gaussian interaction: solving the full set of transport equations for the $\s + \e$ sectors. This can be accomplished numerically (and could be done analytically for interactions of the type $\propto \zeta {\cal F}$), with the background evolution being imposed by hand. To do so, let us define the slow-roll parameter: 
\begin{equation}\label{eq:etaph}
    \eta(N) = \frac{3}{2} \big[ \tanh(5(N-N_1)) - \tanh(5(N-N_2)) \big] = \epsilon_1 - \frac{\epsilon_2}{2}\;,
\end{equation}
where this choice of $\eta(N)$ has been
engineered so that $\epsilon_2 \approx -6$ for $N_1 \lesssim N \lesssim N_2$, but assumes negligibly small value elsewhere. Given that $\epsilon_2 = \rd \ln \epsilon_1/\rd N$, it suffices to set an initial condition for $\epsilon_1$ consistent with an accelerated expansion. In this way, we can simulate the evolution of the system (and the environment) for a scenario beginning with SR inflation for $N < N_1$, followed by USR for $N_1 < N < N_2$, and concluding with another SR period for $N > N_2$ (see fig.~\ref{fig:eps}). While this is inherently a toy model, it nonetheless provides valuable illustrative insights. More details are found in Appendix \ref{ApB}.

\begin{figure}[h!]
     \centering
     \begin{subfigure}[b]{0.48\textwidth}
         \centering
         \hspace*{-1.5cm}
         \includegraphics[width=1.3\textwidth]{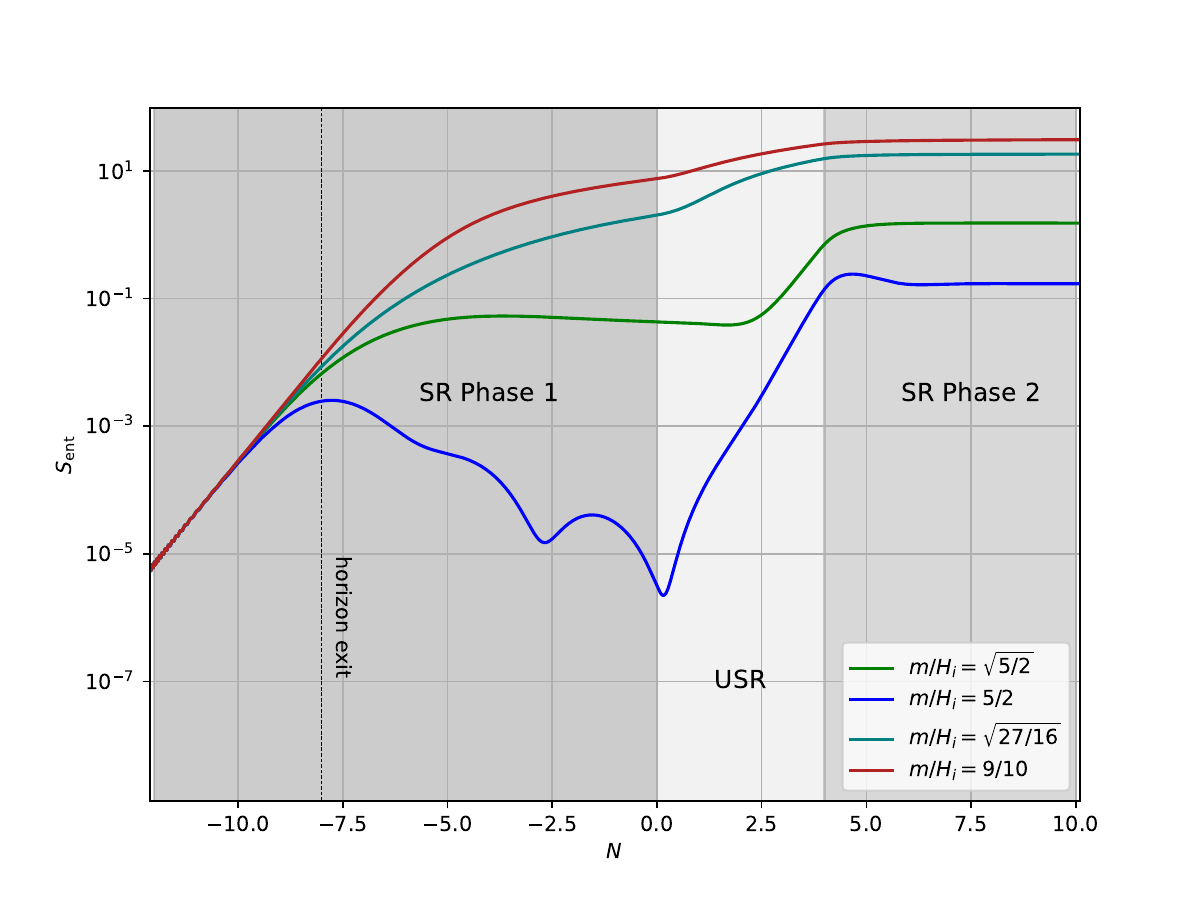}
         \caption{}
     \end{subfigure}
     \hfill
     \begin{subfigure}[b]{0.48\textwidth}
         \centering
         \hspace*{-0.8cm}
         \includegraphics[width=1.3\textwidth]{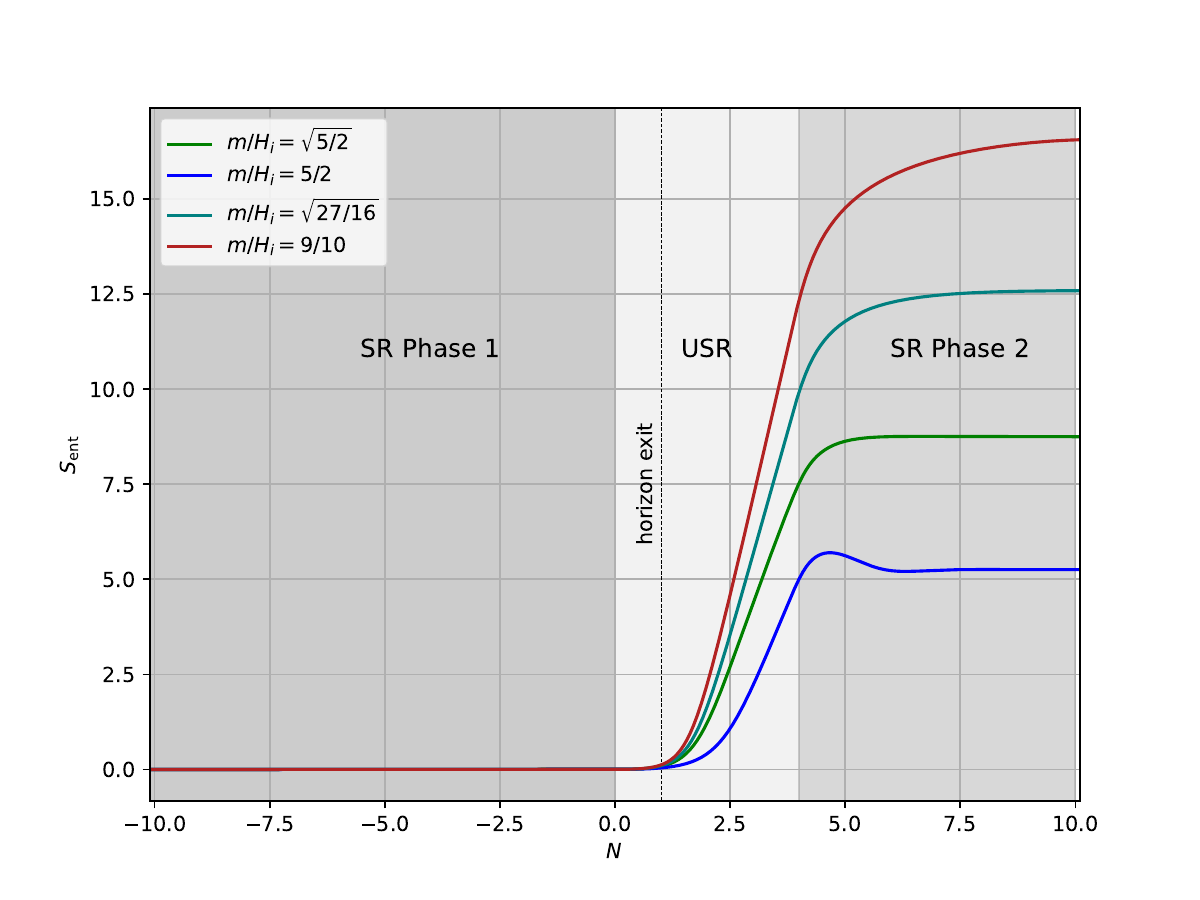}
         \caption{}
     \end{subfigure}
        \caption{Entanglement entropy evolution for the multi-phase inflationary scenario discussed in the text. The left plot depicts the evolution for a mode that exited the horizon during the first SR phase, while the right plot shows the evolution for a mode that crossed the horizon at $N=1$ during USR.}
        \label{fig:sntph}
\end{figure}

In fig.~\ref{fig:purph}, we show the evolution of the purity under the described background, whereas fig.~\ref{fig:sntph} illustrates the entanglement entropy. First, as expected, the transition from SR to USR enhances the efficiency of the decoherence process for modes already outside the horizon. Specifically, for $\nu \in \mathbb{R}$, purity follows a power law that mirrors the one found for a single USR phase. However, soon after the second SR phase kicks in, decoherence decelerates and purity tends to stabilise, even for the cases that already decohere significantly during the first SR phase. Consequently, for sufficiently brief periods of USR, it is plausible that some states may not experience complete decoherence. This observation remains valid even when the mode exits the horizon during the USR phase, a case depicted in the bottom plots of fig.~\ref{fig:purph}. Here, the evolution of purity closely resembles that observed for a single USR phase. However, once again we observe that it tends to stabilise after the transition to the subsequent SR regime.
Notably, for the blue and green curves---associated to parameters leading to recoherence for SR backgrounds---the `final' state is less coherent than when the mode exited the horizon during the first phase. This indicates that the diffusion term(s) responsible for recoherence slowed down the decoherence process during USR. In other words, because of the transition and the short duration of the USR regime, there was insufficient time for the expected USR dynamics to fully manifest, particularly regarding the nature of the diffusion process.

\begin{figure}[h!]
     \centering
     \begin{subfigure}[b]{0.48\textwidth}
         \centering
         \hspace*{-1.5cm}
         \includegraphics[width=1.3\textwidth]{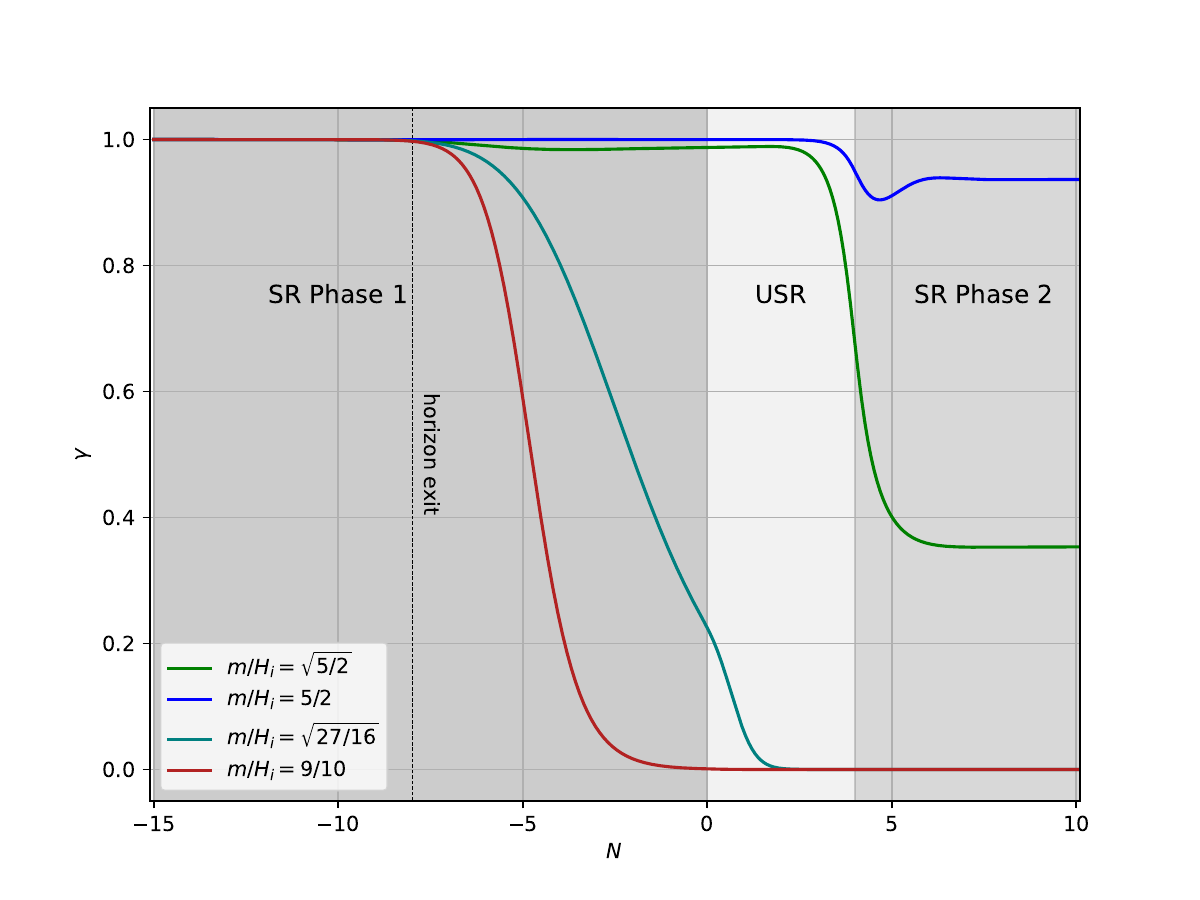}
         \caption{}
     \end{subfigure}
     \hfill
     \begin{subfigure}[b]{0.48\textwidth}
         \centering
         \hspace*{-0.8cm}
         \includegraphics[width=1.3\textwidth]{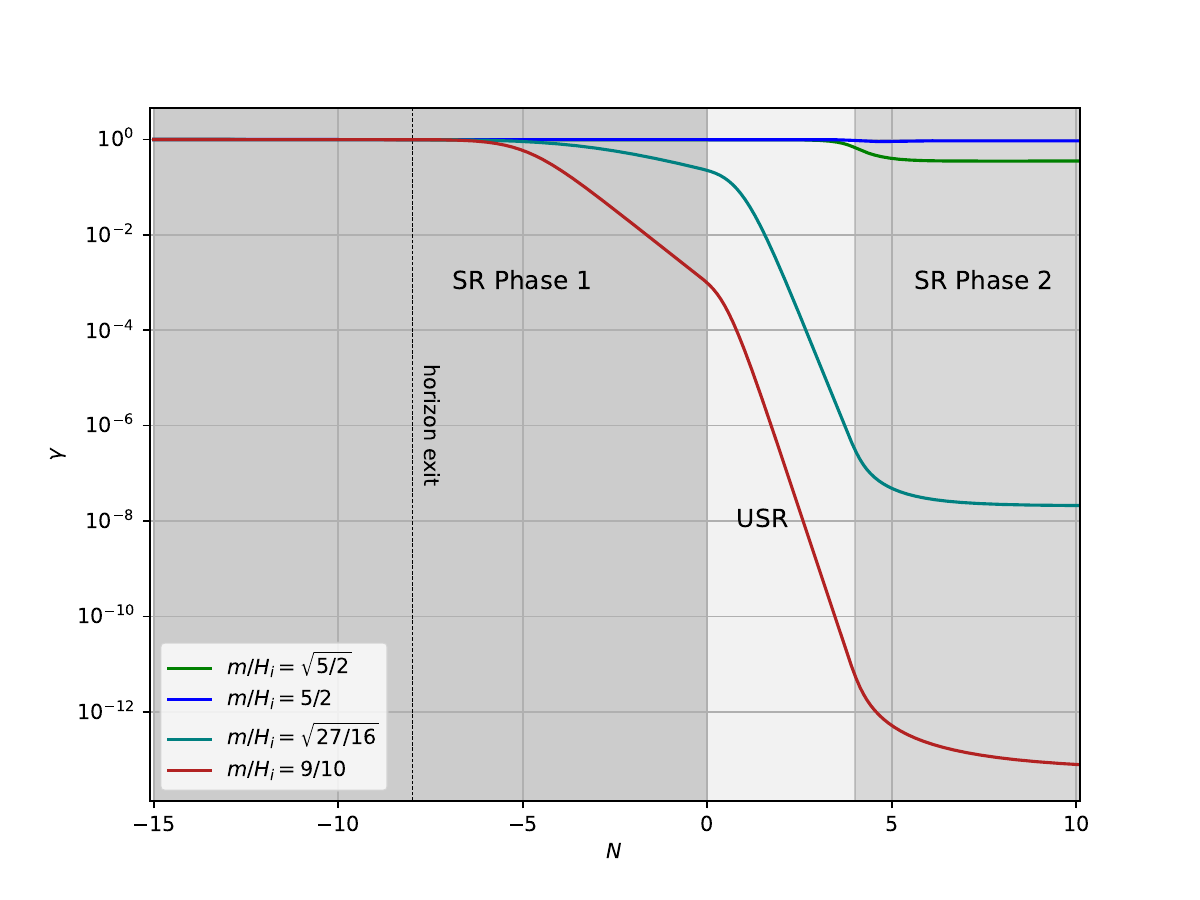}
         \caption{}
     \end{subfigure}\\
     \begin{subfigure}[b]{0.48\textwidth}
         \centering
         \hspace*{-1.5cm}
         \includegraphics[width=1.3\textwidth]{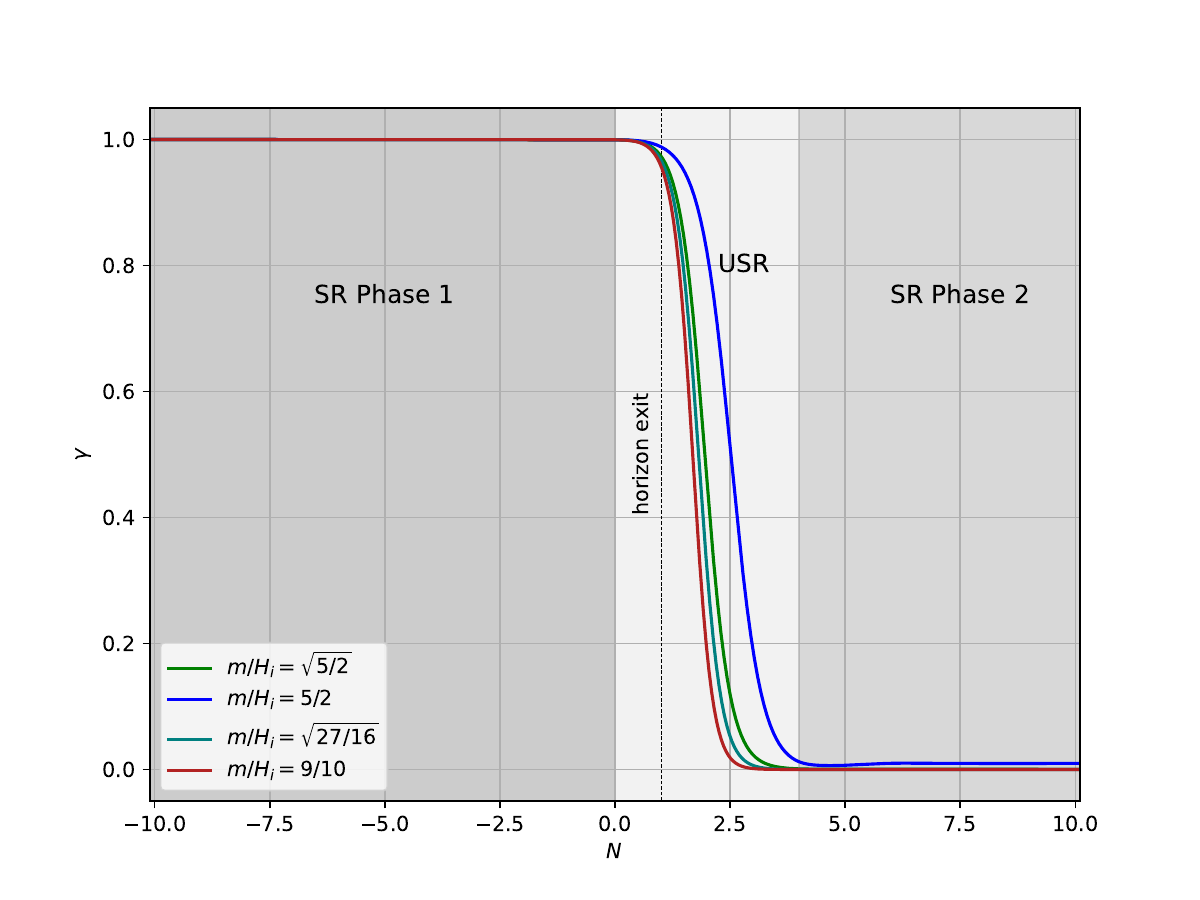}
         \caption{}
     \end{subfigure}
     \hfill
     \begin{subfigure}[b]{0.48\textwidth}
         \centering
         \hspace*{-0.8cm}
         \includegraphics[width=1.3\textwidth]{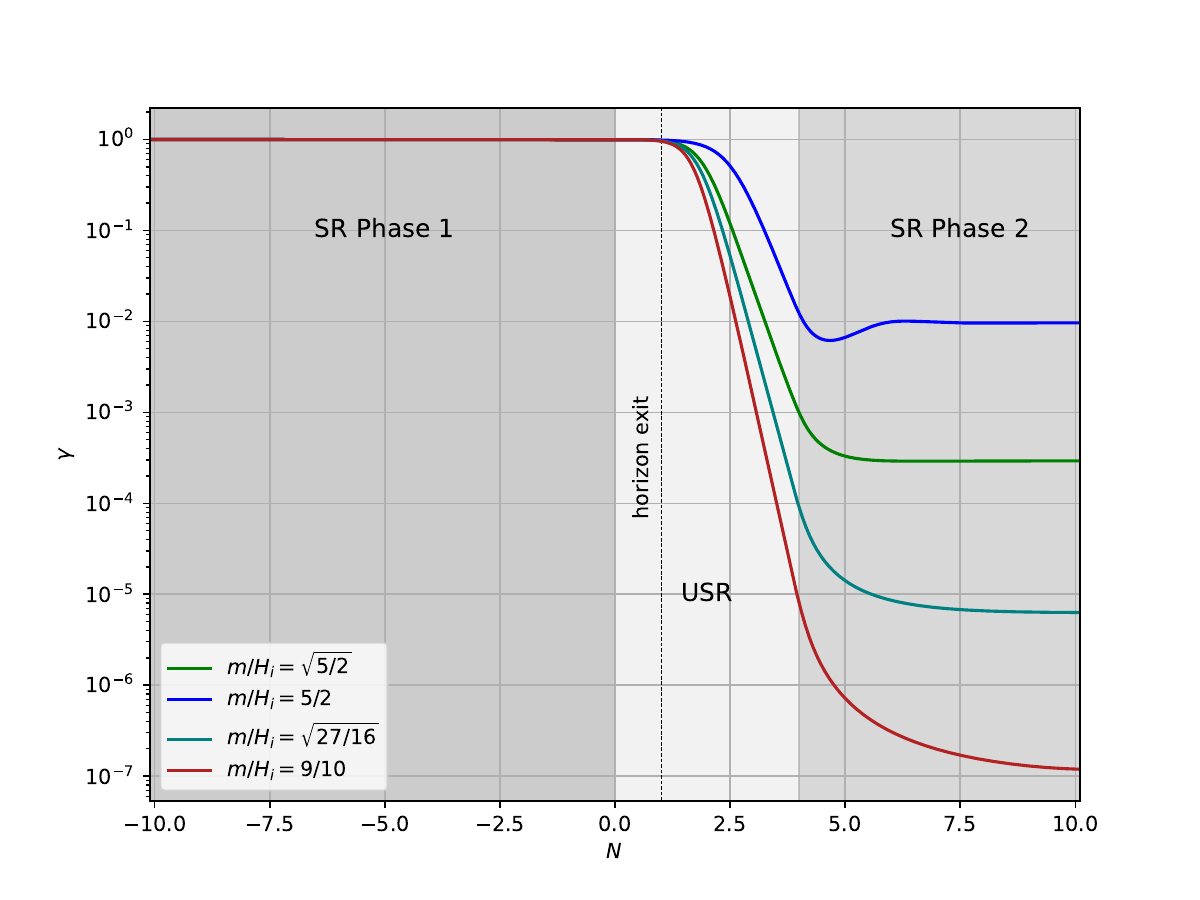}
         \caption{}
     \end{subfigure}
            \caption{Purity evolution for an inflationary scenario consisting of an initial slow-roll (SR) phase, followed by a 4 $e$-fold USR phase, and concluding with a second SR phase. The top plots depict the purity evolution for a mode that crossed the horizon during the first SR phase, indicated by the vertical dotted line at $N = -8$. The bottom plots illustrate the purity evolution for a mode that crossed the horizon during the USR phase, at $N = 1$. Figures on the right are log-scale plots of those on the left. We have considered $\lambda/H_i = 0.1$}
        \label{fig:purph}
\end{figure}

The entropy plots in fig.~\ref{fig:sntph} reinforce this lesson. It is evident that the entanglement entropy rapidly increases when the mode crosses the horizon during USR. However, when the mode crosses the horizon during SR, it takes some time for the green and blue curves to show the expected gradient, with the green curve actually decreasing for approximately half of the duration of the USR period. Thus, non-Markovianity percolated during that period, slowing/delaying the decoherence process. 

It is important to note that the final state, and consequently its purity or entanglement entropy, is highly sensitive to the nature of the phase transitions. In fig.~\ref{fig:ctr}, we illustrate how smoother transitions between phases can reduce the degree of decoherence induced by these transitions. This suggests that a perfectly adiabatic transition between inflationary phases could, in principle, lead to the same level of recoherence observed in a single SR phase. However, in the scenarios studied here, making the transitions smoother effectively shortens the duration of the USR phase. As a result, the reduced decoherence is not unexpected. Moreover, the likelihood of a perfectly adiabatic transition is difficult to assess, since transitioning from an attractor phase to a non-attractor phase typically involves some inherent level of non-adiabaticity.

\section{Discussion}
Nature has carefully hidden the
quantum properties of inflationary perturbations from direct observation. Nevertheless, many aspects of quantum physics provide
interesting consequences for the early universe. There is a possibility that these may manifest themselves in unexpected ways.
In this work, we consider the leading order term in the EFT of inflationary perturbations coupled to a ultraviolet environment of heavy fields.
The system is fully Gaussian, and we generalise previous studies to explore it on a USR non-attractor background. 

Our first main finding is that slow-roll attractors are special in the sense that they typically lead to a ``more'' non-Markovian subsystem \cite{shandera2018open}. By this, we mean that the negative eigenvalue of the dissipator matrix seems always to dominate over the positive one for all masses of the entropic fluctuation. On the other hand, the situation is reversed for the USR solution.
This is somewhat counter-intuitive, because at face value it signals ``less'' non-Markovianity for USR-like flat regions of the scalar potential when compared to ``vanilla'' slow-roll inflation. However, more interestingly, we show that the range of parameters which correspond to some form of gain in coherence of the system mode (namely, recoherence or purity-freezing), is also the same range of parameters for which the entanglement entropy transiently decreases for the system. There is no such behaviour for  USR  and one simply finds that adiabatic mode efficiently decoheres in this case leading to a maximally entangled state. 

The second interesting observation we make regards the behaviour  of the diffusion terms $D_{22}$. As explained in the main text, it is these terms that are primarily responsible for such recoherence and negative entropy growth for SR. Since this term dominates the noise kernel, a natural question that follows is what is the effect of non-Markovian corrections to the noise term in stochastic inflation? Typically, the quantum kicks of sub-horizon modes appear as a white noise in the stochastic Fokker--Planck equation for the probability distribution function long-wavelength modes. However, given how radically exotic phenomena can arise from non-Markovian diffusive terms, it is pertinent to ask if one can find any such curious behaviour when considering non-Markovian corrections to stochastic inflation. 

We end with a tantalizingly provocative question:  In condensed matter experiments, any non-Markovian open quantum system that leads to a short negative growth in entropy does not typically signal any breakdown in the Second Law. Indeed, the entropy required to prepare the subsystem in its initial state such that it can evolve, while interacting  in a non-Markovian manner with the environment, is much greater than any such transient entropy losses. Nevertheless, if we end up in a Hubble patch for which such parameters are preferred that it leads to transient entropy decreases, then it would be Nature that has prepared for the observable universe to start out in this way. An ambitious avenue would be to navigate in the future would be to see if indeed there can be any natural selection law that prefers, or disfavours, such non-Markovian dynamics which lead to such reduction in the entanglement entropy of a given Hubble patch.

\section*{Acknowledgments}
The authors are grateful to Thomas Colas for feedback on an earlier version of this manuscript and for helpful discussions. SB extends additional thanks to Julien Grain and, especially, Vincent Vennin for insightful discussions that improved the treatment of adiabatic transitions.
SB is supported in part by the Higgs Fellowship and by the UK Science and Technology Facilities Council (STFC) Consolidated Grant ``Particle Physics at the Higgs Centre''. JCF and DS are funded by the STFC under grant number ST/X001040/1. XL is supported in part by the Program of China Scholarship Council (Grant No. 202208170014).

\clearpage

\appendix

\section{Analytical solutions for ME coefficients}\label{ApA}

In this appendix we provide the analytical expressions of the integrals required to obtain the ME coefficients. These integrals are derived from eqs.~\eqref{eq:DDs} in the main text. 

First, for operational simplicity, we define the variable $w = -k \tau$. Additionally, we introduce the following functions, valid for USR backgrounds, with the components dependent on $w$ factored out for clarity, 
\begin{equation}
    \bv{v}(w) = e^{i w}\left(1 + \frac{i}{w} \right)\;, \quad \bv{\pi}(w) = e^{i w} \left( 1 + \frac{3i}{w} - \frac{3}{w^2} \right)\;, \quad \bv{u}(w) = w^{1/2} H_{\nu}^{(1)}(w)\;.
\end{equation}
In this way, the coefficients of the TE are written as
    \begin{align}
        D_{12} & =  - \frac{\pi \alpha}{8} \left( \frac{\lambda}{H} \right)^2 \frac{k}{w}\ {\rm Im} \Big[ \bv{\pi}^* (w) \bv{u} (w) \int \frac{\rd w'}{w'} \bv{\pi} (w') \bv{u}^* (w') + \bv{\pi}^* (w) \bv{u}^* (w) \int \frac{\rd w'}{w'} \bv{\pi} (w') \bv{u} (w')      \Big]     \nonumber \\
        \Delta_{12} & = - \frac{\pi \alpha}{8} \left( \frac{\lambda}{H} \right)^2 \frac{k}{w}\ {\rm Re} \Big[ \bv{\pi}^* (w) \bv{u}^* (w) \int \frac{\rd w'}{w'} \bv{\pi} (w') \bv{u} (w') - \bv{\pi}^* (w) \bv{u} (w) \int \frac{\rd w'}{w'} \bv{\pi} (w') \bv{u}^* (w')   \Big] \nonumber \\
        D_{22} & = - \frac{\pi \alpha}{4} \left( \frac{\lambda}{H} \right)^2 \frac{1}{w}\ {\rm Im} \Big[ i \bv{v}^* (w) \bv{u}(w) \int \frac{\rd w'}{w'} \bv{\pi} (w') \bv{u}^* (w') + i \bv{v}^* (w) \bv{u}^*(w) \int \frac{\rd w'}{w'} \bv{\pi} (w') \bv{u}(w')  \Big] \nonumber \\
        \Delta_{22} & = - \frac{\pi \alpha}{4} \left( \frac{\lambda}{H} \right)^2 \frac{1}{w}\ {\rm Re} \Big[ i \bv{v}^*(w) \bv{u}^* (w) \int \frac{\rd w'}{w'} \bv{\pi}(w') \bv{u}(w') - i \bv{v}^* (w) \bv{u}(w) \int \frac{\rd w'}{w'} \bv{\pi}(w') \bv{u}^* (w') \Big]\;. \label{eq:MECs}
    \end{align}

From these equations, we identify two essential integrals:
\begin{align}
    I_1 = \int \frac{\rd w'}{w'} \bv{\pi} (w') \bv{u}^* (w')\;, \qquad I_2 = \int \frac{\rd w'}{w'} \bv{\pi} (w') \bv{u} (w')\;,
\end{align}
where, in practice, we only consider the primitives of the integrals, or equivalently, their upper limits. The lower limits are excluded due to their introduction of spurious contributions, which must be removed by hand to accurately describe the dynamics of the system \cite{Colas:2022hlq,Brahma:2024yor}. With these considerations, we present the result of these integrations for USR backgrounds. The corresponding results for SR backgrounds have been previously reported in the literature \cite{Colas:2022kfu}. In our analysis, we distinguish between the cases where $\nu \in \mathbb{C}$ ($\nu = i \mu$) and  $\nu \in \mathbb{R}$. 

\begin{equation}
\begin{aligned}
I_1^{\mu \in \mathbb{R}} = &\frac{1}{w^{3/2} (9 + 4 \mu^2)} \Bigg\{ 6 e^{i w} \Big[ 2 i w (i + w) \left(1 + \coth(\pi \mu)\right) J_{1 - i \mu}(w)  \\
&\quad - \left(-3 + 2 i \mu + w (3 i + 2 w + 2 \mu)\right) \left(1 + \coth(\pi \mu)\right) J_{-i \mu}(w)  \\
&\quad + \left(2 (1 - i w) w J_{1 + i \mu}(w) + \left(-3 + 3 i w + 2 w^2 - 2 (i + w) \mu\right) J_{i \mu}(w)\right) {\rm{csch}}(\pi \mu) \Big] \\
&+ \frac{2^{-i \mu} w^{1/2 - i \mu}}{\sqrt{\pi}} \Big[ \left(1 + \coth(\pi \mu)\right) \Gamma\left(\frac{1}{2} - i \mu\right)^2 \, {}_2{\tilde F}_2\left(\left\{\frac{1}{2} - i \mu, \frac{1}{2} - i \mu\right\}, \left\{\frac{3}{2} - i \mu, 1 - 2 i \mu\right\}, 2 i w\right) \\
&\quad - 2^{2 i \mu} w^{2 i \mu} {\rm{csch}}(\pi \mu)\ \Gamma\left(\frac{1}{2} + i \mu\right)^2 \, {}_2{\tilde F}_2\left(\left\{\frac{1}{2} + i \mu, \frac{1}{2} + i \mu\right\}, \left\{\frac{3}{2} + i \mu, 1 + 2 i \mu\right\}, 2 i w\right) \Big] \Bigg\}
\end{aligned}
\end{equation}
\begin{equation}
\begin{aligned}
I_2^{\mu \in \mathbb{R}} = &\frac{1}{w^{3/2} (9 + 4 \mu^2)} \Bigg\{ 6 e^{i w} \Big[ 2 i w (i + w) \left(1 + \coth(\pi \mu)\right) J_{1 + i \mu}(w)  \\
&\quad + \left(3 + 2 i \mu + w (-3 i - 2 w + 2 \mu)\right) \left(1 + \coth(\pi \mu)\right) J_{i \mu}(w) \\
&\quad + 2 (1 - i w) w {\rm csch}(\pi \mu) J_{1 - i \mu}(w) + \left(-3 + 2 i \mu + w (3 i + 2 w + 2 \mu)\right) {\rm csch}(\pi \mu) J_{-i \mu}(w)  \Big] \\
&+ \frac{2^{-i \mu} w^{1/2 - i \mu}}{\sqrt{\pi}} \Big[ -{\rm csch}(\pi \mu) \Gamma\left(\frac{1}{2} - i \mu\right)^2 \, {}_2{\tilde F}_2\left(\left\{\frac{1}{2} - i \mu, \frac{1}{2} - i \mu\right\}, \left\{\frac{3}{2} - i \mu, 1 - 2 i \mu\right\}, 2 i w\right) \\
&\quad + 2^{2 i \mu} w^{2 i \mu} \left(1 + \coth(\pi \mu)\right)\ \Gamma\left(\frac{1}{2} + i \mu\right)^2 \, {}_2{\tilde F}_2\left(\left\{\frac{1}{2} + i \mu, \frac{1}{2} + i \mu\right\}, \left\{\frac{3}{2} + i \mu, 1 + 2 i \mu\right\}, 2 i w\right) \Big] \Bigg\}
\end{aligned}
\end{equation}
\begin{equation}
\begin{aligned}
I_1^{\nu \in \mathbb{R}} = &\frac{1}{w^{3/2} (-9 + 4 \nu^2)} \Bigg\{ 6 e^{i w} \Big[ (3i + 3w - 2iw^2 + 2(i + w) \nu) (i + \cot(\pi \nu)) J_{\nu}(w) \\
&\quad - 2w (i + w) (i + \cot(\pi \nu)) J_{1 + \nu}(w)  \\
&\quad + \Big( 2w (i + w) J_{1 - \nu}(w) + (i (-3 + 2\nu) + w (-3 + 2iw + 2\nu)) J_{-\nu}(w) \Big) \csc(\pi \nu) \Big] \\
&\quad + \frac{2^{-\nu} w^{1/2 - \nu}}{\sqrt{\pi}} \Big[ i \csc(\pi \nu)\ \Gamma\left(\frac{1}{2} - \nu\right)^2 \, {}_2{\tilde F}_2\left(\left\{\frac{1}{2} - \nu, \frac{1}{2} - \nu\right\}, \left\{1 - 2 \nu, \frac{3}{2} - \nu\right\}, 2iw\right) \\
&\quad + 4^{\nu} w^{2\nu} \left(1 - i \cot(\pi \nu)\right) \Gamma\left(\frac{1}{2} + \nu\right)^2 \, {}_2{\tilde F}_2\left(\left\{\frac{1}{2} + \nu, \frac{1}{2} + \nu\right\}, \left\{\frac{3}{2} + \nu, 1 + 2 \nu\right\}, 2iw\right) \Big] \Bigg\}
\end{aligned}
\end{equation}
\begin{equation}
\begin{aligned}
I_2^{\nu \in \mathbb{R}} = &\frac{1}{w^{3/2} (-9 + 4 \nu^2)} \Bigg\{ 6 e^{i w} \Big[ (-3 + w (3i + 2w + 2i \nu) - 2 \nu)(1 + i \cot(\pi \nu)) J_{\nu}(w)\\
&\quad + 2w (i + w) (-i + \cot(\pi \nu))  J_{1 + \nu}(w)  \\
&\quad + \Big(-2w (i + w) J_{1 - \nu}(w) + (3i + 3w - 2iw^2 - 2(i + w) \nu) J_{-\nu}(w) \Big) \csc(\pi \nu) \Big] \\
&\quad + \frac{2^{-\nu} w^{1/2 - \nu}}{\sqrt{\pi}} \Big[ -i \csc(\pi \nu) \Gamma\left(\frac{1}{2} - \nu\right)^2 \, {}_2{\tilde F}_2\left(\left\{\frac{1}{2} - \nu, \frac{1}{2} - \nu\right\}, \left\{1 - 2 \nu, \frac{3}{2} - \nu\right\}, 2iw\right) \\
&\quad + 4^{\nu} w^{2\nu} \left(1 + i \cot(\pi \nu)\right) \Gamma\left(\frac{1}{2} + \nu\right)^2 \, {}_2{\tilde F}_2\left(\left\{\frac{1}{2} + \nu, \frac{1}{2} + \nu\right\}, \left\{\frac{3}{2} + \nu, 1 + 2 \nu\right\}, 2iw\right) \Big] \Bigg\}
\end{aligned}
\end{equation}

In the equations above, $J_{\nu}$ denotes the Bessel function of the first kind, while ${}_p {\tilde F}_q$ is the regularized generalised hypergeometric function. Then, the ME coefficients are found by plugging these expressions into eqs.~\eqref{eq:MECs}. Even though the resulting expressions are rather long, below we present the supper-Hubble approximation of the coefficients. For $\nu = i\mu$, we have got that:
\begin{align}
 D_{12}^{\mu \in \mathbb{R}} & \approx -\frac{\lambda^2}{H^2}\frac{(1 + \coth(\pi\mu)) kw e^{-\pi\mu}}{\pi \mu (9 + 4\mu^2)(49 + 4\mu^2)}  \Big(2\pi(21 + 4\mu^2)\cosh(\pi\mu) \nonumber \\
&\quad + \mu \sinh(\pi\mu) \left[(w/2)^{2i\mu}(21 + 4\mu(2i + \mu))\Gamma(-i\mu)^2 + (w/2)^{-2i\mu}(21 + 4\mu(-2i + \mu))\Gamma(i\mu)^2\right]\Big) \nonumber \\
\Delta_{12}^{\mu \in \mathbb{R}} & \approx \frac{16\lambda^2}{H^2} \frac{kw}{441 + 232\mu^2 + 16\mu^4} \nonumber \\
D_{22}^{\mu \in \mathbb{R}} & \approx \frac{\lambda^2}{H^2} 
    \frac{e^{-\pi \mu} \left(1 + \coth(\pi \mu)\right)}{\pi} \left( 
        \frac{6 \pi \cosh(\pi \mu)}{9 \mu + 4 \mu^3} + \left( 
            \frac{(w/2)^{2 i \mu} \Gamma(-i \mu)^2}{3 - 2 i \mu} + 
            \frac{(w/2)^{-2 i \mu} \Gamma(i \mu)^2}{3 + 2 i \mu}
        \right) \sinh(\pi \mu) 
    \right) \nonumber \\
\Delta_{22}^{\mu \in \mathbb{R}} & \approx -\frac{4\lambda^2}{H^2} \frac{49 - 128w^2 + 200\mu^2 + 16\mu^4}{(9 + 4\mu^2)(49 + 200\mu^2 + 16\mu^4)}\;,
\end{align}
whereas for $\nu \in \mathbb{R}$:
\begin{align}
  D_{12}^{\nu \in \mathbb{R}} & \approx  \frac{\pi}{8} \frac{k\lambda^2}{H^2} \left(\frac{2^{3 + 2\nu} w^{-2\nu + 1} \csc(\pi\nu) \Gamma(\nu)}{\pi(-21 - 8\nu + 4\nu^2) \Gamma(1 - \nu)} - \frac{16w(-21 + 4\nu^2) \cot(\pi\nu)}{\pi\nu(441 - 232\nu^2 + 16\nu^4)} + \frac{2^{3 - 2\nu} w^{2\nu + 1} \csc^2(\pi\nu)}{(-21 + 8\nu + 4\nu^2) \Gamma(1 + \nu)^2}\right) \nonumber \\
  \Delta_{12}^{\nu \in \mathbb{R}} & \approx  \frac{16\lambda^2}{H^2} \frac{kw}{441 - 232\nu^2 + 16\nu^4} \nonumber \\
   D_{22}^{\nu \in \mathbb{R}} & \approx -\pi \frac{\lambda^2}{H^2} \left(\frac{6 \cot(\pi\nu)}{9\pi\nu - 4\pi\nu^3} - \frac{4^\nu w^{-2\nu} \csc(\pi\nu) \Gamma(\nu)}{\pi(3 + 2\nu) \Gamma(1 - \nu)} + \frac{4^{-\nu} w^{2\nu} \csc^2(\pi\nu)}{(-3 + 2\nu) \Gamma(1 + \nu)^2}\right) \nonumber \\
  \Delta_{22}^{\nu \in \mathbb{R}} & \approx  -\frac{\lambda^2}{H^2} \left(\frac{4}{9 - 4\nu^2} + \frac{512w^2}{-441 + 1996\nu^2 - 944\nu^4 + 64\nu^6}\right)\;.
\end{align}

\section{Transport equations for the full system}\label{ApB}
\begin{figure}[h!]
     \centering
     \begin{subfigure}[b]{0.48\textwidth}
         \centering
         \hspace*{-1.5cm}
         \includegraphics[width=1.3\textwidth]{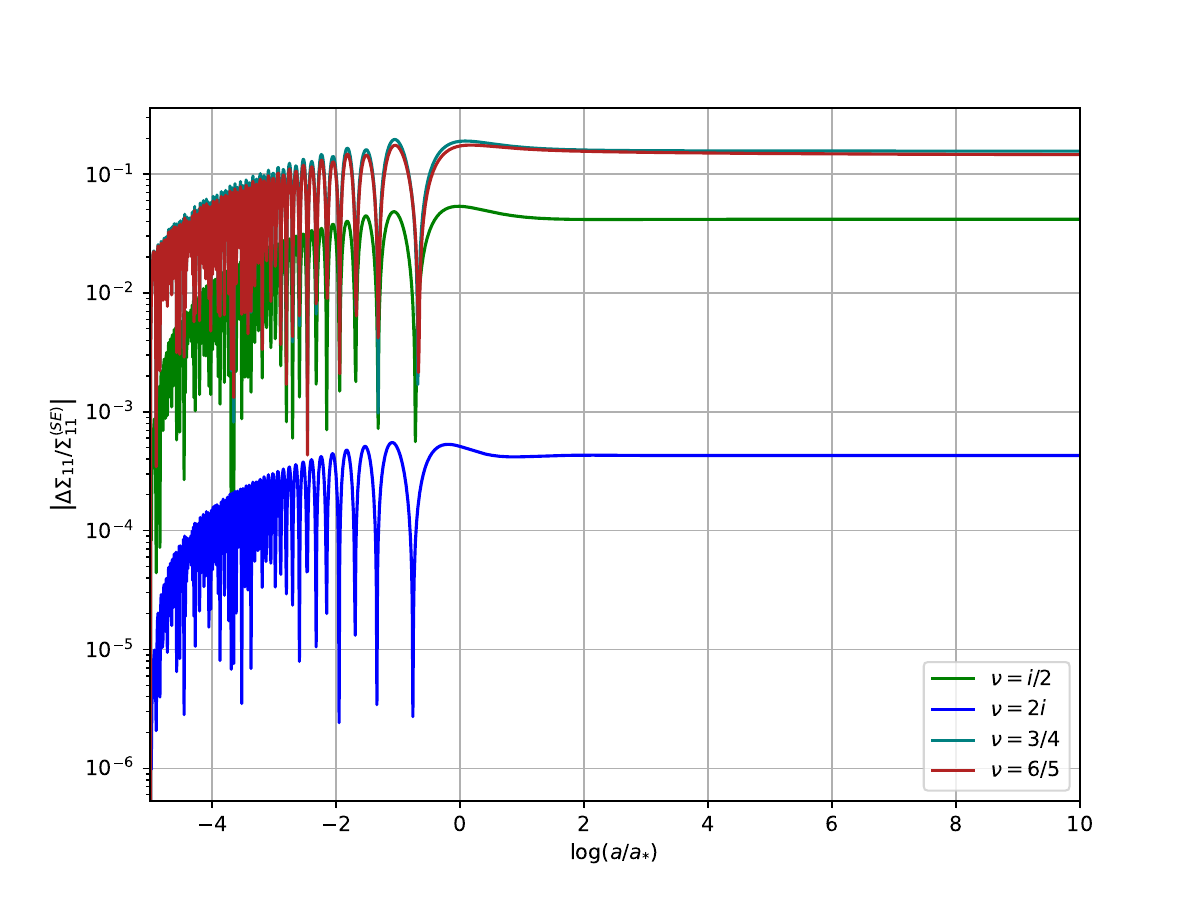}
         \caption{}
     \end{subfigure}
     \hfill
     \begin{subfigure}[b]{0.48\textwidth}
         \centering
         \hspace*{-0.7cm}
         \includegraphics[width=1.3\textwidth]{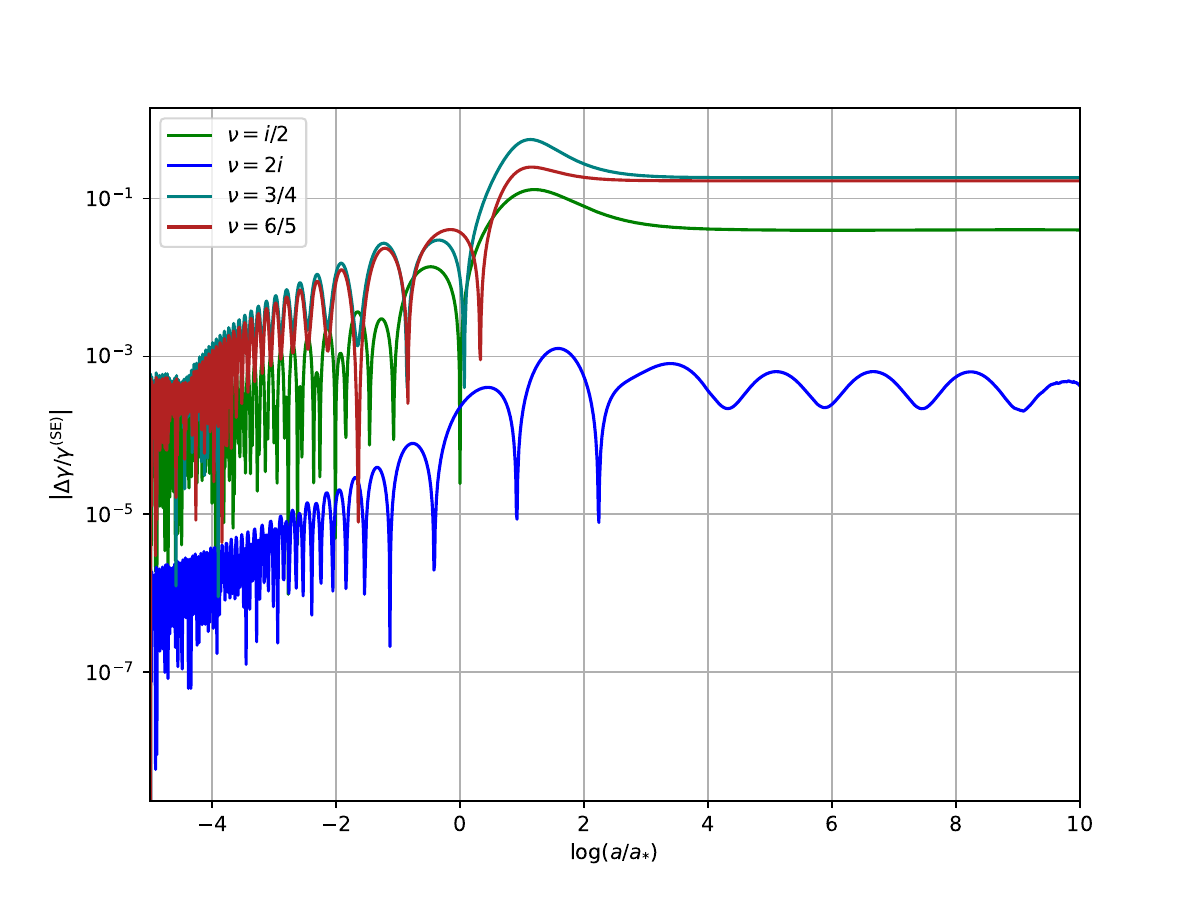}
         \caption{}
     \end{subfigure}
        \caption{Deviation in $\Sigma_{11}$ (left) and purity $\gamma$ (right) between results obtained using the TCL$_2$ master equation and those derived from the complete $\s + \e$ system of transport equations. We have set $\lambda/H = 0.1$.}
        \label{fig:compse}
\end{figure}

In the main body of this work, we have primarily focused on an effective description of the system's evolution. This approach facilitates a more straightforward interpretation of the dynamics and aligns more closely with the type of theory one might formulate in realistic scenarios. However, the Lagrangian under consideration enables us to solve the dynamics for the full theory, encompassing both the system and its environment. To achieve this, the most efficient approach involves utilizing the complete set of transport equations, which allows us to examine not only the $\s$-$\s$ correlations but also the $\e$-$\e$ and $\s$-$\e$ correlations.
Given that the total evolution is unitary (open effective dynamics emerges only when we focus solely on the system), the set of transport equations takes the simple form \cite{Colas:2022hlq, Colas:2024xjy, Brahma:2024yor}: 
\begin{equation}\label{eq:FSTE}
    \frac{\rd}{\rd \tau}\Sigma^{(\s+\e)} = 2 \Omega \h{H}^{(\s+\e)}\Sigma^{(\s+\e)} - 2 \Sigma^{(\s+\e)} \h{H}^{(\s+\e)}\Omega\;,
\end{equation}
where $\Omega$ is a block-diagonal matrix with antisymmetric entries, and $\h{H}^{(\s+\e)}$ is the Hamiltonian of the full system. Explicitly, these matrices are given by:
\begin{align}
\Omega = \begin{pmatrix}
    0 & 1 & 0 & 0 \\
    -1& 0 & 0 & 0 \\
    0 & 0 & 0 & 1 \\
    0 & 0 & -1 & 0
\end{pmatrix}\;, \qquad 
   \h{H}^{(\s+\e)} = \frac{1}{2}
     \begin{pmatrix}
k^2 & z'/z & 0 & 0 \\
z'/z & 1 & -\lambda a & 0 \\
0 & -\lambda a & k^2 + m^2 a^2 & a'/a \\
0 & 0 & a'/a & 1
\end{pmatrix}\;.\label{eq:Hes}
\end{align}
The differing couplings to gravity of the $\s$ and $\e$ fields are manifest in the $12$ and $34$ entries of the Hamiltonian. In principle, this should be taken into account in order to impose appropriate initial conditions. Here, let us focus on USR backgrounds, as the SR case has already been explored in the literature. For $\s-\s$ entries, initial conditions must be consistent with eqs.~\eqref{eq:BDUSR}, which describe Bunch-Davies initial states, which we have considered throughout this work. In contrast, for $\e-\e$ correlations, it is suitable to use initial conditions suitable for SR. We also assume that initial $\s-\e$ correlations are null. Solving the set of transport equations under these assumptions, we found a good agreement with the results obtained using an EFT-TCL approach. We quantify the degree of agreement through the deviation functions:
$$\left | \frac{\Delta \Sigma_{11}}{\Sigma_{11}^{(\s\e)}}\right| = \left| 1 - \frac{\Sigma_{11}^{\rm (TCL)}}{\Sigma_{11}^{(\s\e)}} \right|\;, \qquad \left | \frac{\Delta \gamma}{\gamma^{(\s\e)}}\right| = \left| 1 - \frac{\gamma^{\rm (TCL)}}{\gamma^{(\s\e)}} \right|\;,$$
where quantities with the superscript $(\s\e)$ have been obtained through the full set of transport equations, eq.~\eqref{eq:FSTE}. These functions are shown in fig.~\ref{fig:compse}, which in general show good overall agreement between the two methods.

\begin{figure}[h!]
    \centering
    \includegraphics[width=0.8\textwidth]{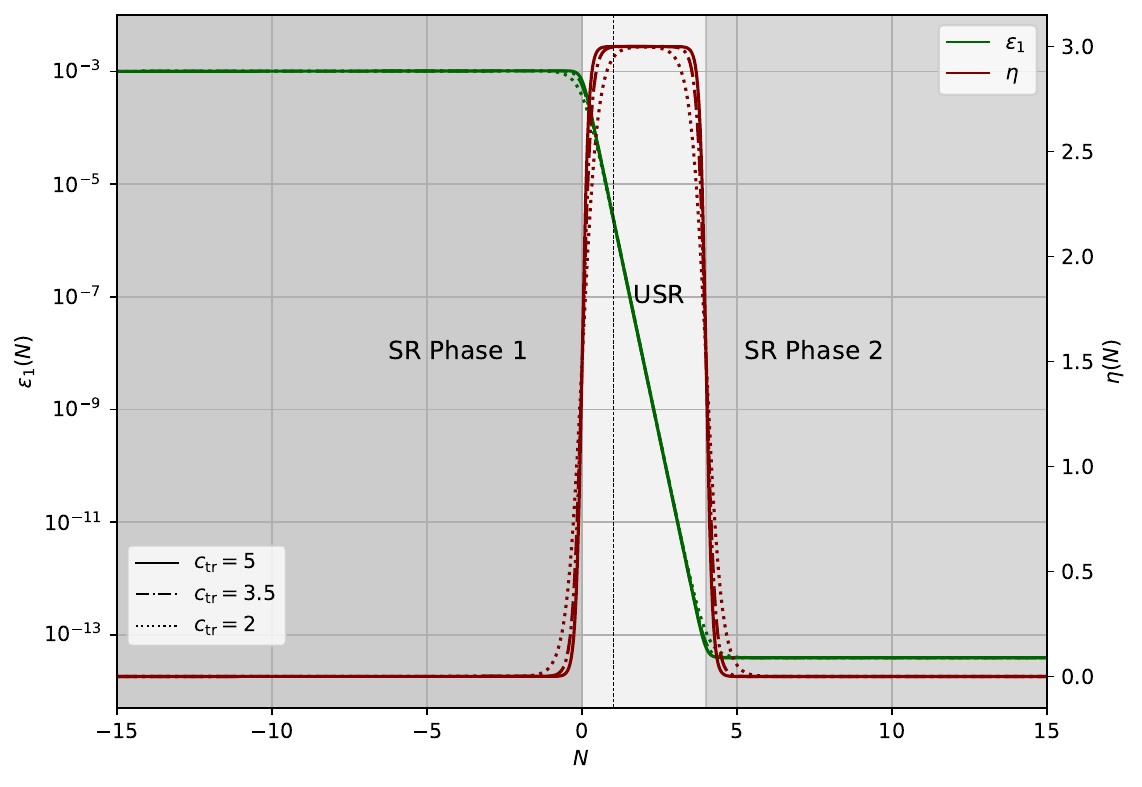}
    \caption{Slow-roll parameters $\epsilon_1$ and $\eta$ used to simulate the evolution of inflationary perturbations on a SR-USR-SR scenario. Roughly, the USR period occurs between $N\simeq 0$ and $4$. The case $c_{\rm tr} = 5$ corresponds to eq.~\eqref{eq:etaph}, which determines the background considered for figs.~\ref{fig:purph} and \ref{fig:sntph}.}
    \label{fig:eps}
\end{figure}

Finally, let us emphasize that, provided that this Hamiltonian applies across different (inflationary) backgrounds, the set of transport equations is expected to remain consistent in structure, allowing for the description of the evolution of the correlations by imposing the appropriate evolution of $z'/z$ and $a'/a$. This approach was implemented in the final part of Sec.~\ref{sec:QSE}, where we looked at a sequence of inflationary phases of the form SR-USR-SR. This was achieved through eq.~\eqref{eq:etaph}, and noting that $z'/z = 1 + \epsilon_1 - \eta$. The corresponding evolution of the slow-roll parameters of interest is shown in fig.~\ref{fig:eps}.

\begin{figure}[h!]
    \centering
    \includegraphics[width=0.8\textwidth]{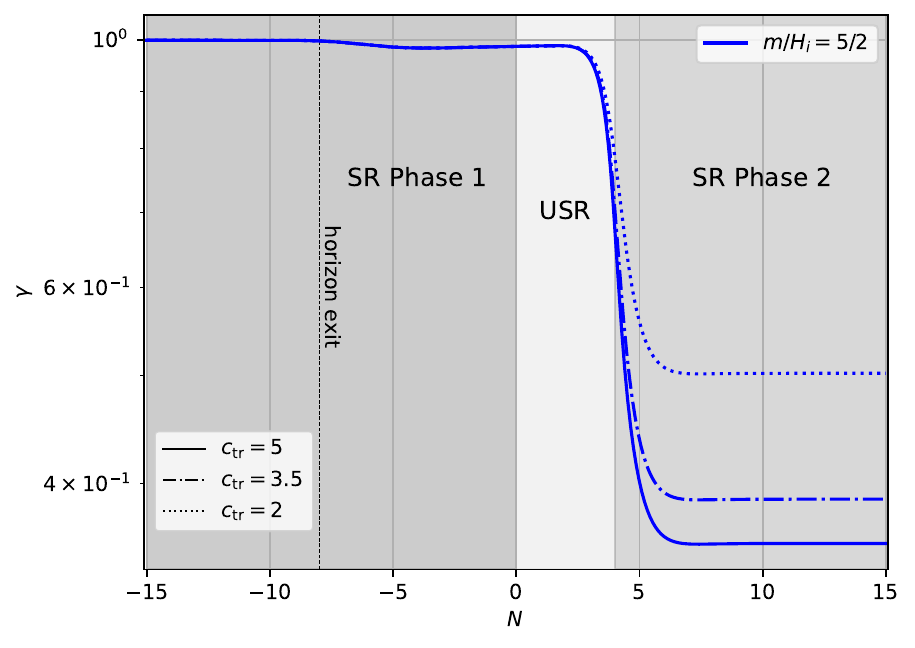}
    \caption{Purity evolution for $m/H_i = 5/2$ and $\lambda/H_i = 0.1$ for a SR-USR-SR scenario with varying transitions smoothness. }
    \label{fig:ctr}
\end{figure}

One can also look at how the apparent final value of purity varies with the smoothness of the transitions. To do so, we express the slow-roll parameter $\eta$ as
\begin{equation}\label{eq:etactr}
    \eta(N) = \frac{3}{2}\left[ \tanh(c_{\rm tr}(N-N_1)) - \tanh(c_{\rm tr}(N-N_2))\right]\;,
\end{equation}
with $c_{\rm tr}$ controls the smoothness of the transition. Fig.~\ref{fig:ctr} illustrates the evolution of purity for $m/H_i = 5/2$ (and $\lambda/H_i = 0.1$) with horizon exit during the first slow-roll phase, for different choices of $c_{\rm tr}$. The plot clearly demonstrates that smoother transitions result in greater purity recovery during the final SR phase. As discussed in the main text, this is at least in part a consequence of reducing the time of the system under the non-attractor background. In fact, further decreasing $c_{\rm tr}$ prevents the realization of an USR phase altogether.
\newpage
\printbibliography

@article{Jefferson:2017sdb,
    author = "Jefferson, Ro and Myers, Robert C.",
    title = "{Circuit complexity in quantum field theory}",
    eprint = "1707.08570",
    archivePrefix = "arXiv",
    primaryClass = "hep-th",
    doi = "10.1007/JHEP10(2017)107",
    journal = "JHEP",
    volume = "10",
    pages = "107",
    year = "2017"
}

@article{Balasubramanian:2011wt,
    author = "Balasubramanian, Vijay and McDermott, Michael B. and Van Raamsdonk, Mark",
    title = "{Momentum-space entanglement and renormalization in quantum field theory}",
    eprint = "1108.3568",
    archivePrefix = "arXiv",
    primaryClass = "hep-th",
    reportNumber = "UPR-1233-T",
    doi = "10.1103/PhysRevD.86.045014",
    journal = "Phys. Rev. D",
    volume = "86",
    pages = "045014",
    year = "2012"
}

@article{Ribes-Metidieri:2024vjn,
    author = "Ribes-Metidieri, Patricia and Agullo, Ivan and Bonga, B\'eatrice",
    title = "{Inflation does not create entanglement in local observables}",
    eprint = "2409.16366",
    archivePrefix = "arXiv",
    primaryClass = "gr-qc",
    month = "9",
    year = "2024"
}

@article{Agullo:2024har,
    author = "Agullo, Ivan and Bonga, B\'eatrice and Mart\'\i{}n-Mart\'\i{}nez, Eduardo and Nadal-Gisbert, Sergi and Perche, T. Rick and Polo-G\'omez, Jos\'e and Ribes-Metidieri, Patricia and de S. L. Torres, Bruno",
    title = "{The multimode nature of spacetime entanglement in QFT}",
    eprint = "2409.16368",
    archivePrefix = "arXiv",
    primaryClass = "quant-ph",
    month = "9",
    year = "2024"
}

@article{Sou:2022nsd,
    author = "Sou, Chon Man and Tran, Duc Huy and Wang, Yi",
    title = "{Decoherence of cosmological perturbations from boundary terms and the non-classicality of gravity}",
    eprint = "2207.04435",
    archivePrefix = "arXiv",
    primaryClass = "hep-th",
    doi = "10.1007/JHEP04(2023)092",
    journal = "JHEP",
    volume = "04",
    pages = "092",
    year = "2023"
}

@article{Sou:2024tjv,
    author = "Sou, Chon Man and Wang, Junqi and Wang, Yi",
    title = "{Cosmological Bell tests with decoherence effects}",
    eprint = "2405.07141",
    archivePrefix = "arXiv",
    primaryClass = "hep-th",
    doi = "10.1088/1475-7516/2024/10/084",
    journal = "JCAP",
    volume = "10",
    pages = "084",
    year = "2024"
}

@misc{glavan_notes,
  author       = {Glavan, Drazˇen and Prokopec, Tomislav},
  title        = {A {PEDESTRIAN} {INTRODUCTION} {TO} {NON}-{EQUILIBRIUM} {QFT}.},
  year         = {2017},
  note         = {Lecture notes},
  howpublished = {University of Utrecht},
  url          = {https://webspace.science.uu.nl/~proko101/LecturenotesNSTP530M2014.pdf},
}

@book{breuer2002theory,
  title={The theory of open quantum systems},
  author={Breuer, Heinz-Peter and Petruccione, Francesco},
  year={2002},
  publisher={Oxford University Press, USA}
}

@article{Colas:2022hlq,
    author = "Colas, Thomas and Grain, Julien and Vennin, Vincent",
    title = "{Benchmarking the cosmological master equations}",
    eprint = "2209.01929",
    archivePrefix = "arXiv",
    primaryClass = "hep-th",
    doi = "10.1140/epjc/s10052-022-11047-9",
    journal = "Eur. Phys. J. C",
    volume = "82",
    number = "12",
    pages = "1085",
    year = "2022"
}

@article{Colas:2022kfu,
    author = "Colas, Thomas and Grain, Julien and Vennin, Vincent",
    title = "{Quantum recoherence in the early universe}",
    eprint = "2212.09486",
    archivePrefix = "arXiv",
    primaryClass = "gr-qc",
    doi = "10.1209/0295-5075/acdd94",
    journal = "EPL",
    volume = "142",
    number = "6",
    pages = "69002",
    year = "2023"
}

@article{Colas:2024xjy,
    author = "Colas, Thomas and de Rham, Claudia and Kaplanek, Greg",
    title = "{Decoherence out of fire: purity loss in expanding and contracting universes}",
    eprint = "2401.02832",
    archivePrefix = "arXiv",
    primaryClass = "hep-th",
    reportNumber = "Imperial/TP/2024/CdR/01; Imperial/TP/2024/GK/01",
    doi = "10.1088/1475-7516/2024/05/025",
    journal = "JCAP",
    volume = "05",
    pages = "025",
    year = "2024"
}

@article{Burgess:2024eng,
    author = "Burgess, C. P. and Colas, Thomas and Holman, R. and Kaplanek, Greg and Vennin, Vincent",
    title = "{Cosmic Purity Lost: Perturbative and Resummed Late-Time Inflationary Decoherence}",
    eprint = "2403.12240",
    archivePrefix = "arXiv",
    primaryClass = "gr-qc",
    month = "3",
    year = "2024"
}

@article{Salcedo:2024smn,
    author = "Salcedo, Santiago Agui and Colas, Thomas and Pajer, Enrico",
    title = "{The Open Effective Field Theory of Inflation}",
    eprint = "2404.15416",
    archivePrefix = "arXiv",
    primaryClass = "hep-th",
    month = "4",
    year = "2024"
}

@article{Brahma:2021mng,
    author = "Brahma, Suddhasattwa and Berera, Arjun and Calder\'on-Figueroa, Jaime",
    title = "{Universal signature of quantum entanglement across cosmological distances}",
    eprint = "2107.06910",
    archivePrefix = "arXiv",
    primaryClass = "hep-th",
    doi = "10.1088/1361-6382/aca066",
    journal = "Class. Quant. Grav.",
    volume = "39",
    number = "24",
    pages = "245002",
    year = "2022"
}

@article{Brahma:2022yxu,
    author = "Brahma, Suddhasattwa and Berera, Arjun and Calder\'on-Figueroa, Jaime",
    title = "{Quantum corrections to the primordial tensor spectrum: open EFTs \& Markovian decoupling of UV modes}",
    eprint = "2206.05797",
    archivePrefix = "arXiv",
    primaryClass = "hep-th",
    doi = "10.1007/JHEP08(2022)225",
    journal = "JHEP",
    volume = "08",
    pages = "225",
    year = "2022"
}

@article{Brahma:2023hki,
    author = "Brahma, Suddhasattwa and Calder\'on-Figueroa, Jaime and Hassan, Moatasem and Mi, Xuan",
    title = "{Momentum-space entanglement entropy in de Sitter spacetime}",
    eprint = "2302.13894",
    archivePrefix = "arXiv",
    primaryClass = "hep-th",
    doi = "10.1103/PhysRevD.108.043522",
    journal = "Phys. Rev. D",
    volume = "108",
    number = "4",
    pages = "043522",
    year = "2023"
}

@article{Brahma:2020zpk,
    author = "Brahma, Suddhasattwa and Alaryani, Omar and Brandenberger, Robert",
    title = "{Entanglement entropy of cosmological perturbations}",
    eprint = "2005.09688",
    archivePrefix = "arXiv",
    primaryClass = "hep-th",
    doi = "10.1103/PhysRevD.102.043529",
    journal = "Phys. Rev. D",
    volume = "102",
    number = "4",
    pages = "043529",
    year = "2020"
}

@article{Burgess:2022rdo,
    author = "Burgess, C. P. and Kaplanek, Greg",
    title = "{Gravity, Horizons and Open EFTs}",
    eprint = "2212.09157",
    archivePrefix = "arXiv",
    primaryClass = "hep-th",
    reportNumber = "CERN-TH-2022-175; Imperial/TP/2022/GK/03, CERN-TH-2022-175, Imperial/TP/2022/GK/03",
    month = "12",
    year = "2022"
}

@article{Assassi:2013gxa,
    author = "Assassi, Valentin and Baumann, Daniel and Green, Daniel and McAllister, Liam",
    title = "{Planck-Suppressed Operators}",
    eprint = "1304.5226",
    archivePrefix = "arXiv",
    primaryClass = "hep-th",
    doi = "10.1088/1475-7516/2014/01/033",
    journal = "JCAP",
    volume = "01",
    pages = "033",
    year = "2014"
}

@article{Tolley:2007nq,
    author = "Tolley, Andrew J. and Wesley, Daniel H.",
    title = "{Scale-invariance in expanding and contracting universes from two-field models}",
    eprint = "hep-th/0703101",
    archivePrefix = "arXiv",
    reportNumber = "DAMTP-2007-23",
    doi = "10.1088/1475-7516/2007/05/006",
    journal = "JCAP",
    volume = "05",
    pages = "006",
    year = "2007"
}

@article{Chen:2009zp,
    author = "Chen, Xingang and Wang, Yi",
    title = "{Quasi-Single Field Inflation and Non-Gaussianities}",
    eprint = "0911.3380",
    archivePrefix = "arXiv",
    primaryClass = "hep-th",
    doi = "10.1088/1475-7516/2010/04/027",
    journal = "JCAP",
    volume = "04",
    pages = "027",
    year = "2010"
}

@article{Caldeira:1982iu,
    author = "Caldeira, A. O. and Leggett, A. J.",
    title = "{Path integral approach to quantum Brownian motion}",
    doi = "10.1016/0378-4371(83)90013-4",
    journal = "Physica A",
    volume = "121",
    pages = "587--616",
    year = "1983"
}

@article{Dekker77,
  title = {Quantization of the linearly damped harmonic oscillator},
  author = {Dekker, H.},
  journal = {Phys. Rev. A},
  volume = {16},
  issue = {5},
  pages = {2126--2134},
  numpages = {0},
  year = {1977},
  month = {Nov},
  publisher = {American Physical Society},
  doi = {10.1103/PhysRevA.16.2126},
  url = {https://link.aps.org/doi/10.1103/PhysRevA.16.2126}
}

@article{Brahma:2024yor,
    author = "Brahma, Suddhasattwa and Calder\'on-Figueroa, Jaime and Luo, Xiancong",
    title = "{Time-convolutionless cosmological master equations: Late-time resummations and decoherence for non-local kernels}",
    eprint = "2407.12091",
    archivePrefix = "arXiv",
    primaryClass = "hep-th",
    month = "7",
    year = "2024"
}

@article{Hollowood:2017bil,
    author = "Hollowood, T. J. and McDonald, J. I.",
    title = "{Decoherence, discord and the quantum master equation for cosmological perturbations}",
    eprint = "1701.02235",
    archivePrefix = "arXiv",
    primaryClass = "gr-qc",
    doi = "10.1103/PhysRevD.95.103521",
    journal = "Phys. Rev. D",
    volume = "95",
    number = "10",
    pages = "103521",
    year = "2017"
}

@article{Burgess:2014eoa,
    author = "Burgess, C. P. and Holman, R. and Tasinato, G. and Williams, M.",
    title = "{EFT Beyond the Horizon: Stochastic Inflation and How Primordial Quantum Fluctuations Go Classical}",
    eprint = "1408.5002",
    archivePrefix = "arXiv",
    primaryClass = "hep-th",
    reportNumber = "CERN-PH-TH-2014-142",
    doi = "10.1007/JHEP03(2015)090",
    journal = "JHEP",
    volume = "03",
    pages = "090",
    year = "2015"
}

@article{Joos:1984uk,
    author = "Joos, E. and Zeh, H. D.",
    title = "{The Emergence of classical properties through interaction with the environment}",
    doi = "10.1007/BF01725541",
    journal = "Z. Phys. B",
    volume = "59",
    pages = "223--243",
    year = "1985"
}

@article{horhammer_2008,
	title = {Information and Entropy in Quantum Brownian Motion},
    eprint = "0710.1716",
    archivePrefix = "arXiv",
    primaryClass = "quant-ph",
	volume = {133},
	issn = {1572-9613},
	url = {https://doi.org/10.1007/s10955-008-9640-x},
	doi = {10.1007/s10955-008-9640-x},
	pages = {1161--1174},
	number = {6},
	journaltitle = {Journal of Statistical Physics},
	shortjournal = {Journal of Statistical Physics},
	author = {Hörhammer, Christian and Büttner, Helmut},
	date = {2008-12-01},
}

@article{Lally:2019dit,
    author = "Lally, Sapphire and Werren, Nicholas and Al-Khalili, Jim and Rocco, Andrea",
    title = "{Master equation for non-Markovian quantum Brownian motion: The emergence of lateral coherences}",
    eprint = "1907.05874",
    archivePrefix = "arXiv",
    primaryClass = "quant-ph",
    doi = "10.1103/PhysRevA.105.012209",
    journal = "Phys. Rev. A",
    volume = "105",
    number = "1",
    pages = "012209",
    year = "2022"
}

@article{PhysRevE.66.036102,
  title = {Statistical thermodynamics of quantum Brownian motion: Construction of perpetuum mobile of the second kind},
  author = {Nieuwenhuizen, Th. M. and Allahverdyan, A. E.},
  journal = {Phys. Rev. E},
  volume = {66},
  issue = {3},
  pages = {036102},
  numpages = {52},
  year = {2002},
  month = {Sep},
  publisher = {American Physical Society},
  doi = {10.1103/PhysRevE.66.036102},
  url = {https://link.aps.org/doi/10.1103/PhysRevE.66.036102}
}

@article{Serafini_2005,
doi = {10.1088/1464-4266/7/4/R01},
url = {https://dx.doi.org/10.1088/1464-4266/7/4/R01},
year = {2005},
month = {feb},
publisher = {},
volume = {7},
number = {4},
pages = {R19},
author = {A Serafini and M G A Paris and F Illuminati and S De Siena},
title = {Quantifying decoherence in continuous variable systems},
journal = {Journal of Optics B: Quantum and Semiclassical Optics}
}

@article{Li_2018,
   title={Concepts of quantum non-Markovianity: A hierarchy},
   volume={759},
   ISSN={0370-1573},
   url={http://dx.doi.org/10.1016/j.physrep.2018.07.001},
   DOI={10.1016/j.physrep.2018.07.001},
   journal={Physics Reports},
   publisher={Elsevier BV},
   author={Li, Li and Hall, Michael J.W. and Wiseman, Howard M.},
   year={2018},
   month=oct, pages={1–51} }

@article{Hall_2014,
   title={Canonical form of master equations and characterization of non-Markovianity},
   volume={89},
   ISSN={1094-1622},
   url={http://dx.doi.org/10.1103/PhysRevA.89.042120},
   DOI={10.1103/physreva.89.042120},
   number={4},
   journal={Physical Review A},
   publisher={American Physical Society (APS)},
   author={Hall, Michael J. W. and Cresser, James D. and Li, Li and Andersson, Erika},
   year={2014},
   month=apr }

@article{Boyanovsky:2015tba,
    author = "Boyanovsky, D.",
    title = "{Effective field theory during inflation: Reduced density matrix and its quantum master equation}",
    eprint = "1506.07395",
    archivePrefix = "arXiv",
    primaryClass = "astro-ph.CO",
    doi = "10.1103/PhysRevD.92.023527",
    journal = "Phys. Rev. D",
    volume = "92",
    number = "2",
    pages = "023527",
    year = "2015"
}

@book{joos2014decoherence,
  title={Decoherence and the Appearance of a Classical World in Quantum Theory},
  author={Joos, E. and Zeh, H.D. and Kiefer, C. and Giulini, D.J.W. and Kupsch, J. and Stamatescu, I.O.},
  isbn={9783662053294},
  year={2014},
  publisher={Springer Berlin Heidelberg}
}

@article{Bowen:2024emo,
    author = "Bowen, Brenden and Agarwal, Nishant and Kamal, Archana",
    title = "{Open system dynamics in interacting quantum field theories}",
    eprint = "2403.18907",
    archivePrefix = "arXiv",
    primaryClass = "hep-th",
    month = "3",
    year = "2024"
}

@article{Bhattacharyya:2024duw,
    author = "Bhattacharyya, Arpan and Brahma, Suddhasattwa and Haque, S. Shajidul and Lund, Jacob S. and Paul, Arpon",
    title = "{The early universe as an open quantum system: complexity and decoherence}",
    eprint = "2401.12134",
    archivePrefix = "arXiv",
    primaryClass = "hep-th",
    doi = "10.1007/JHEP05(2024)058",
    journal = "JHEP",
    volume = "05",
    pages = "058",
    year = "2024"
}

@article{Alicki:2023rfv,
    author = "Alicki, Robert and Barenboim, Gabriela and Jenkins, Alejandro",
    title = "{Quantum thermodynamics of de Sitter space}",
    eprint = "2307.04800",
    archivePrefix = "arXiv",
    primaryClass = "gr-qc",
    reportNumber = "IFIC/23-31",
    doi = "10.1103/PhysRevD.108.123530",
    journal = "Phys. Rev. D",
    volume = "108",
    number = "12",
    pages = "123530",
    year = "2023"
}

@article{Kading:2023mdk,
    author = {K\"ading, Christian and Pitschmann, Mario and Voith, Caroline},
    title = "{Dilaton-induced open quantum dynamics}",
    eprint = "2306.10896",
    archivePrefix = "arXiv",
    primaryClass = "hep-ph",
    doi = "10.1140/epjc/s10052-023-11939-4",
    journal = "Eur. Phys. J. C",
    volume = "83",
    number = "8",
    pages = "767",
    year = "2023"
}

@article{Burgess:2022nwu,
    author = "Burgess, C. P. and Holman, R. and Kaplanek, Greg and Martin, Jerome and Vennin, Vincent",
    title = "{Minimal decoherence from inflation}",
    eprint = "2211.11046",
    archivePrefix = "arXiv",
    primaryClass = "hep-th",
    reportNumber = "CERN-TH-2022-174; Imperial/TP/2022/GK/02",
    doi = "10.1088/1475-7516/2023/07/022",
    journal = "JCAP",
    volume = "07",
    pages = "022",
    year = "2023"
}

@article{Kading:2022hhc,
    author = {K\"ading, Christian and Pitschmann, Mario},
    title = "{Density Matrix Formalism for Interacting Quantum Fields}",
    eprint = "2210.06991",
    archivePrefix = "arXiv",
    primaryClass = "hep-th",
    doi = "10.3390/universe8110601",
    journal = "Universe",
    volume = "8",
    number = "11",
    pages = "601",
    year = "2022"
}

@article{Kaplanek:2022xrr,
    author = "Kaplanek, Greg and Tjoa, Erickson",
    title = "{Effective master equations~for two accelerated qubits}",
    eprint = "2207.13750",
    archivePrefix = "arXiv",
    primaryClass = "quant-ph",
    reportNumber = "Imperial/TP/2022/GK/01",
    doi = "10.1103/PhysRevA.107.012208",
    journal = "Phys. Rev. A",
    volume = "107",
    number = "1",
    pages = "012208",
    year = "2023"
}

@article{Colas:2024ysu,
    author = "Colas, Thomas and Grain, Julien and Kaplanek, Greg and Vennin, Vincent",
    title = "{In-in formalism for the entropy of quantum fields in curved spacetimes}",
    eprint = "2406.17856",
    archivePrefix = "arXiv",
    primaryClass = "hep-th",
    month = "6",
    year = "2024"
}

@inproceedings{Colas:2024lse,
    author = "Colas, Thomas",
    title = "{Open Effective Field Theories for cosmology}",
    booktitle = "{58th Rencontres de Moriond on Cosmology}",
    eprint = "2405.09639",
    archivePrefix = "arXiv",
    primaryClass = "astro-ph.CO",
    month = "5",
    year = "2024"
}

@phdthesis{Micheli:2023qnc,
    author = "Micheli, Amaury",
    title = "{Entanglement and decoherence in cosmology and in analogue gravity experiments}",
    reportNumber = "tel-04260012, 2023UPASP078",
    school = "IJCLab, Orsay",
    year = "2023"
}

@phdthesis{Colas:2023wxa,
    author = "Colas, Thomas",
    title = "{Open Effective Field Theories for primordial cosmology : dissipation, decoherence and late-time resummation of cosmological inhomogeneities}",
    reportNumber = "tel-04195628, 2023UPASP075",
    school = "Institut d'astrophysique spatiale, France, AstroParticule et Cosmologie, France, APC, Paris",
    year = "2023"
}

@article{Ning:2023ybc,
    author = "Ning, Sirui and Sou, Chon Man and Wang, Yi",
    title = "{On the decoherence of primordial gravitons}",
    eprint = "2305.08071",
    archivePrefix = "arXiv",
    primaryClass = "hep-th",
    doi = "10.1007/JHEP06(2023)101",
    journal = "JHEP",
    volume = "06",
    pages = "101",
    year = "2023"
}

@article{Kaplanek:2019dqu,
    author = "Kaplanek, Greg and Burgess, C. P.",
    title = "{Hot Accelerated Qubits: Decoherence, Thermalization, Secular Growth and Reliable Late-time Predictions}",
    eprint = "1912.12951",
    archivePrefix = "arXiv",
    primaryClass = "hep-th",
    doi = "10.1007/JHEP03(2020)008",
    journal = "JHEP",
    volume = "03",
    pages = "008",
    year = "2020"
}

@article{Boyanovsky:2015jen,
    author = "Boyanovsky, D.",
    title = "{Effective field theory during inflation. II. Stochastic dynamics and power spectrum suppression}",
    eprint = "1511.06649",
    archivePrefix = "arXiv",
    primaryClass = "astro-ph.CO",
    doi = "10.1103/PhysRevD.93.043501",
    journal = "Phys. Rev. D",
    volume = "93",
    pages = "043501",
    year = "2016"
}

@article{Cao:2022kjn,
    author = "Cao, Shuyang and Boyanovsky, Daniel",
    title = "{Nonequilibrium dynamics of axionlike particles: The quantum master equation}",
    eprint = "2212.05161",
    archivePrefix = "arXiv",
    primaryClass = "astro-ph.CO",
    doi = "10.1103/PhysRevD.107.063518",
    journal = "Phys. Rev. D",
    volume = "107",
    number = "6",
    pages = "063518",
    year = "2023"
}

@article{Burgess:2015ajz,
    author = "Burgess, C. P. and Holman, R. and Tasinato, G.",
    title = "{Open EFTs, IR effects \textbackslash{}\& late-time resummations: systematic corrections in stochastic inflation}",
    eprint = "1512.00169",
    archivePrefix = "arXiv",
    primaryClass = "gr-qc",
    doi = "10.1007/JHEP01(2016)153",
    journal = "JHEP",
    volume = "01",
    pages = "153",
    year = "2016"
}

@article{Prudhoe:2022pte,
    author = "Prudhoe, Sean and Shandera, Sarah",
    title = "{Classifying the non-time-local and entangling dynamics of an open qubit system}",
    eprint = "2201.07080",
    archivePrefix = "arXiv",
    primaryClass = "quant-ph",
    doi = "10.1007/JHEP02(2023)007",
    journal = "JHEP",
    volume = "02",
    pages = "007",
    year = "2023"
}

@article{Assassi:2012et,
    author = "Assassi, Valentin and Baumann, Daniel and Green, Daniel",
    title = "{Symmetries and Loops in Inflation}",
    eprint = "1210.7792",
    archivePrefix = "arXiv",
    primaryClass = "hep-th",
    reportNumber = "SITP-12-36",
    doi = "10.1007/JHEP02(2013)151",
    journal = "JHEP",
    volume = "02",
    pages = "151",
    year = "2013"
}

@article{Morikawa:1989xz,
    author = "Morikawa, Masahiro",
    title = "{Dissipation and Fluctuation of Quantum Fields in Expanding Universes}",
    reportNumber = "Print-89-0564 (BRITISH COLUMBIA)",
    doi = "10.1103/PhysRevD.42.1027",
    journal = "Phys. Rev. D",
    volume = "42",
    pages = "1027--1034",
    year = "1990"
}

@article{Calzetta:1995ys,
    author = "Calzetta, E. and Hu, B. L.",
    title = "{Quantum fluctuations, decoherence of the mean field, and structure formation in the early universe}",
    eprint = "gr-qc/9505046",
    archivePrefix = "arXiv",
    reportNumber = "IASSNS-HEP-95-38, UMD-PP-95-083",
    doi = "10.1103/PhysRevD.52.6770",
    journal = "Phys. Rev. D",
    volume = "52",
    pages = "6770--6788",
    year = "1995"
}

@article{DaddiHammou:2022itk,
    author = "Daddi Hammou, Aoumeur and Bartolo, Nicola",
    title = "{Cosmic decoherence: primordial power spectra and non-Gaussianities}",
    eprint = "2211.07598",
    archivePrefix = "arXiv",
    primaryClass = "astro-ph.CO",
    doi = "10.1088/1475-7516/2023/04/055",
    journal = "JCAP",
    volume = "04",
    pages = "055",
    year = "2023"
}

@book{banerjee2018open,
  title={Open Quantum Systems: Dynamics of Nonclassical Evolution},
  author={Banerjee, S.},
  isbn={9789811331824},
  series={Texts and Readings in Physical Sciences},
  url={https://books.google.co.uk/books?id=s-11DwAAQBAJ},
  year={2018},
  publisher={Springer Nature Singapore}
}

@article{hollowood2017decoherence,
    author = "Hollowood, T. J. and McDonald, J. I.",
    title = "{Decoherence, discord and the quantum master equation for cosmological perturbations}",
    eprint = "1701.02235",
    archivePrefix = "arXiv",
    primaryClass = "gr-qc",
    doi = "10.1103/PhysRevD.95.103521",
    journal = "Phys. Rev. D",
    volume = "95",
    number = "10",
    pages = "103521",
    year = "2017"
}

@article{starobinsky1980new,
  title={A new type of isotropic cosmological models without singularity},
  author={Starobinsky, Alexei A},
  journal={Physics Letters B},
  volume={91},
  number={1},
  pages={99--102},
  year={1980},
  publisher={Elsevier}
}

@article{fang1980entropy,
  title={Entropy generation in the early universe by dissipative processes near the Higgs phase transition},
  author={Fang, LZ},
  journal={Physics Letters B},
  volume={95},
  number={1},
  pages={154--156},
  year={1980},
  publisher={Elsevier}
}

@article{Guth:1980zm,
    author = "Guth, Alan H.",
    editor = "Fang, Li-Zhi and Ruffini, R.",
    title = "{The Inflationary Universe: A Possible Solution to the Horizon and Flatness Problems}",
    reportNumber = "SLAC-PUB-2576",
    doi = "10.1103/PhysRevD.23.347",
    journal = "Phys. Rev. D",
    volume = "23",
    pages = "347--356",
    year = "1981"
}

@article{Sato:1981qmu,
    author = "Sato, Katsuhiko",
    title = "{First-order phase transition of a vacuum and the expansion of the Universe}",
    doi = "10.1093/mnras/195.3.467",
    journal = "Mon. Not. Roy. Astron. Soc.",
    volume = "195",
    number = "3",
    pages = "467--479",
    year = "1981"
}

@article{Linde:1981mu,
    author = "Linde, Andrei D.",
    editor = "Fang, Li-Zhi and Ruffini, R.",
    title = "{A New Inflationary Universe Scenario: A Possible Solution of the Horizon, Flatness, Homogeneity, Isotropy and Primordial Monopole Problems}",
    reportNumber = "LEBEDEV-81-229",
    doi = "10.1016/0370-2693(82)91219-9",
    journal = "Phys. Lett. B",
    volume = "108",
    pages = "389--393",
    year = "1982"
}

@article{Albrecht:1982wi,
    author = "Albrecht, Andreas and Steinhardt, Paul J.",
    editor = "Fang, Li-Zhi and Ruffini, R.",
    title = "{Cosmology for Grand Unified Theories with Radiatively Induced Symmetry Breaking}",
    reportNumber = "UPR-0185T",
    doi = "10.1103/PhysRevLett.48.1220",
    journal = "Phys. Rev. Lett.",
    volume = "48",
    pages = "1220--1223",
    year = "1982"
}

@article{Mukhanov:1982nu,
    author = "Mukhanov, Viatcheslav F. and Chibisov, G. V.",
    title = "{The Vacuum energy and large scale structure of the universe}",
    journal = "Sov. Phys. JETP",
    volume = "56",
    pages = "258--265",
    year = "1982"
}

@article{senatore2010loops,
    author = "Senatore, Leonardo and Zaldarriaga, Matias",
    title = "{On Loops in Inflation}",
    eprint = "0912.2734",
    archivePrefix = "arXiv",
    primaryClass = "hep-th",
    doi = "10.1007/JHEP12(2010)008",
    journal = "JHEP",
    volume = "12",
    pages = "008",
    year = "2010"
}

@article{Starobinsky:1979ty,
    author = "Starobinsky, Alexei A.",
    editor = "Khalatnikov, I. M. and Mineev, V. P.",
    title = "{Spectrum of relict gravitational radiation and the early state of the universe}",
    journal = "JETP Lett.",
    volume = "30",
    pages = "682--685",
    year = "1979"
}

@article{crossley2017effective,
    author = "Crossley, Michael and Glorioso, Paolo and Liu, Hong",
    title = "{Effective field theory of dissipative fluids}",
    eprint = "1511.03646",
    archivePrefix = "arXiv",
    primaryClass = "hep-th",
    reportNumber = "MIT-CTP-4734",
    doi = "10.1007/JHEP09(2017)095",
    journal = "JHEP",
    volume = "09",
    pages = "095",
    year = "2017"
}

@article{shandera2018open,
    author = "Shandera, Sarah and Agarwal, Nishant and Kamal, Archana",
    title = "{Open quantum cosmological system}",
    eprint = "1708.00493",
    archivePrefix = "arXiv",
    primaryClass = "hep-th",
    doi = "10.1103/PhysRevD.98.083535",
    journal = "Phys. Rev. D",
    volume = "98",
    number = "8",
    pages = "083535",
    year = "2018"
}

@article{burgess2008decoherence,
    author = "Burgess, Cliff P. and Holman, R. and Hoover, D.",
    title = "{Decoherence of inflationary primordial fluctuations}",
    eprint = "astro-ph/0601646",
    archivePrefix = "arXiv",
    reportNumber = "MCMASTER-06-01",
    doi = "10.1103/PhysRevD.77.063534",
    journal = "Phys. Rev. D",
    volume = "77",
    pages = "063534",
    year = "2008"
}

@book{calzetta2009nonequilibrium,
  title={Nonequilibrium quantum field theory},
  author={Calzetta, Esteban A and Hu, Bei-Lok B},
  year={2009},
  publisher={Cambridge University Press}
}

@article{martin2018non,
    author = "Martin, J\'er\^ome and Vennin, Vincent",
    title = "{Non Gaussianities from Quantum Decoherence during Inflation}",
    eprint = "1805.05609",
    archivePrefix = "arXiv",
    primaryClass = "astro-ph.CO",
    doi = "10.1088/1475-7516/2018/06/037",
    journal = "JCAP",
    volume = "06",
    pages = "037",
    year = "2018"
}

@article{banerjee2023thermalization,
    author = "Banerjee, Subhashish and Choudhury, Sayantan and Chowdhury, Satyaki and Knaute, Johannes and Panda, Sudhakar and Shirish, K.",
    title = "{Thermalization in quenched open quantum cosmology}",
    eprint = "2104.10692",
    archivePrefix = "arXiv",
    primaryClass = "hep-th",
    doi = "10.1016/j.nuclphysb.2023.116368",
    journal = "Nucl. Phys. B",
    volume = "996",
    pages = "116368",
    year = "2023"
}

@article{Andersen:2021lii,
    author = "Andersen, Jens O. and Eriksson, Magdalena and Tranberg, Anders",
    title = "{Stochastic inflation from quantum field theory and the parametric dependence of the effective noise amplitude}",
    eprint = "2111.14503",
    archivePrefix = "arXiv",
    primaryClass = "hep-ph",
    doi = "10.1007/JHEP02(2022)121",
    journal = "JHEP",
    volume = "02",
    pages = "121",
    year = "2022"
}

@article{Boyanovsky:2015xoa,
    author = "Boyanovsky, D.",
    title = "{Effective Field Theory out of Equilibrium: Brownian quantum fields}",
    eprint = "1503.00156",
    archivePrefix = "arXiv",
    primaryClass = "hep-ph",
    doi = "10.1088/1367-2630/17/6/063017",
    journal = "New J. Phys.",
    volume = "17",
    number = "6",
    pages = "063017",
    year = "2015"
}

@article{Choudhury:2018ppd,
    author = "Choudhury, Sayantan and Panda, Sudhakar",
    title = "{Cosmological Spectrum of Two-Point Correlation Function from Vacuum Fluctuation of Stringy Axion Field in De Sitter Space: A Study of the Role of Quantum Entanglement}",
    eprint = "1809.02905",
    archivePrefix = "arXiv",
    primaryClass = "hep-th",
    doi = "10.3390/universe6060079",
    journal = "Universe",
    volume = "6",
    number = "6",
    pages = "79",
    year = "2020"
}

@article{Kading:2022jjl,
    author = {K\"ading, Christian and Pitschmann, Mario},
    title = "{New method for directly computing reduced density matrices}",
    eprint = "2204.08829",
    archivePrefix = "arXiv",
    primaryClass = "hep-th",
    doi = "10.1103/PhysRevD.107.016005",
    journal = "Phys. Rev. D",
    volume = "107",
    number = "1",
    pages = "016005",
    year = "2023"
}

@article{Alicki:2023tfz,
    author = "Alicki, Robert and Barenboim, Gabriela and Jenkins, Alejandro",
    title = "{The irreversible relaxation of inflation}",
    eprint = "2307.04803",
    archivePrefix = "arXiv",
    primaryClass = "gr-qc",
    reportNumber = "IFIC/23-29",
    month = "7",
    year = "2023"
}

@article{Creminelli:2023aly,
    author = "Creminelli, Paolo and Kumar, Soubhik and Salehian, Borna and Santoni, Luca",
    title = "{Dissipative inflation via scalar production}",
    eprint = "2305.07695",
    archivePrefix = "arXiv",
    primaryClass = "hep-th",
    doi = "10.1088/1475-7516/2023/08/076",
    journal = "JCAP",
    volume = "08",
    pages = "076",
    year = "2023"
}

@article{Keefe:2024cia,
    author = "Keefe, A. and Agarwal, N. and Kamal, A.",
    title = "{Quantifying spectral signatures of non-Markovianity beyond the Born-Redfield master equation}",
    eprint = "2405.01722",
    archivePrefix = "arXiv",
    primaryClass = "quant-ph",
    month = "5",
    year = "2024"
}

@article{Seery:2007we,
    author = "Seery, David",
    title = "{One-loop corrections to a scalar field during inflation}",
    eprint = "0707.3377",
    archivePrefix = "arXiv",
    primaryClass = "astro-ph",
    doi = "10.1088/1475-7516/2007/11/025",
    journal = "JCAP",
    volume = "11",
    pages = "025",
    year = "2007"
}

@article{nambu1988stochastic,
  title={Stochastic stage of an inflationary universe model},
  author={Nambu, Yasusada and Sasaki, Misao},
  journal={Physics Letters B},
  volume={205},
  number={4},
  pages={441--446},
  year={1988},
  publisher={Elsevier}
}

@article{nambu1989stochastic,
  title={Stochastic approach to chaotic inflation and the distribution of universes},
  author={Nambu, Yasusada and Sasaki, Misao},
  journal={Physics Letters B},
  volume={219},
  number={2-3},
  pages={240--246},
  year={1989},
  publisher={Elsevier}
}

@article{kandrup1989stochastic,
  title={Stochastic inflation as a time-dependent random walk},
  author={Kandrup, Henry E},
  journal={Physical Review D},
  volume={39},
  number={8},
  pages={2245},
  year={1989},
  publisher={APS}
}

@article{nakao1988stochastic,
  title={Stochastic dynamics of new inflation},
  author={Nakao, Ken-ichi and Nambu, Yasusada and Sasaki, Misao},
  journal={Progress of Theoretical Physics},
  volume={80},
  number={6},
  pages={1041--1068},
  year={1988},
  publisher={Oxford University Press}
}

@article{mollerach1991stochastic,
  title={Stochastic inflation in a simple two-field model},
  author={Mollerach, Silvia and Matarrese, Sabino and Ortolan, Antonello and Lucchin, Francesco},
  journal={Physical Review D},
  volume={44},
  number={6},
  pages={1670},
  year={1991},
  publisher={APS}
}

@article{linde1994big,
  title={From the Big Bang theory to the theory of a stationary universe},
  author={Linde, Andrei and Linde, Dmitri and Mezhlumian, Arthur},
  journal={Physical Review D},
  volume={49},
  number={4},
  pages={1783},
  year={1994},
  publisher={APS}
}

@article{starobinsky1994equilibrium,
  title={Equilibrium state of a self-interacting scalar field in the de Sitter background},
  author={Starobinsky, Alexei A and Yokoyama, Jun’ichi},
  journal={Physical Review D},
  volume={50},
  number={10},
  pages={6357},
  year={1994},
  publisher={APS}
}

@article{Bhattacharya:2023twz,
    author = "Bhattacharya, Sourav and Joshi, Nitin",
    title = "{Decoherence and entropy generation at one loop in the inflationary de Sitter spacetime for Yukawa interaction}",
    eprint = "2307.13443",
    archivePrefix = "arXiv",
    primaryClass = "hep-th",
    doi = "10.1088/1475-7516/2024/04/078",
    journal = "JCAP",
    volume = "04",
    pages = "078",
    year = "2024"
}

@article{Bhattacharya:2022wpe,
    author = "Bhattacharya, Sourav and Joshi, Nitin and Kaushal, Shagun",
    title = "{Decoherence and entropy generation in an open quantum scalar-fermion system with Yukawa interaction}",
    eprint = "2206.15045",
    archivePrefix = "arXiv",
    primaryClass = "hep-th",
    doi = "10.1140/epjc/s10052-023-11357-6",
    journal = "Eur. Phys. J. C",
    volume = "83",
    number = "3",
    pages = "208",
    year = "2023"
}

@article{Bhattacharya:2023xvd,
    author = "Bhattacharya, Sourav and Joshi, Nitin and Roy, Kinsuk",
    title = "{Resummation of local and non-local scalar self energies via the Schwinger-Dyson equation in de Sitter spacetime}",
    eprint = "2310.19436",
    archivePrefix = "arXiv",
    primaryClass = "hep-th",
    month = "10",
    year = "2023"
}

@article{Chandran:2018wwc,
    author = "Chandran, S. Mahesh and Shankaranarayanan, S.",
    title = "{Divergence of entanglement entropy in quantum systems: Zero-modes}",
    eprint = "1810.03888",
    archivePrefix = "arXiv",
    primaryClass = "quant-ph",
    doi = "10.1103/PhysRevD.99.045010",
    journal = "Phys. Rev. D",
    volume = "99",
    number = "4",
    pages = "045010",
    year = "2019"
}

@article{brandenberger1990classical,
  title={Classical perturbations from decoherence of quantum fluctuations in the inflationary universe},
  author={Brandenberger, Robert and Laflamme, Raymond and Miji{\'c}, Milan},
  journal={Modern Physics Letters A},
  volume={5},
  number={28},
  pages={2311--2317},
  year={1990},
  publisher={World Scientific}
}

@article{barvinsky1999decoherence,
    author = "Barvinsky, A. O. and Kamenshchik, A. Yu. and Kiefer, C. and Mishakov, I. V.",
    title = "{Decoherence in quantum cosmology at the onset of inflation}",
    eprint = "gr-qc/9812043",
    archivePrefix = "arXiv",
    reportNumber = "FREIBURG-THEP-98-26",
    doi = "10.1016/S0550-3213(99)00208-4",
    journal = "Nucl. Phys. B",
    volume = "551",
    pages = "374--396",
    year = "1999"
}

@article{lombardo2005influence,
    author = "Lombardo, Fernando C.",
    editor = "Elze, Hans-Thomas",
    title = "{Influence functional approach to decoherence during inflation}",
    eprint = "gr-qc/0412069",
    archivePrefix = "arXiv",
    doi = "10.1590/S0103-97332005000300005",
    journal = "Braz. J. Phys.",
    volume = "35",
    pages = "391--396",
    year = "2005"
}

@article{lombardo2005decoherence,
    author = "Lombardo, Fernando C. and Lopez Nacir, Diana",
    title = "{Decoherence during inflation: The Generation of classical inhomogeneities}",
    eprint = "gr-qc/0506051",
    archivePrefix = "arXiv",
    doi = "10.1103/PhysRevD.72.063506",
    journal = "Phys. Rev. D",
    volume = "72",
    pages = "063506",
    year = "2005"
}

@article{martineau2007decoherence,
    author = "Martineau, Patrick",
    title = "{On the decoherence of primordial fluctuations during inflation}",
    eprint = "astro-ph/0601134",
    archivePrefix = "arXiv",
    doi = "10.1088/0264-9381/24/23/006",
    journal = "Class. Quant. Grav.",
    volume = "24",
    pages = "5817--5834",
    year = "2007"
}

@article{prokopec2007decoherence,
    author = "Prokopec, Tomislav and Rigopoulos, Gerasimos I.",
    title = "{Decoherence from Isocurvature perturbations in Inflation}",
    eprint = "astro-ph/0612067",
    archivePrefix = "arXiv",
    reportNumber = "ITP-UU-06-51, SPIN-06-41",
    doi = "10.1088/1475-7516/2007/11/029",
    journal = "JCAP",
    volume = "11",
    pages = "029",
    year = "2007"
}

@article{sharman2007decoherence,
    author = "Sharman, Jonathan W. and Moore, Guy D.",
    title = "{Decoherence due to the Horizon after Inflation}",
    eprint = "0708.3353",
    archivePrefix = "arXiv",
    primaryClass = "gr-qc",
    doi = "10.1088/1475-7516/2007/11/020",
    journal = "JCAP",
    volume = "11",
    pages = "020",
    year = "2007"
}

@article{campo2008decoherence,
    author = "Campo, David and Parentani, Renaud",
    title = "{Decoherence and entropy of primordial fluctuations. I: Formalism and interpretation}",
    eprint = "0805.0548",
    archivePrefix = "arXiv",
    primaryClass = "hep-th",
    doi = "10.1103/PhysRevD.78.065044",
    journal = "Phys. Rev. D",
    volume = "78",
    pages = "065044",
    year = "2008"
}

@article{anastopoulos2013master,
    author = "Anastopoulos, C. and Hu, B. L.",
    title = "{A Master Equation for Gravitational Decoherence: Probing the Textures of Spacetime}",
    eprint = "1305.5231",
    archivePrefix = "arXiv",
    primaryClass = "gr-qc",
    doi = "10.1088/0264-9381/30/16/165007",
    journal = "Class. Quant. Grav.",
    volume = "30",
    pages = "165007",
    year = "2013"
}

@article{Nelson:2016kjm,
    author = "Nelson, Elliot",
    title = "{Quantum Decoherence During Inflation from Gravitational Nonlinearities}",
    eprint = "1601.03734",
    archivePrefix = "arXiv",
    primaryClass = "gr-qc",
    doi = "10.1088/1475-7516/2016/03/022",
    journal = "JCAP",
    volume = "03",
    pages = "022",
    year = "2016"
}

@article{Oppenheim:2022xjr,
    author = "Oppenheim, Jonathan and Sparaciari, Carlo and \v{S}oda, Barbara and Weller-Davies, Zachary",
    title = "{Gravitationally induced decoherence vs space-time diffusion: testing the quantum nature of gravity}",
    eprint = "2203.01982",
    archivePrefix = "arXiv",
    primaryClass = "quant-ph",
    doi = "10.1038/s41467-023-43348-2",
    journal = "Nature Commun.",
    volume = "14",
    number = "1",
    pages = "7910",
    year = "2023"
}

@article{Sharifian:2023jem,
    author = "Sharifian, Mohammad and Zarei, Moslem and Abdi, Mehdi and Bartolo, Nicola and Matarrese, Sabino",
    title = "{Open quantum system approach to the gravitational decoherence of spin-1/2 particles}",
    eprint = "2309.07236",
    archivePrefix = "arXiv",
    primaryClass = "gr-qc",
    doi = "10.1103/PhysRevD.109.043510",
    journal = "Phys. Rev. D",
    volume = "109",
    number = "4",
    pages = "043510",
    year = "2024"
}

@article{Berera:1995ie,
    author = "Berera, Arjun",
    title = "{Warm inflation}",
    eprint = "astro-ph/9509049",
    archivePrefix = "arXiv",
    reportNumber = "PSU-TH-159",
    doi = "10.1103/PhysRevLett.75.3218",
    journal = "Phys. Rev. Lett.",
    volume = "75",
    pages = "3218--3221",
    year = "1995"
}

@article{Maldacena:2002vr,
    author = "Maldacena, Juan Martin",
    title = "{Non-Gaussian features of primordial fluctuations in single field inflationary models}",
    eprint = "astro-ph/0210603",
    archivePrefix = "arXiv",
    doi = "10.1088/1126-6708/2003/05/013",
    journal = "JHEP",
    volume = "05",
    pages = "013",
    year = "2003"
}

@article{Bernardeau:2001qr,
    author = "Bernardeau, F. and Colombi, S. and Gaztanaga, E. and Scoccimarro, R.",
    title = "{Large scale structure of the universe and cosmological perturbation theory}",
    eprint = "astro-ph/0112551",
    archivePrefix = "arXiv",
    reportNumber = "SACLAY-T01-142",
    doi = "10.1016/S0370-1573(02)00135-7",
    journal = "Phys. Rept.",
    volume = "367",
    pages = "1--248",
    year = "2002"
}

@article{Susskind:2014rva,
    author = "Susskind, Leonard",
    title = "{Computational Complexity and Black Hole Horizons}",
    eprint = "1403.5695",
    archivePrefix = "arXiv",
    primaryClass = "hep-th",
    doi = "10.1002/prop.201500092",
    journal = "Fortsch. Phys.",
    volume = "64",
    pages = "24--43",
    year = "2016",
    note = "[Addendum: Fortsch.Phys. 64, 44--48 (2016)]"
}

@article{Nielsen:2005mkt,
    author = "Nielsen, Michael A.",
    title = "{A geometric approach to quantum circuit lower bounds}",
    eprint = "quant-ph/0502070",
    archivePrefix = "arXiv",
    doi = "10.26421/QIC6.3-2",
    journal = "Quant. Inf. Comput.",
    volume = "6",
    number = "3",
    pages = "213--262",
    year = "2006"
}

@article{Ryu:2006bv,
    author = "Ryu, Shinsei and Takayanagi, Tadashi",
    title = "{Holographic derivation of entanglement entropy from AdS/CFT}",
    eprint = "hep-th/0603001",
    archivePrefix = "arXiv",
    reportNumber = "NSF-KITP-06-11",
    doi = "10.1103/PhysRevLett.96.181602",
    journal = "Phys. Rev. Lett.",
    volume = "96",
    pages = "181602",
    year = "2006"
}

@article{Boyanovsky:1996rw,
    author = "Boyanovsky, D. and Cormier, D. and de Vega, H. J. and Holman, R.",
    title = "{Out-of-equilibrium dynamics of an inflationary phase transition}",
    eprint = "hep-ph/9610396",
    archivePrefix = "arXiv",
    reportNumber = "PITT-96-7777-7, CMU-HEP-96-11, LPTHE-96-19, DOE-ER-40682-122, PITT-96-7777.7, DOR-ER-40682-122",
    doi = "10.1103/PhysRevD.55.3373",
    journal = "Phys. Rev. D",
    volume = "55",
    pages = "3373--3388",
    year = "1997"
}

@article{Bartolo:2004if,
    author = "Bartolo, N. and Komatsu, E. and Matarrese, Sabino and Riotto, A.",
    title = "{Non-Gaussianity from inflation: Theory and observations}",
    eprint = "astro-ph/0406398",
    archivePrefix = "arXiv",
    reportNumber = "DFPD-04-A-12",
    doi = "10.1016/j.physrep.2004.08.022",
    journal = "Phys. Rept.",
    volume = "402",
    pages = "103--266",
    year = "2004"
}

@article{Cheung:2007st,
    author = "Cheung, Clifford and Creminelli, Paolo and Fitzpatrick, A. Liam and Kaplan, Jared and Senatore, Leonardo",
    title = "{The Effective Field Theory of Inflation}",
    eprint = "0709.0293",
    archivePrefix = "arXiv",
    primaryClass = "hep-th",
    reportNumber = "IC-2007-032",
    doi = "10.1088/1126-6708/2008/03/014",
    journal = "JHEP",
    volume = "03",
    pages = "014",
    year = "2008"
}

@article{Dias:2012qy,
    author = "Dias, Mafalda and Ribeiro, Raquel H. and Seery, David",
    title = "{The \ensuremath{\delta}N formula is the dynamical renormalization group}",
    eprint = "1210.7800",
    archivePrefix = "arXiv",
    primaryClass = "astro-ph.CO",
    doi = "10.1088/1475-7516/2013/10/062",
    journal = "JCAP",
    volume = "10",
    pages = "062",
    year = "2013"
}

@article{Weinberg:2005vy,
    author = "Weinberg, Steven",
    title = "{Quantum contributions to cosmological correlations}",
    eprint = "hep-th/0506236",
    archivePrefix = "arXiv",
    reportNumber = "UTTG-01-05",
    doi = "10.1103/PhysRevD.72.043514",
    journal = "Phys. Rev. D",
    volume = "72",
    pages = "043514",
    year = "2005"
}

@article{Weinberg:2006ac,
    author = "Weinberg, Steven",
    title = "{Quantum contributions to cosmological correlations. II. Can these corrections become large?}",
    eprint = "hep-th/0605244",
    archivePrefix = "arXiv",
    reportNumber = "UTTG-0306",
    doi = "10.1103/PhysRevD.74.023508",
    journal = "Phys. Rev. D",
    volume = "74",
    pages = "023508",
    year = "2006"
}

@article{Anderson:2005hi,
    author = "Anderson, Paul R. and Molina-Paris, Carmen and Mottola, Emil",
    title = "{Short distance and initial state effects in inflation: Stress tensor and decoherence}",
    eprint = "hep-th/0504134",
    archivePrefix = "arXiv",
    reportNumber = "LA-UR-04-6142",
    doi = "10.1103/PhysRevD.72.043515",
    journal = "Phys. Rev. D",
    volume = "72",
    pages = "043515",
    year = "2005"
}

\end{document}